# Proceedings of the 35th Annual Condensed Matter and Materials Meeting

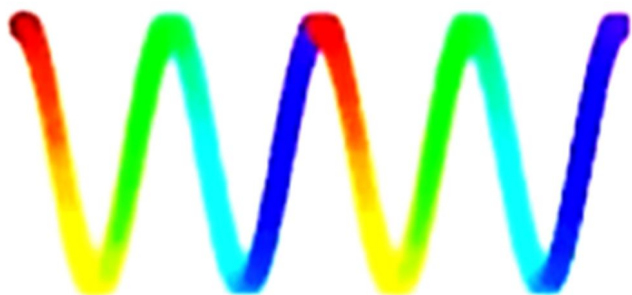

2011

Charles Sturt University
Wagga Wagga NSW
1st – 4th February 2011

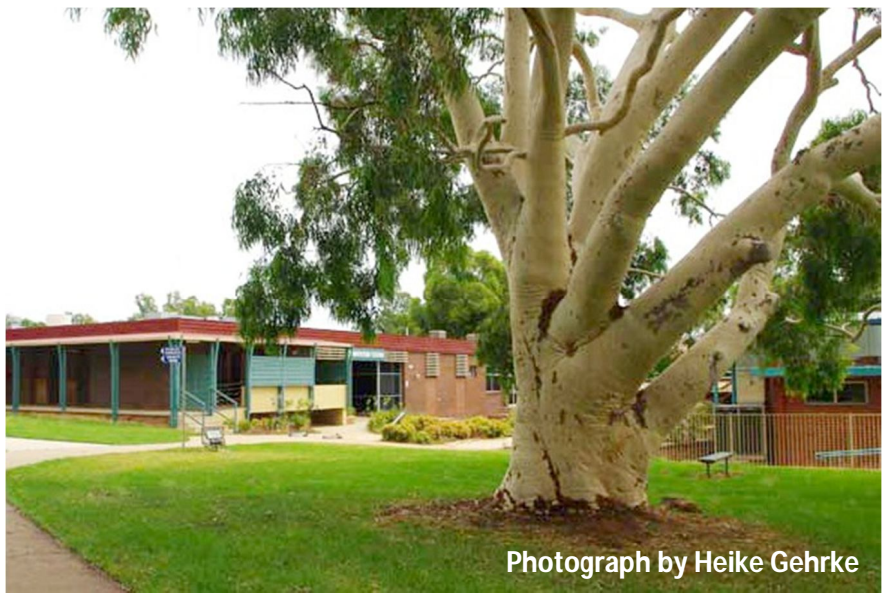

Photograph by Heike Gehrke

ISBN: 978-0-646-55969-8

Editorial Note – 35th Annual Condensed Matter and Materials Meeting – Wagga Wagga NSW, 2011.

# Proceedings of Wagga 2011

The 35th Annual Condensed Matter and Materials Meeting

ISBN: 978-0-646-55969-8

Editor: Adam Micolich

The 35th Annual Condensed Matter and Materials Meeting was held at the Charles Sturt University campus in Wagga Wagga NSW from the 1st to the 4th of February 2011. The conference was attended by 92 delegates from a range of universities across Australia and New Zealand, and countries as distant as Turkey.

There were a total of 9 invited and 21 contributed talks during the three days of scientific sessions, as well as two poster sessions with a total of 49 poster presentations. The conference also featured a number of very enjoyable social sessions including the annual trivia quiz, with "*No Idea*"* winning the Lindsay Davis cup in this year's competition.

All presenters were invited to submit a manuscript for publication in the conference proceedings. The length limits where six pages for invited papers and four pages for contributed papers.

Each manuscript was reviewed by two anonymous referees who worked to a set of guidelines made available by the editor. Each accepted publication therefore satisfies the requirements for classification as a refereed conference publication (E1).

The organizers would like to thank the 21 reviewers for their time and effort in reviewing manuscripts, which resulted in 18 papers being accepted for publication. The accepted manuscripts are available at the online publication section of the Australian Institute of Physics national web site (http://www.aip.org.au/).

**Wagga 11 Organising Committee:** Jaan Oitmaa, Chris Hamer, Clemens Ulrich, Adam Micolich, Marion Stevens-Kalceff, Oleg Sushkov, Alex Hamilton, Michelle Simmons.

**Correspondence:** mico@phys.unsw.edu.au

**Date:** 28 June 2011

Team members: Maciej Bartkowiak, Paul Gubbens, Chris Hamer, Stephen Harker, Annamieke Mulders, Jeff Sellar, Supitcha Supansomboon and Clemens Ulrich.

# Proceedings Contents

# WW2011   SCIENTIFIC PROGRAM

## Wednesday Morning, 2 February

| | | | |
|---|---|---|---|
| **08:45 – 09:00** | | **Opening:** *J. Oitmaa, UNSW* | |
| **09:00 – 10:30** | **W-I** | **Chairperson:** *G.A. Stewart, Australian Defense Force Academy* | |
| 09:00 – 09:30 | W1 | 100 Years of Superconductivity, 25 Years of HTS<br>*J.L. Tallon, MacDiarmid Institute, Lower Hutt, New Zealand* | *INVITED* |
| 09:30 – 10:00 | W2 | Superconductivity: From Zero Resistance to Terahertz Devices<br>*J.C. Macfarlane, CSIRO Materials Science and Engineering* | |
| 10:00 – 10:30 | W3 | The Australian Synchrotron and condensed matter science<br>*D.J. Cookson, Australian Synchrotron* | *INVITED* |
| **10:30 – 10:50** | | **Morning tea** | |
| **10:50 – 12:20** | **W-II** | **Chairperson:** T. Soehnel, University of Auckland | |
| 10:50 – 11:20 | W4 | Compound Semiconductor Nanowires for Next Generation Optoelectronic Devices<br>*C. Jagadish, The Australian National University* | *INVITED* |
| 11:20 – 11:40 | W5 | Terahertz generation from high index GaAs planes at different angles of incidence<br>*K. Radhanpura, University of Wollongong* | |
| 11:40 – 12:00 | W6 | Nitrogen Doping and In-situ Heat Treatment of Carbon Nitride Thin Films<br>*D.W.M. Lau, University of Melbourne* | |
| 12:00 – 12.20 | W7 | Single dopant transport spectroscopy in silicon<br>*J. Verduijn, S. Rogge, Delft University of Technology, UNSW* | |
| **12:20 – 14.00** | | **Lunch** | |
| **14:00 – 15:30** | **W-III** | **Chairperson:** *M.B. Cortie, University of Technology Sydney* | |
| 14:00 – 14:30 | W8 | Engineered quantum systems<br>*G. Milburn, University of Queensland* | *INVITED* |
| 14:30 – 14:50 | W9 | Vacancies and Void Formation near Si/SiO$_2$ Interface<br>*R. Weed, The Australian National University* | |
| 14:50 – 15:10 | W10 | Nd-Eu magnetic interactions in Nd$^{3+}$:EuCl$_3$.6H$_2$O<br>*R.L. Ahlefeldt, The Australian National University* | |
| 15:10 – 15:30 | W11 | Closing the gap: The influence of relativistic effects on the band structure of HgSe and HgTe<br>*S. Biering, Massey University Albany* | |
| **15:30 – 16:00** | | **Afternoon Tea** | |
| **16:00 – 18:00** | | **Poster Session:** WP1 – WP25 | |
| **18:30 – 22:00** | | **Conference Dinner** | |

# Thursday Morning, 3 February

**09:00 – 10:30**    **T-I**    **Chairperson:** *O.P. Sushkov, UNSW*

| | | | |
|---|---|---|---|
| 09:00 – 09:30 | T1 | Magnetic domain wall dynamics: from inkblots to spin torque | *INVITED* |
| | | P.J. Metaxas, University of Western Australia | |
| 09:30 – 09:50 | T2 | Inelastic Neutron Scattering and EPR Studies of Cobalt Dimers | |
| | | R.A. Mole, ANSTO, The Bragg Institute | |
| 09:50 – 10:10 | T3 | Structural and magnetic phase separation in $PrMn_2Ge_{2-x}Si_x$ compounds | |
| | | J.L. Wang, S.J. Kennedy, University of Wollongong, ANSTO | |
| 10:10 – 10:30 | T4 | Temperature dependence of the spontaneous remagnetization in $Nd_{60}Fe_{30}Al_{10}$ and $Nd_{60}Fe_{20}Co_{10}Al_{10}$ bulk amorphous ferromagnets | |
| | | S.J. Collocott, CSIRO Materials Science and Engineering | |

**10:30 – 10:50**    **Morning tea**

**10:50 – 12:30**    **T-II**    **Chairperson:** *K.-D. Liss, The Bragg Institute / ANSTO*

| | | | |
|---|---|---|---|
| 10:50 – 11:20 | T5 | Engineering graphene growth | *INVITED* |
| | | N. Medhekar, Monash University | |
| 11:20 – 11:40 | T6 | 101 uses for the nitrogen-vacancy centre in diamond | |
| | | N. Manson, Australian National University | |
| 11:40 – 12:00 | T7 | Hard-ball modelling of BCC to closest-packed transition in nanoscale shape memory alloy actuators | |
| | | M.B. Cortie, University of Technology Sydney | |
| 12:00 – 12.30 | T8 | Structural variety in brownmillerite-type materials | *INVITED* |
| | | H. Krüger, The Australian National University, University of Innsbruck | |

**12:30 – 14.00**    **Lunch**

**14:00 – 15:30**    **T-III**    **Chairperson:** *A.J. Hill, CSIRO Materials Science and Engineering*

| | | |
|---|---|---|
| 14:00 – 14:20 | T9 | The structure of Yttria-Stabilized Zirconia: A combined medium energy photoemission and ab-initio investigation |
| | | G. Cousland, The University of Sydney |
| 14:20 – 14:40 | T10 | Positron Annihilation Lifetime Spectra of Radiation Damage, Neutral Zircon Crystals |
| | | J. Roberts, The Australian National University |
| 14:40 – 15:00 | T11 | Experimental study of diffusion and clustering in aluminum alloys |
| | | M.D.H. Lay, CSIRO Material Science and Engineering |
| 15:00 – 15:20 | T12 | Neutrons and Li-Ion Batteries |
| | | N. Sharma, ANSTO, The Bragg Institute |

**15:20 – 15:50**    **Afternoon Tea**

**16:00 – 18:00**    **Poster Session: TP1 – TP24**

**18:00 – 19:00**    **Dinner**

**20:00 – 22:30**    **Trivia Quiz, Conference Centre**
                               *Quizmaster: Trevor Finlayson, University of Melbourne*

# Friday Morning, 4 February

**09:00 – 10:30**  **F-I**  **Chairperson:** *S.J. Collocott, CSIRO Materials Science and Engineering*

09:00 – 09:30  F1  Advanced resonant X-ray diffraction applied to the study of ordering phenomena in complex oxides  *INVITED*
<u>A.M. Mulders</u>, *Australian Defense Force Academy @UNSW*

09:30 – 09:50  F2  $Cu_5SbO_6$ – Synchrotron, Neutron Diffraction Studies and Magnetic Properties
<u>T. Söhnel</u>, *The University of Auckland*

09:50 – 10:10  F3  Comparison investigation for flux pinning of Titanium and Zirconium doped $Y_1B_2C_3O_{7-\delta}$ films prepared by TFA-MOD
<u>Q. Li</u>, *University of Wollongong*

10:10 – 10:30  F4  Diffuse scattering from PZN ($PbZn_{1/3}Nb_{2/3}O_3$)
<u>R.E. Whitfield</u>, *The Australian National University*

**10:30 – 10:50**  **Morning tea**

**10:50 – 12:00**  **F-II**  **Chairperson:** *R.A. Lewis, University of Wollongong*

10:50 – 11:20  F5  Multilayered Water-Based Organic Photovoltaics
*A. Stapleton, <u>P.C. Dastoor</u>, University of Newcastle*  *INVITED*

11:20 – 11:40  F6  Study on the interface between organic and inorganic semiconductors
<u>A.-U. Rehman</u>, *Zhejiang University, China*

11:40 – 12:00  F7  Slow photon photocatalytic enhancement in titania inverse opal photonic crystals
*V. Jovic, <u>G.I.N Waterhouse</u>, The University of Auckland, Auckland*

**12:00 – 12:20**  **Presentations and Closing**

**12:20 – 14.00**  **Lunch**

# POSTER SESSION: Wednesday 2 February

WP1  A.A. Abiona, W.J. Kemp, A.P. Byrne, M. Ridgway, and H. Timmers
*Possible Pb-vacancy pairing in germanium: Dependence on doping and orientation*

WP2  J.G. Bartholomew, S. Marzban, M.J. Sellars, and R.-P. Wang
*Coherence properties of rare earth ion doped thin films*

WP3  M. Bartkowiak, G.J. Kearley, M. Yethiraj, and A M Mulders
*Ab initio determination of the structure of the ferroelectric phase of $SrTi^{18}O_3$*

WP4  T.J. Bastow, C.R. Hutchinson, A. Deschamps, and A.J. Hill
*Precipitate growth in a mechanically stressed (deformed) Al(Cu,Li,Mg,Ag) alloy observed by $^7Li$, $^{27}Al$, and $^{63}Cu$ NMR and XRD*

WP5  J. Bertinshaw, T. Saerbeck, A. Nelson, M. James, V. Nagarajan, F.Klose, and C. Ulrich
*Studying multiferroic $BiFeO_3$ and ferromagnetic $La_{0.67}Sr_{0.33}MnO_3$ tunnel junctions with Raman spectroscopy and neutron scattering techniques*

WP6  J.D. Cashion, W.P. Gates, T.L. Greaves, and O. Dorjkhaidav
*Identification of $Fe^{3+}$ site coordinations in NAu-2 Nontronite*

WP7  W. Chen and O.P. Sushkov
*Fermi arc – hole pocket dichotomy: effect of spin fluctuation in underdoped cuprates*

WP8  E. Constable and R.A. Lewis
*Continuous-wave terahertz spectroscopy as a non-contact, non-destructive method for characterising semiconductors*

WP9  M. de los Reyes, K.R. Whittle, M. Mitchell, S.E. Ashbrook, and G.R Lumpkin
*Pyrochlore-fluorite transition in $Y_2Sn_{2-x}Zr_xO_7$ - implications for stability*

WP10 B. Deviren, S. Akbudak, and M. Keskin
*Mixed spin-1 and spin-3/2 Ising system with two alternative layers of a honeycomb lattice within the effective-field theory*

WP11 J.B. Dunlop, T.R. Finlayson, and P. Gwan
*Condensed matter and materials trivia*

WP12 C. Feng, H. Li, G. Du, Z. Guo, N. Sharma, V.K. Peterson, and H. Liu
*Non-stoichiometric Mn doping in olivine lithium iron phosphate: structure and electrochemical properties*

WP13 T.R. Finlayson, S. Danilkin, A.J. Studer, and R.E. Whitfield
*Anomalous precursive behaviour for the martensitic material $Ni_{0.625}Al_{0.375}$*

WP14 L.G. Gladkis, H. Timmers, J.M. Scarvell, P.N. Smith
*Reliable shape information on prosthesis wear debris particles from atomic force microscopy*

# POSTER SESSION: Thursday 3 February

TP1  J. Leslie, B. Hillman, and P. Kluth
*Proximity effect on ion track etching in amorphous $SiO_2$*

TP2  T. Li, O.P. Sushkov, and U. Zuelicke
*Spin dynamics and Zeeman splitting of holes in a GaAs point contact*

TP3  K.-D. Liss, D.D. Qu, M. Reid, and J. Shen
*On the atomic anisotropy of thermal expansion in bulk metallic glass*

TP4  Y. Liu and H. Timmers
*Possible lubrication and temperature effects in the microscratching of polyethylene terephthalate*

TP5  A. E. Malik, W.D. Hutchison, K. Nishimura, and R.G. Elliman
*Studies of magnetic nanoparticles formed in $SiO_2$ by ion implantation*

TP6  J. Mao, Z. Guo, and H. Liu
*Effect of hydrogen back pressure on de/rehydrogenation behaviour of $LiBH_4$-$MgH_2$ system and the role of additive toward to enhanced hydrogen sorption properties*

TP7  A.P. Micolich, K. Storm, G. Nylund, and L. Samuelson
*Chemical control of gate length in lateral wrap-gated InAs nanowire FETs*

TP8  J. Oitmaa and A. Brooks Harris
*High temperature thermodynamics of the multiferroic $Ni_3V_2O_8$*

TP9  J. Oitmaa and C.J. Hamer,
*Does the quantum compass model in 3D have a phase transition?*

TP10  J. Oitmaa and O.P. Sushkov
*Scaling of critical temperature and ground state magnetization near a quantum phase transition*

TP11  L.J. Rogers, K.R. Ferguson, and N.B. Manson
*Strain to selectively excite certain orientations of NV centres in diamond*

TP12  S. Sakarya, P.C.M. Gubbens, A. Yaouanc, P. Dalmas de R´eotier, D. Andreica, A. Amato, U. Zimmermann, N. H. van Dijk, E. Brück, Y. Huang, T. Gortenmulder, A. D. Hillier, and P. J. C. King
*Ambient and high pressure μSR measurements on the ferromagnetic superconductor $UGe_2$*

TP13  A. Sokolova
*Small Angle Scattering: instrumentation and applications to study various materials at the nanoscale*

TP14  G.A. Stewart, H. Salama, A. Mulders, D. Scott, and H.StC. O'Neill
*Thermal hysteresis of the $^{169}$Tm quadrupole interaction in orthorhombic thulium manganite*

TP15  S. Supansomboon, A. Dowd, and M.B. Cortie
*Phase relationships in the $PtAl_2$-$AuAl_2$ system*

TP16  W. X. Tang, C. X. Zheng, Z. Y. Zhou, D. E. Jesson, J. Tersoff
*Surface dynamics during Langmuir evaporation of GaAs*

TP17  G.J.Troup, D.R.Hutton, J.Boas, A.Casini, M.Picollo, and Robyn Slogget
*From radiation damage, through minerals and gemstones, to art with EPR*

TP18  J.L. Wang, A.J. Studer, S.J. Campbell, S.J. Kennedy, R. Zeng, and S.X. Dou
*Magnetic structures of $Pr_{1-x}Lu_xMn_2Ge_2$ compounds (x = 0.2 and 0.4)*

TP19  J.A. Warner, L.G. Gladkis, A. E. Kiss, J. Young P.N. Smith, J. Scarvell, and H.Timmers
*Polymer particle production and dispersion in Knee prostheses*

TP20  K.R. Whittle, D.P. Riley, M.G. Blackford, R.D. Aughterson, S. Moricca, G.R. Lumpkin, and N.J. Zaluzec
*$M_{(n+1)}Ax_n$ Phases are they tolerant/resistant to damage*

TP21  W. Xie, H. Ju, J.I. Mardel, A.J. Hill, J.E. McGrath, and B.D. Freeman
*The role of free volume in the tradeoff between high water permeability and high permeability selectivity of polymeric desalination membranes*

TP22  P. Zhang, Z. Guo, and H. Liu
*Electrospinning technology used to synthesize nanomaterials for lithium ion batteries*

TP23  C.X. Zheng, Z.Y. Zhou, W.X. Tang, D.E. Jesson, J. Tersoff, and B. A. Joyce
*Design and application of a III-V surface electron microscope*

TP24  C. Zhong, J.Z. Wang, S.L. Chou, K. Konstantinov and H.K. Liu
*Spray pyrolysis prepared hollow spherical CuO/C: synthesis, characterization, and its application in lithium-ion batteries*

# Terahertz Generation from High Index GaAs Planes at Different Angles of Incidence

K. Radhanpura, S. Hargreaves and R. A. Lewis

*Institute of Superconducting and Electronics Materials, University of Wollongong, Wollongong, NSW 2522, Australia.*

Generation of terahertz radiation from zinc-blende $\bar{4}3m$ crystal planes is presented. Comparison of theory with experimental measurements for high index (11*N*) GaAs crystals for normal and non-normal incidence of incoming radiation gives insight into different mechanisms involved in the generation of terahertz radiation from these crystals.

**1.  Introduction**

Terahertz (THz) frequency radiation, ranging from 0.1-10 THz, can be obtained by excitation of ultra-short near-infrared (NIR) pulses on semiconductor materials. The generation of THz field without any external bias can be attributed to mechanisms such as transient current (TC) effects and optical rectification (OR) effects. The linear TC effect includes the surface-field (SF) effect which can be understood as acceleration of the charge carriers due to band bending by Fermi-level pinning near the semiconductor surface [1] and the Photo-Dember (PD) effect in which the dipole is generated due to different rates of diffusion of electrons and holes from the surface of the emitter crystal [2]. On the other hand, the nonlinear OR effect can be understood as difference frequency mixing of two or more frequency components. Since there are a number of frequency components present in the incident NIR beam, difference frequency mixing tends to generate the frequency in the range of THz radiation which can be either second order bulk OR effect [3] or higher-order surface-electric-field induced OR effect [4].

For some semiconductor crystals, THz radiation can be generated due to both TC and OR effects. Different experimental geometries can be employed in order to generate THz radiation from semiconductor crystals. In the case of transmission geometry [3], the incident NIR radiation is in the direction of the normal to the emitter crystal and THz field is detected in the straight-through direction. On the other hand in the case of reflection geometry [2], the NIR beam is incident at 45° angle to the normal to the emitter crystal and generated THz radiation is detected in the specular-reflection direction. For TC emitters, the generated terahertz electric field is in the direction of the surface normal of the emitter crystal so the TC effect does not play any role in the THz generation in the transmission geometry. In the case of reflection geometry, in addition to the OR effect, transient currents can also play a role in THz generation due to a component of the surface field in the direction of detection.

The effect of rotation of the emitter crystal about its surface normal, known as the azimuthal angle dependence, can be helpful to distinguish between linear and nonlinear effects. The OR field tends to vary with the angular rotation of the emitter crystal, which is not the case for transient current emitters. In this paper, the generation of a THz field from high index (11*N*) GaAs crystals, where *N* ranges from 0 to 5, is presented for two orthogonal polarization components. The measurements are shown for GaAs *A* face (Ga-rich) and *B* face (As-rich) in both transmission and reflection geometries. By analysing these results we distinguish different crystallographic planes and obtain knowledge about the surface properties of the emitter crystal.

## 2. Theory for Optical Rectification

For an emitter exhibiting THz generation through both bulk and surface optical rectification, the THz field in the far field approximation can be given as [5]

$$\begin{bmatrix} E_p^{THz} \\ E_s^{THz} \end{bmatrix} = \begin{bmatrix} P_{y''}^{bulk} + P_{y''}^{surf} \\ P_{z''}^{bulk} + P_{z''}^{surf} \end{bmatrix} Z_0, \quad (1)$$

where $\hat{x}''$ is the direction of the surface normal, $\hat{y}''$ is the direction of the projection of the excitation beam onto the surface and $\hat{z}'' = \hat{x}'' \times \hat{y}''$. $Z_0$ is a proportionality factor which depends on experimental factors such as distance between emitter and detector. For the special case when the polarization angle for excitation radiation and the incident angle of excitation radiation with respect to the surface normal of emitter crystal are $0°$, the polarization components for bulk and surface OR in terms of azimuthal angle $\theta$ can be written as [5]:

$$\begin{bmatrix} P_{y''}^{bulk} \\ P_{z''}^{bulk} \end{bmatrix} = \frac{d_{14} E_0^2}{\sqrt{2}(N^2+2)^{\frac{3}{2}}} \begin{bmatrix} 3(N^2-1)\sin\theta - 3(N^2+1)\sin 3\theta \\ (N^2-1)\cos\theta - 3(N^2+1)\cos 3\theta \end{bmatrix} \quad (2)$$

$$\begin{bmatrix} P_{y''}^{surf} \\ P_{z''}^{surf} \end{bmatrix} = -\frac{\gamma F_0 E_0^2}{2\sqrt{2}(N^2+2)^2} \begin{bmatrix} 3(N^2-1)\sin\theta + N(N^2+5)\sin 3\theta \\ N(N^2-1)\cos\theta + N(N^2+5)\cos 3\theta \end{bmatrix} \quad (3)$$

where $d_{14}$ is second order susceptibility tensor component, $E_0$ is amplitude of electric field of incident NIR radiation, $\theta$ is the azimuthal angle, Miller indices $h = 1$, $k = 1$, $l = N$; $\gamma$ is third order susceptibility tensor component and $F_0$ is amplitude of surface field of the GaAs emitter crystal.

It can be seen from Equations (2) and (3) that bulk and surface optical rectification components include terms $\sin\theta$, $\cos\theta$, $\sin 3\theta$ and $\cos 3\theta$. Thus, for semiconductor materials exhibiting THz field generation through these mechanisms, three cycle dependence for THz signal can be estimated as we rotate them about the surface normal.

## 3. Experimental Procedure and Results

Terahertz Time Domain Spectroscopy (THz-TDS) has been used to obtain the temporal behaviour of the terahertz field. THz is generated by ultrashort NIR pulses (<12 fs) with the centre wavelength of 790 nm. The GaAs crystal planes (110), (111), (112), (113), (114) and (115) with $A$ and $B$ faces have been investigated as THz emitters in both transmission and reflection geometry. Electro-optic detector (110) ZnTe has been used for detection of the THz field. The THz signal is measured as the differential voltage across the photodiode pair, generated due to rotation of the polarization in the ZnTe detector.

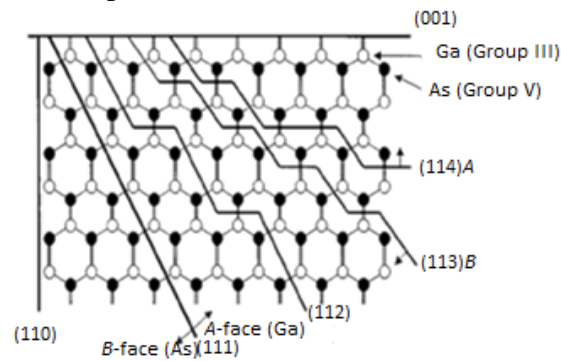

Fig. 1. Different crystal planes for GaAs (11$N$) with $A$ (Ga-rich) and $B$ (As-rich) faces.

The pump beam is horizontally (*p*) polarized. The components of the horizontal (*p*) and vertical (*s*) polarization of the generated THz field have been separated using a wire-grid polarizer. Azimuthal angle dependence has been obtained for all GaAs (11*N*) crystals by rotating them about the surface normal. Variation in the THz peak to peak (p-p) signal has been drawn against the azimuthal angle $\theta$ for both *p* and *s* polarizations. Fig. 1 represents different crystal planes for GaAs (11*N*) crystals with Ga-rich and As-rich faces.

3.1   Transmission geometry ($0^o$ angle of incidence)

As discussed earlier, in this geometry the transient current effect does not play any role in the THz generation. So optical rectification is alone responsible for THz generation in this case, which is evident from the three maxima around the zero line for all emitter crystals as shown in Fig. 2. Here we have shown results for (112) and (114) crystal faces only. By comparing the results for (112) *B* and (114) *B* it can be seen that for (11*N*) crystals the overall THz signal is reduced as the value of the Miller index *N* increases.

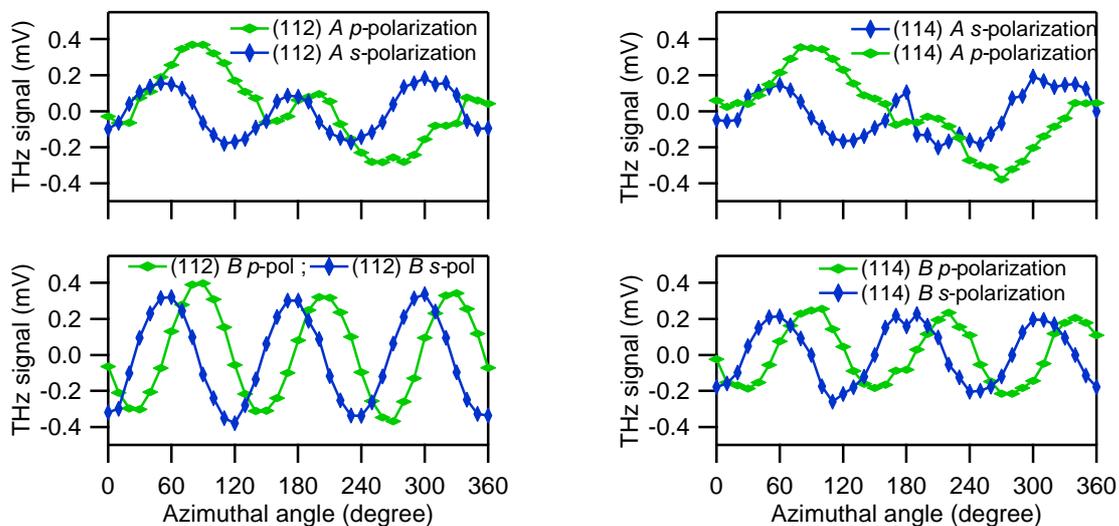

Fig. 2. Transmission geometry: Horizontal (*p*) and vertical (*s*) polarization components for GaAs (112) *A*, (112) *B*, (114) *A* and (114) *B* faces in transmission geometry. Each point represents the p-p THz signal measured at angle $\theta$.

When the measured data is compared with the polarization components obtained from theory for optical rectification as given by equations (1), (2) and (3), it can be shown that both bulk and surface OR mechanisms are responsible for the generation of THz radiation from GaAs (111), (112), (113), (114) and (115) *A* and *B* faces in transmission geometry [5,6]. On the other hand, bulk OR is alone responsible for THz generation from (110) GaAs crystal planes (not shown here). Once we know the contribution of the bulk effect from (110) GaAs crystal, it can be used to calculate the surface field present on GaAs semiconductor surface by comparing experimental data with the theoretical curves for bulk and surface OR fields. The calculated surface field is the same on both Ga-rich and As-rich surfaces within experimental error [5].

3.2.   Reflection geometry ($45^o$ angle of incidence)

As shown in Fig. 3, in the case of reflection geometry, the THz is generated through the linear surface-field effect as well as nonlinear bulk and surface OR effects. Again only results for (112) and (114) crystal planes are shown here. The contribution from the linear effect is evident from the offset observed for horizontal polarized components for all crystal faces. This offset is around 1.6 mV for (112) faces and 0.6 mV for (114) faces. The nonlinear effects are again understood from the three cycle dependence ($3\theta$ dependence) of the generated THz signal for both polarization components. In the case of the *A* face, the surface field is in phase with the bulk field and hence the overall signal is enhanced for the *A* face. On the other hand, for the *B* face, the surface field is out of phase with the bulk field and hence the overall signal is reduced. These features are clearly visible from the *p*- and *s*- polarized components of the *A* and *B* faces as shown in Fig. 3.

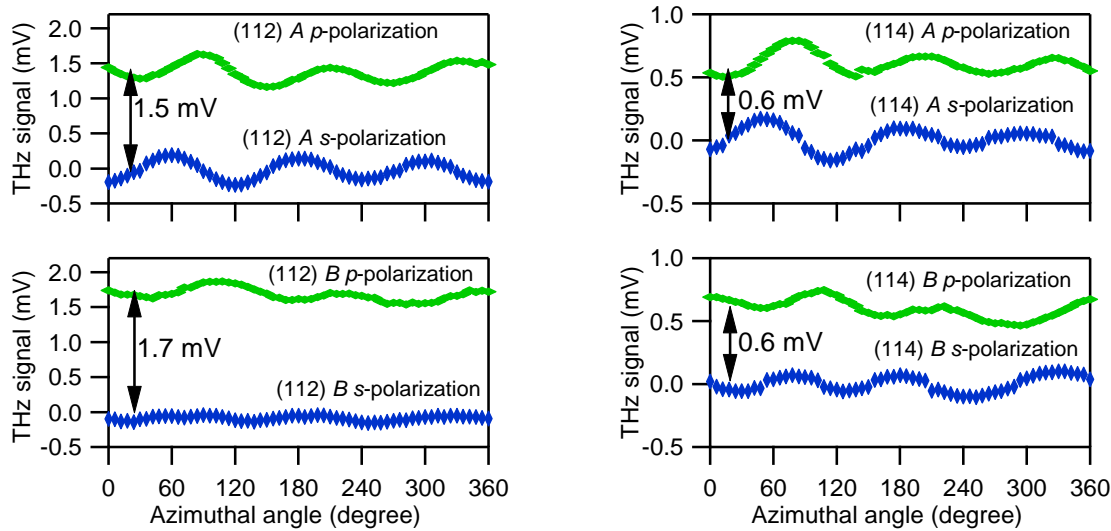

Fig. 3. Reflection geometry: Horizontal (*p*) and vertical (*s*) polarization components for GaAs (112) *A*, (112) *B*, (114) *A* and (114) *B* faces in reflection geometry. The effect of transient current has been represented in terms of an offset from the zero-line. The offsets are roughly 1.5 mV, 1.7 mV, 0.6 mV and 0.6 mV for (112) *A*, (112) *B,* (114) *A*, and (114) B respectively.

## 4. Conclusion

The azimuthal angle dependence has been measured for the THz signal emitted from GaAs (11*N*) samples for Ga-rich and As-rich faces. In the case of transmission geometry, the surface field can be calculated by using the bulk field which is deduced from (110) crystal planes. In the case of reflection geometry, it is possible to estimate the contribution of the transient current from the horizontally polarized components of the terahertz signal. It is also helpful to distinguish between the two faces of GaAs crystals. Such results can be obtained using high index crystal planes only. Alternatively, the optical rectification can be used to identify the crystallographic directions of the emitter crystal.

**Acknowledgments**

This work is supported by the Australian Research Council and by the University of Wollongong. We are thankful to M. Henini for provision of the GaAs samples used here.

## Modulated High-Temperature Phases in Brownmillerites $Ca_2(Fe_{1-x}Al_x)_2O_5$

H. Krüger[a,b]

[a] *Research School of Chemistry, The Australian National University, Canberra, Australia*
[b] *Institute of Mineralogy and Petrography, University of Innsbruck, Austria.*

Samples of the title solid solution series (in the range $0 < x < 0.292$) have been investigated using single-crystal X-ray diffraction at room temperature and at high temperatures. Above a phase transition [960(5) K at $x = 0.073(3)$ and 875(5) K at $x = 0.281(4)$] modulated phases are formed. Depending on the composition either commensurate or incommensurate superstructures occur. All of them can be described using the superspace group *Imma(00γ)s00*.

1. **Introduction**

Brownmillerite ($Ca_2AlFeO_5$) is one of the four main phases in Portland cement clinkers and therefore its structure, phase transitions and properties are of high interest. The brownmillerite structure type is adopted by many compounds and is among the most frequently studied oxygen-deficient perovskites (for an overview of these materials, see [1] and citations therein). Brownmillerites are a class of oxygen-deficient perovskites, in which the oxygen vacancies are ordered in a way such that chains of tetrahedrally coordinated cations are formed. Their general formula is $A_2B_2O_5$ ($A_2BB'O_5$), where corner-linked [$BO_6$] octahedra assemble perovskite-like layers, alternating with sheets containing *zweier* single chains of [$BO_4$] tetrahedra. The octahedral and tetrahedral layers are connected to form a three-dimensional framework, with the *A* ions in interstitial voids. The majority of these compounds adopt the space groups *Pnma* and *I2mb*, as shown in Fig. 1a and 1b, respectively.

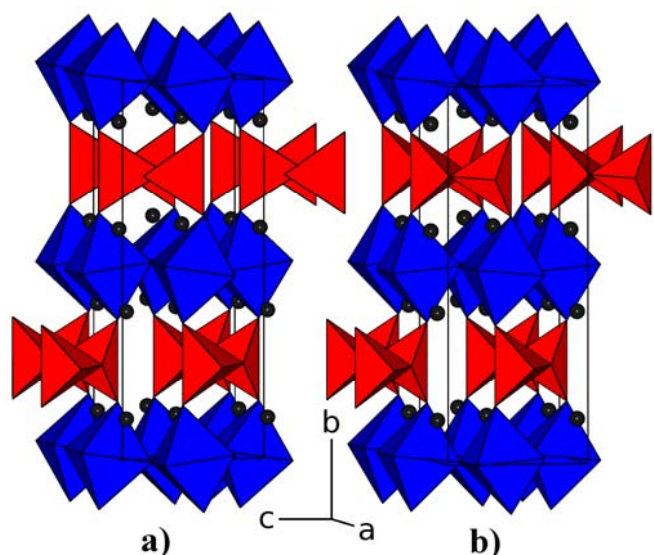

Fig. 1. Brownmillerite structures: space group *Pnma* (a), exhibiting alternating layers of R and L-type chains. Space group *I2mb* (b) showing exclusively one type of tetrahedral chains. The *A* cations are shown as black spheres.

Differences between the two variants can be found in the orientation of the tetrahedral chains: in *Pnma* two possible configurations of the tetrahedral chains (R and L-type) alternate from one to the next layer. Because of the absence of a centre of symmetry only one type of chain is present in space group *I2mb*. A third frequently reported space group *Imma* requires disorder of R and L-type chains and seems to be an approximation of various superstructures due to ordering of the chains within the layers. Many of such superstructures have been found (see [2] and references therein). As the satellite reflections (arising from an intra-layer order) are rather weak, they are not easy to detect, especially in powder data.

The brownmillerite solid solution series $Ca_2(Fe_{1-x}Al_x)_2O_5$ shows a morphotropic phase transition at ca. $x = 0.28$ from space group *Pnma* to *I2mb* with increasing Al-content. High-temperature (HT) phase transitions in the brownmillerite solid solutions series have been known for a long time and were investigated with differential thermal analysis (DTA) [3], neutron [4] and X-ray powder diffraction [5,6] methods. The space groups reported for the HT phases were *Imma* or *I2mb* [4-6]. Using *in-situ* HT single-crystal X-ray diffraction (XRD) we found the HT structures of $Ca_2Fe_2O_5$ [7] and $Ca_2Al_2O_5$ [8] to be incommensurately modulated. Structural models were derived and refined in superspace group *Imma(00γ)s00* (for details on the superspace approach see [9]). A more detailed single-crystal XRD and HT electron microscopy study on the phase transition of $Ca_2Fe_2O_5$ [10] revealed a temperature range of ca. 25 K, where the *Pnma* and the modulated HT phase coexist. The present study reports new HT single-crystal XRD results on the phase transitions of crystals with the compositions of $x = 0.073(3)$, $x = 0.175(4)$ and $x = 0.281(4)$. Furthermore, samples with $x = 0.121(5)$, $x = 0.171(5)$, $x = 0.229(5)$ and $x = 0.292(4)$ have been examined at room temperature.

## 2. Sample preparation

Single crystals have been synthesised using a $CaCl_2$ flux [11]. Details on the synthesis can be found in [6].

## 3. Single-crystal X-ray diffraction

For the HT experiments the crystals have been embedded in 0.1 mm quartz glass mark-tubes (Hilgenberg GmbH, Malsfeld, Germany). The XRD data collections were carried out using a STOE IPDS-2 image plate diffractometer. The chemical composition (Fe:Al ratio) of the samples was determined by structure refinements [12] using room-temperature data. Temperature calibration and computer-controlled series of HT experiments were performed as described in [13]. Each experiment covered a range of at least 20° in ω (with steps of 1.5°). The detector distance was set to 120 mm and the exposure time was 5 min per frame. Data integration for Figs. 2-4 was performed using the same profile parameters and the same scaling factor. The software *integrate* [14] was utilised for the data reduction process. The satellite reflections of the HT modifications were indexed with a modulation wave vector of **q** = γ **c*** [9,14].

## 4. Results

The Al-content $x$ of all samples was established by crystal-structure refinements using room-temperature XRD data. The results are listed in Table 1. The distribution of Al on the octahedral and tetrahedral sites is not equal. As the results in Tab. 1 show, approximately 3/4 of the Al occupies the tetrahedral site. The preference for the tetrahedral site was reported earlier [6]. The morphotropic phase transition occurs between $0.229 < x < 0.281$. The value published by Redhammer *et al.* is 0.28 for single-crystals, but higher (0.325-0.35) for powders (see [6] for more details and references).

All samples examined at HT show a phase transition to modulated structures, as evident by the occurrence of satellite reflections. As the HT modifications show an *I*-centred lattice, phases adopting space group *Pnma* at room temperature lose their *h+k+l=2n+1* reflections at the transition. Consequently, monitoring the intensities of satellites (where present) and $h+k+l=2n+1$ reflections gives a powerful tool to observe the coexistence of the phases [10].

### 4.1  $Ca_2Fe_{0.927}Al_{0.073}O_5$

XRD data of the sample with $x = 0.073(3)$ has been collected in the ranges of 890-970 K (steps of 5 K), 980-1080 K (steps of 20 K), subsequently from 1050-990 K (steps of 20 K) and finally from 970-890 K (steps of 5 K). The temperature-dependent intensity changes of selected reflections are shown in Fig. 2. The satellite reflections of the modulated HT phase appear at 960 K. From 990 K their intensities remain constant. The $h+k+l=2n+1$ reflections disappear in the transition, as the lattice becomes *I*-centred. Initially strong reflections of this group can be observed up to 1000 K. Consequently, reflections of both phases can be observed over an interval of 40 K. The **q**-vector shows a γ-value of 0.653(1) at 965 K. With higher temperatures the value decreases [0.642(1) at 1080 K]. While cooling γ exhibits a hysteresis, the value stays almost constant down to 965 K [0.644(1)].

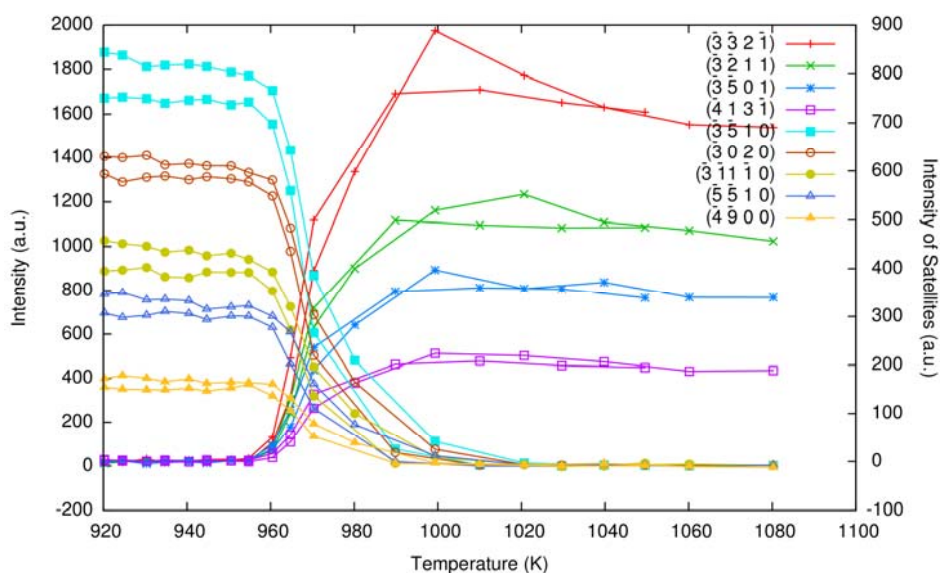

Fig. 2. *Pnma-Imma(00γ)s00* phase transition of $Ca_2Fe_{0.927}Al_{0.073}O_5$. Selected *P*-lattice reflections and satellites show the coexistence of the phases. The heating-cooling cycle does not show a significant hysteresis.

### 4.2  $Ca_2(Fe_{0.825}Al_{0.175})_2O_5$

A crystal with $x = 0.175(4)$ was investigated at the following temperatures: 875-942 K (5 K steps), 953-923 K (10 K steps), 914-872 K (3 K steps) and 865 K. While heating the transformation sets in between 890 and 894 K (first satellites can be observed at 894 K). The reverse transition (while cooling) shows a small hysteresis, satellites can be observed down to 885 K. At the first appearance of the satellite reflections the **q**-vector shows a commensurate value of γ = 2/3. Additional experiments show that this value stays unchanged up to 1053 K. Measurements at 1064 and 1073 K show smaller γ-values significantly deviating from 2/3: 0.654(1) and 0.651(1), respectively. This suggests the existence of a commensurate-incommensurate transition between 1053 and 1064 K. The γ-value shows a hysteresis, while cooling the value stays incommensurate down to at least 1013 K.

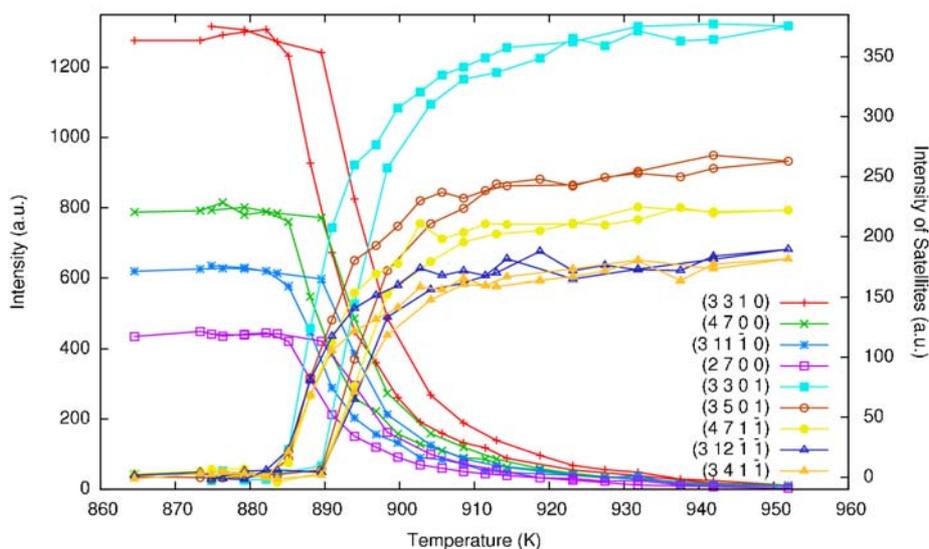

Fig. 3. The *Pnma-Imma(00γ)s00* phase transition of $Ca_2(Fe_{0.825}Al_{0.175})_2O_5$: temperature-dependent changes of selected reflections show a hysteresis of ca. 5 K.

### 4.3   $Ca_2(Fe_{0.719}Al_{0.281})_2O_5$

The temperature ranges covered by the experiments are: 922-863 K and 876-965 K (steps of 4 K). Crystals with the composition $x = 0.281(4)$ adopt space group *I2mb* at ambient conditions. The transformation to a commensurate superstructure ($\gamma = 2/3$) starts at ca. 870 K. The **q**-vector stays unchanged at least up to 1083 K. Unlike the *I2mb* compounds with $x = 0.175$ and $x = 1$ [8], no significant hysteresis in the phase transition temperature can be detected.

Fig. 4 shows the intensities of selected reflections. In contrast to the compounds $x = 0.147$ and $x = 0.175$, this composition does not show the disappearing of $h+k+l=2n+1$ reflections. Due to the *I*-centred lattice they are already extinct at ambient temperatures. However, some $h+k+l=2n$ reflections show significant changes of their intensities across the phase transition. As this type of modulation (square-wave modulation of occupancies) cannot change its amplitude, the temperature range in which the satellite intensities are increasing can be interpreted as a phase-coexistence range [8].

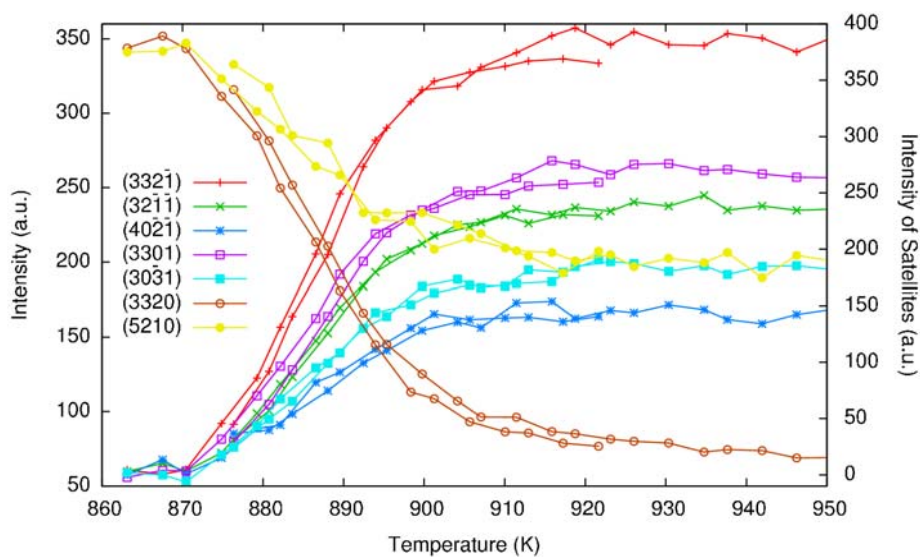

Fig. 4. *I2mb-Imma(00γ)s00* phase transition of $Ca_2(Fe_{0.719}Al_{0.281})_2O_5$: changes of the intensities of selected satellites and main reflections.

## 5   Discussion

All samples examined at high temperature exhibit a wide (>20 K) range of phase co-existence at the transition as described for $Ca_2Fe_2O_5$ [10]. The phase transition temperatures we found are lower than the ones reported by [6]. This is attributed to the methods used to determine them. The method used in [6] was to monitor a disappearing *P*-reflection [$h+k+l=2n+1$, e.g. (131)]. In this study the first appearance of satellite reflections is used, which will give lower temperatures because of the phase coexistence. However, the values given in Tab. 1 roughly resemble the same slope as given in [6] but shifted to lower temperatures.

The Al end-member (synthesised at 2.5 GPa and 1273 K) shows a large hysteresis in the phase transition temperature [8]. Furthermore, the rate of satellite intensity change shows a difference between the transitions at cooling and heating. Hence the assumed phase coexistence range is 40 K for cooling and 20 K for heating [8]. The **q**-vector of the HT structure shows an incommensurate $\gamma$-value close to 0.59. No significant trend of $\gamma$ could be observed, because the transition temperature is too close to the experimental limit.

Interestingly, the magnitude of the modulation wave vector of both end-members is smaller (Tab. 1) than that of the intermediate samples. Crystals with $x = 0.175$ and $x = 0.281$ show larger and commensurate values. Generally, the value $\gamma$ seems to decrease with higher temperatures for all samples (except $x = 0.281$, where no decrease could be observed up to the maximum temperature of the experimental setup, which is approx. 1080 K).

A hysteresis in the phase transition temperature was observed for only two samples of the system: $x = 0.0175$ and $x = 1.0$ [8] (see Tab. 1). All other samples do not show any hysteresis in the XRD experiments. However, it has to be noted that the temperature was changed in a step-wise way between the individual XRD experiments, which are performed at a constant temperature (small changes may occur during the image plate readout, due to air draught inside the diffractometer). Obviously, these are not optimal conditions for quantitative and comparable determination of temperature hysteresis in phase transitions. Using DTA at a much faster and well-defined rate of 5 K/min we found a hysteresis of 24 K for the iron end-member [10], which does not show any hysteresis in XRD experiments. Therefore it can be expected that all samples exhibit a hysteresis (using DTA), which is larger than that found in XRD.

Table 1. Aluminium content (*x*), space group (SG), phase transition (PT) temperatures (first appearance of the satellite reflections), hysteresis (by XRD), **q**-vector at PT and amount of Al on the octahedral and tetrahedral site (given in % of a full occupied site).

| Al content *x* | SG at RT | PT Temp. [K] | hysteresis [K] | $\gamma$ | %Al Oct. | %Al Tetr. | Reference |
|---|---|---|---|---|---|---|---|
| 0.000 | *Pnma* | 960 | 0 | 0.607(1) | 0.0(0) | 0.0(0) | [10] |
| 0.073(3) | *Pnma* | 960 | 0 | 0.653(1) | 4.1(4) | 10.6(5) | |
| 0.121(5) | *Pnma* | | | | 5.3(7) | 18.9(7) | |
| 0.171(5) | *Pnma* | | | | 7.2(7) | 27.0(8) | |
| 0.175(4) | *Pnma* | 890 | 5 | 2/3 | 9.9(6) | 25.0(6) | |
| 0.229(5) | *Pnma* | | | | 12.6(7) | 33.3(7) | |
| 0.281(4) | *I2mb* | 875 | 0 | 2/3 | 14.3(6) | 41.9(6) | |
| 0.292(4) | *I2mb* | | | | 14.6(5) | 43.8(6) | |
| 1.000 | *I2mb* | 1065 | 100 | 0.590(2) | 100.0(0) | 100.0(0) | [8] |

## 6. Conclusions

Basically, the findings of this study agree with the results reported by [6] in the following points: the composition of the morphotropic phase transition, the composition dependence of the HT phase transition temperature of the *Pnma* structure and the fact that Al prefers the tetrahedral site. However, the solid-solution series shows modulated HT structures, which can all be described in superspace group *Imma(00γ)s00*. The HT phases and phase transitions of the investigated samples exhibit interesting features which need further investigation. In particular more single-crystal samples have to be studied *in situ* at high temperatures. Targets of these future studies will be the relationship between composition, temperature and modulation wave vector, as well as the hysteresis observed in some of the samples. To obtain better data on the temperature hysteresis dedicated DTA experiments are planned.

**Acknowledgments**
HK acknowledges financial support of the Austrian Science Fund (FWF): J2996-N19. HK also thanks G. Redhammer for providing the sample material, and R. Whitfield for proofreading the manuscript.

# Crystal Structure of $Cu_5SbO_6$ – Synchrotron and Neutron Diffraction Studies

E. Rey[a], P.-Z. Si[a,b], and T. Söhnel[a]

[a] *Department of Chemistry, The University of Auckland, Auckland, New Zealand.*
[b] *Department of Material Science & Engineering, China Jiliang University, Hangzhou 310018, China.*

One very interesting compound in the system Cu/Sb/O is the mixed-valence $Cu_5SbO_6$ which crystallizes in the high temperature modification as a modified Delafossite structure type. Single crystals of $Cu_5SbO_6$ has been synthesized by chemical vapour transport methods and characterised using single crystal X-ray diffraction and powder neutron and synchrotron diffraction.

## 1. Introduction

In the ternary copper-antimony-oxygen system four different compounds have been synthesised so far: $CuSb_2O_6$, $Cu_9Sb_4O_{19}$, $Cu_4SbO_{4.5}$ and $Cu_5SbO_6$, but only one crystal structure of $CuSb_2O_6$ has been resolved up to now [1-7]. $CuSb_2O_6$ has had its properties well studied, for its interesting structural and magnetic properties. At room temperature $CuSb_2O_6$ has a monoclinic distorted trirutile structure and a tetragonal trirutile structure at T < 380 K [8]. Because of the Jahn-Teller distortion of $Cu^{2+}$ in an octahedral coordination sphere, the monoclinic structure is unusual compared to analogous antimonates such as $ZnSb_2O_6$ and $FeSb_2O_6$ which are tetragonal trirutile structures [4]. In this structure, antimony is +5 and copper is in the +2 oxidation state [7]. $Cu_9Sb_4O_{19}$ was synthesised by reacting CuO and $CuSb_2O_6$ under 10 kbar and 1000-1100 °C with a piston-in-cylinder apparatus [5], and 10 bar at 1100 °C and 10 kbar at 1000-1100 °C oxygen pressure. The high pressure was proposed to suppress the decomposition of Cu(I) to Cu(II) in the system [3]. The elemental composition of $Cu_9Sb_4O_{19}$ was confirmed with X-ray fluorescence, and in air it is stable to 945 °C after which it breaks down to $CuSb_2O_6$ and $Cu_4SbO_{4.5}$ [5]. The final known ternary copper antimonite is $Cu_4SbO_{4.5}$ / $Cu_5SbO_6$. $Cu_4SbO_{4.5}$ was first reported in a series of papers describing its synthesis and powder X-ray characterisation [2, 4]. Kol'tsova and Chastukhin solved the phase diagram in the temperature range between 700°C and 1150°C and re-evaluated the composition of $Cu_4SbO_{4.5}$ to $Cu_5SbO_6$ while examining the Cu-Sb-O system in air [7].

The main perspective of this project is to describe the ternary phase $Cu_5SbO_6$, obtain the crystal structures of $Cu_5SbO_6$ of all possible modifications. For the preparation a number of different methods will be applied, ranging from classical high-temperature sintering to chemical transport reactions for the preparation of single crystals. For the identification of the obtained compounds we used X-ray single crystal and powder diffraction, high resolution neutron and Synchrotron powder diffraction.

## 2. Experimental

Powder samples of $Cu_5SbO_6$ were prepared by mixing 10 CuO + $Sb_2O_3$ over the range 900 – 1120°C in 10 K steps, for 48 h with intermediate grinding and then air quenched. These samples were analysed by powder X-ray diffraction. Chemical vapour transport was used to synthesize single crystals of $Cu_5SbO_6$. A mixture of 10 CuO + 1.7 $Sb_2O_3$ (excess of $Sb_2O_3$) were pre-heated in a sealed evacuated quartz tube, with $PdCl_2$ as transport agent. The mixture was placed in one end (source) and nothing in the other end (sink). The quartz tube was placed in a two-zone furnace which heated the chemical side to 900°C and the sink side to 800°C.

Single crystal X-ray diffraction were collected on a Bruker SMART II diffractometer (MoKa, λ = 0.71073 Å). High resolution synchrotron X-ray diffraction pattern were collected on the powder diffraction beam-line at the Australian Synchrotron (λ = 0.68950 Å). Neutron diffraction patterns at room temperature and 4 K were measured using the high resolution powder diffraction diffractometer Echidna at ANSTO's OPAL facility as Lucas Heights.

## 3. Results

### 3.1. Crystal Structure of $Cu_5SbO_6$

Single crystal XRD on crystals of $Cu_5SbO_6$ formed by the chemical transport revealed the structure of the high temperature form of $Cu_5SbO_6$. The new structure type is a modified and distorted Delafossite structure, which contains a mixture of $Cu^+$ and $Cu^{2+}$ in the same compound. It contains layers of $Cu^{2+}O_6$ and $Sb^{5+}O_6$ octahedra with $Cu^{1+}$ between them, this is shown in Figure 1 below. The $Cu_2SbO_6$ - layers can be described as $(Cu^{2+}_{2/3}Sb^{5+}_{1/3})O_2$ in accordance to the pure Delafossite structure type ($CuFeO_2$), which contains $FeO_2$ layers. In $Cu_5SbO_6$ the magnetically active $CuO_2$ layer is diluted in an ordered fashion with non-magnetic $Sb^{5+}$ forming $Cu^{2+}$-$Cu^{2+}$ pairs. This layer could also be observed in the Na containing compound $Na_3Cu_2SbO_6$ [8]. The $Cu^+$ cations which are 'sandwiched' between them show the typical 2-fold dumbbell coordination sphere.

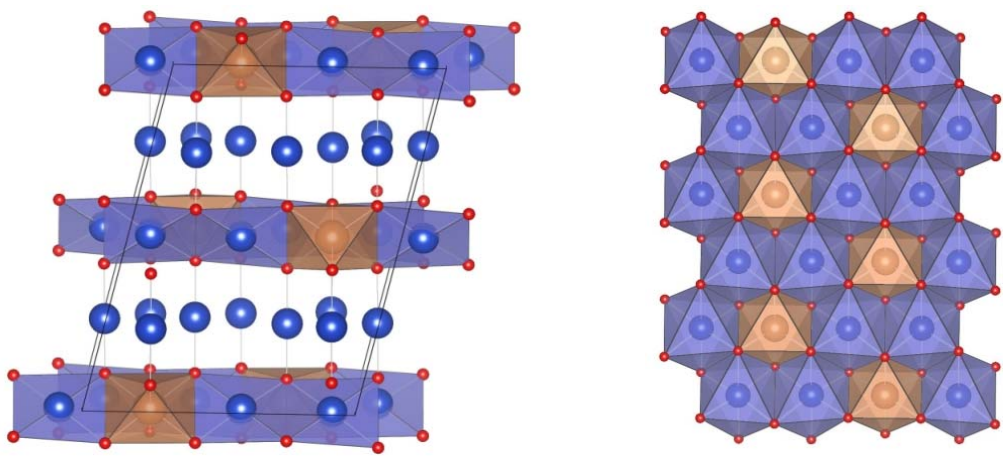

Fig 1. Left: Side view of the crystal structure of $Cu_5SbO_6$; Right: View of the $Cu_2SbO_6$ layer with likely Cu-O bonds including longer distances

The $SbO_6$ octahedra is regular with 1.958(9) to 2.034(7) Å bond distances, while the $CuO_6$ octahedra have four short bonds of 1.982(9) to 2.028(8) Å, and two longer bonds of 2.327(8) and 2.368 Å. A comparison of the crystal structure with just the four Cu-O bonds, and one including the longer ones are shown in Figure 2 below, which shows just the $Cu_2SbO_6$ layer. The crystallographic data are summarised in Table 1.

### 3.2 X-ray powder diffraction

X-ray powder diffraction for phase analysis of $Cu_5SbO_6$ was performed at various temperatures in air. This revealed the existence of at least two definite modifications of $Cu_5SbO_6$, one at temperatures below about 1000°C, and another at temperatures higher than about 1100°C, as well as an intermediate region between. It turned out, that the oxygen partial pressure used during the preparations seems to have an influence on the formation of a certain modification at a certain temperature. To examine the influence of the oxygen in the system a mixture of CuO and Sb$_2$O$_3$ (10:1) and the previously prepared low temperature modification were sealed in an evacuated quartz tube and reheated at 950°C. Heating the CuO and Sb$_2$O$_3$ powders in a sealed evacuated quartz vessel at 950 °C produced a XRD pattern similar to the high temperature form. This suggests that the O$_2$ pressure generated by the reaction 10 CuO$_{(s)}$ + Sb$_2$O$_{3(s)}$ -> 2 Cu$_5$SbO$_{6(s)}$ + ½ O$_{2(g)}$ prevented the formation of the low temperature version. The XRD of the low temperature Cu$_5$SbO$_6$ powder which was heated again at its formation temperature in a vacuum showed a change towards the high temperature phase. The process of the formation of the high-temperature modification in a closed system is not reversible at low temperatures in air.

Table 1. Fractional coordinates for Cu$_5$SbO$_6$; Space group *C2/c (No. 15)*, a = 8.9372(11) Å, β = 5.5967(6) Å, c = 11.8667(14) Å and β = 103.697(9)°; R$_1$(obs) = 4.84 %, wR$_2$(all) = 13.77%

| Atom | Wyckoff position | x | Y | z | U$_{ani}$ |
|---|---|---|---|---|---|
| Sb | 4c | 1/4 | ¼ | 0 | 0.0076(4) |
| Cu(1) | 8f | 0.08325(9) | 0.24851(11) | 0.49810(7) | 0.0098(4) |
| Cu(2) | 8f | 0.17134(10) | 0.0973(2) | 0.25700(7) | 0.0146(4) |
| Cu(3) | 4e | 0 | 0.6271(3) | 1/4 | 0.0150(4) |
| O(1) | 8f | 0.1176(7) | 0.1224(9) | 0.0967(5) | 0.0122(11) |
| O(2) | 8f | 0.4452(7) | 0.1218(8) | 0.0911(5) | 0.0113(11) |
| O(3) | 8f | 0.2243(7) | 0.0732(8) | 0.4189(5) | 0.0098(10) |

The high resolution measurements using the powder diffraction beam-line at the AS could clearly distinguish between the high temperature modification and the low temperature modification. As it turns out at no temperature between 900°C and 1120°C neither the low nor the high temperature modification exists in pure form. The content of the LT-modification decreases with increasing temperature but it was not completely converted even if the temperature is very close to the melting temperature at 1150°C. Rietveld refinements on these Synchrotron data and powder neutron diffraction data on both, LT-Cu$_5$SbO$_6$ and HT-Cu$_5$SbO$_6$, confirm an ordering (high temp. modification) / disordering (low temp. modification) effect of the Sb$^{5+}$ and Cu$^{2+}$ ions in the brucite-like layers between the Cu$^{1+}$ atoms. It also shows that Cu$_4$SbO$_{4.5}$ can now be identified as the low-temperature modification of Cu$_5$SbO$_6$. Neutron powder diffraction data seem to verify the incorporation of addition oxygen in the LT-modification, whereas no additional oxygen can be found in the HT-modification. The refinement results are summarised in Table 2.

Table 2. Fractional coordinates for low-temperature Cu$_5$SbO$_6$ at 900°C; Space group *C2/c (No. 15)*, a = 2.97701(6) Å, b = 5.5952(1) Å, c = 11.5195(2) Å and b = 90.151(2)°; R$_F$ = 3.64 %, R$_{BRAGG}$ = 5.70%, R$_P$ = 7.07 %

| Atom | Wyckoff position | x | y | z | B$_{ani}$ | SOF |
|---|---|---|---|---|---|---|
| Sb | 4a | 0 | 1/2 | 0 | 0.12(7) | 1/3 |
| Cu(1) | | 0 | 1/2 | 0 | | 2/3 |
| Cu(2) | 4e | 0 | 0.8568(6) | 1/4 | 0.74(8) | 1 |
| O(1) | 8f | 0.503(3) | 0.3581(6) | 0.0903(2) | 0.83(7) | 1 |
| O(2) | 4e | 0 | 0.534(8) | 0.25 | 0.83(7) | 0.038(4) |

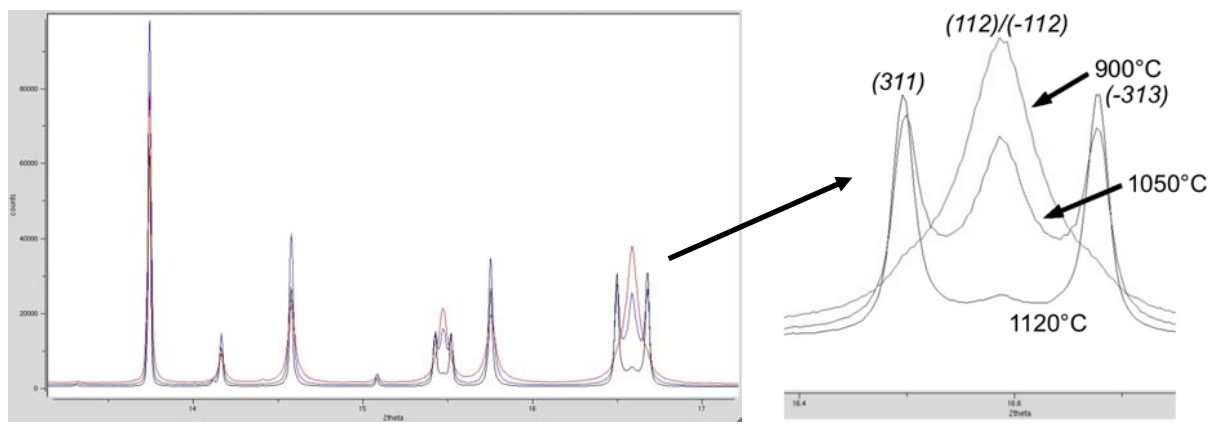

Fig 2. Left: Section of synchrotron high resolution powder patterns of 900°C, 1050°C and 1120°C samples; Right: (311)/(-313) reflexes (HT-Mod.) and (112)/(-112) reflexes (LT-Mod.) at about $2\theta = 16.6°$.

## 4.  Conclusions

Single crystals of the high temperature modification were prepared by chemical vapour transport of CuO and $Sb_2O_3$ oxides with $PdCl_2$ as transport agent and the crystals structure could be solved which can be described as a modified Delafossite structure. In this structure layers of ordered edge sharing $Cu^{2+}O_6$ and $Sb^{5+}O_6$ octahedra are formed. The $Cu^{1+}$ are incorporated between these planes. Powders of $Cu_5SbO_6$ at various temperatures were prepared and analyzed by x-ray, synchrotron and neutron powder diffraction. It can be found that two distinct modifications of $Cu_5SbO_6$ exist and both of which can be formed by the reaction of CuO and $Sb_2O_3$ in air. Both modifications are present in the whole temperature range between 900°C and 1150°C.

**Acknowledgments**

The authors would like to thank Dr Kia Wallwork for her assistance at the powder diffraction beam-line of the Australian Synchrotron and Dr Maxim Avdeev, Dr Chris Ling and Dr Neeraj Sharma for the help at the ECHIDA beam-line at ANSTO, Sydney. The financial support from the Australian Institute of Nuclear Science and Engineering (AINSE) and the Australian Synchrotron is gratefully acknowledged. One of us (TS) wishes to acknowledge the University of Auckland for the support of an Early Career Excellence Award.

# Slow Photon Photocatalytic Enhancement using Titania Inverse Opal Photonic Crystals

V. Jovic, T. Söhnel, J.B. Metson and G.I.N. Waterhouse

*Department of Chemistry, The University of Auckland, Auckland, New Zealand.*

Titania inverse opal photonic crystals possessing photonic band gaps (PBGs) in the UV-Vis region were successfully fabricated using the colloidal crystal template approach. SEM, XRD, BET, Ti L-edge NEXAFS and optical measurements revealed these materials possess a 3-dimensionally ordered macroporous (3DOM) structure comprising air spheres in a nanocrystalline anatase $TiO_2$ matrix. The photocatalytic performance of the titania inverse opals under UV irradiation was evaluated, and slow photon photocatalytic enhancement of ethanol decomposition was observed for a $TiO_2$ inverse opal with a PBG at 344 nm.

1. **Introduction**
Photonic crystals are highly ordered materials that possess a periodically modulated refractive index in one, two or three dimensions, with periods typically on the length scale of visible light [1, 2]. Periodicity causes incoherent Bragg diffraction from lattice planes resulting in the creation of a photonic band gap (PBG), a narrow range of frequencies for which light propagation in the photonic crystal are forbidden. Opals and inverse opals (IO's) are important classes of photonic crystals. Opals comprise a *face-centered cubic* (*fcc*) arrangement of monodisperse solid spheres ($\phi_{solid}$ = 74%, $\phi_{void}$ = 26%), whereas inverse opals comprise a *fcc* arrangement of air spheres in a solid matrix ($\phi_{solid}$ < 26%, $\phi_{void}$ > 74%). The optical properties of opal and inverse opal photonic crystals can be described by the modified Bragg's Law expression below:

$$\lambda_{max} = 1.633 D \sqrt{n_{avg}^2 - \sin^2 \theta}$$

where $\lambda_{max}$ is the PBG position (in nm) for first order Bragg diffraction on *fcc*(111) planes, *D* is the sphere diameter (in nm), $n_{avg}$ is the average refractive index of the photonic crystal and *θ* is the incident angle of light with respect to the surface normal of *fcc*(111) planes [3, 4]. At the edges of the PBG, light travels with a strongly reduced group velocity ($v_g$) resulting in the creation of slow photons [5, 6]. At the 'red edge' of the PBG, the incident light is localised on the dielectric material. If the 'red edge' of the PBG for titania inverse opals could be coupled to the electronic absorption band of $TiO_2$ (~388 nm), this should in theory enhance UV light absorption by the semiconductor, which in turn should enhance the photocatalytic activity of $TiO_2$ via increasing electron-hole pair formation. Inverse opal architectures have other desirable properties which make them ideal for catalytic and photocatalytic applications, such as large surface area and inherent macroporosity.

This project is aimed at the fabrication of $TiO_2$ inverse opal thin films and powders via the 'bottom up' colloidal crystal template approach. Here, the structural, optical and photocatalytic properties of a series of titania inverse opal photonic crystals are reported. The photocatalytic activities of the samples under UV irradiation have been evaluated against the photooxidation of gas phase ethanol.

## 2. Experimental

Batches of monodisperse PMMA colloids with diameters in the range 175-450 nm were synthesized by the free-radical-initiated emulsion polymerization of methyl methacrylate (MMA) [3, 7]. The PMMA suspensions were then transferred to 50 mL polypropylene centrifuge tubes and centrifuged at 1500 rpm for 24 h to form 3-dimensional PMMA colloidal crystals (synthetic opals). After centrifugation, the supernatant was decanted off and the PMMA colloidal crystal templates were dried in air for 3 weeks.

PMMA colloidal crystal thin films were fabricated on glass microscope slides using a flow controlled vertical deposition method [3, 8]. A PMMA colloid suspension (200 mL, 5wt % PMMA in $H_2O$) was poured into a 200 mL Duran glass petri dish. Glass microscope slides were vertically immersed in the PMMA colloidal suspension. A peristaltic pump, operating at a speed of 0.2 mLmin$^{-1}$, was used to slowly pump out the PMMA colloidal suspension. As the meniscus moved down the glass slide, a colloidal crystal thin film self-assembled at the falling meniscus with an *fcc* (111) plane parallel to the glass substrate.

$TiO_2$ inverse opal powders and thin films were prepared by the colloidal crystal template technique [3]. A titania sol was prepared by mixing ethanol (25 mL), titanium(IV) propoxide (25 mL), HCl (5 mL) and milli-Q water (10 mL). The titania sol was then applied drop-wise over the lightly crushed colloidal crystal template in a Buchner funnel under a strong vacuum. The infiltrated templates were left to hydrolyze in air at room temperature for 24 h, then carefully calcined at 300°C (2 h), then 400°C (2 h), to remove the PMMA template and crystallize the $TiO_2$ network, respectively. Titania inverse opal thin films were prepared by diluting the titania sol 10-fold with ethanol and infiltrating the PMMA colloidal crystal thin films using capillary action.

## 3. Results

SEM micrographs of PMMA colloidal crystal templates of two different sizes (Samples 2 and 6) can be seen in Figure 1 (i) and (ii) below. The corresponding $TiO_2$ inverse opals are shown in Figures (iii) and (iv). From the SEM micrographs, sphere sizes were determined to be 215 and 452 nm, respectively. The corresponding inverse opal pore sizes calculated from these micrographs were 188 and 355 nm, respectively.

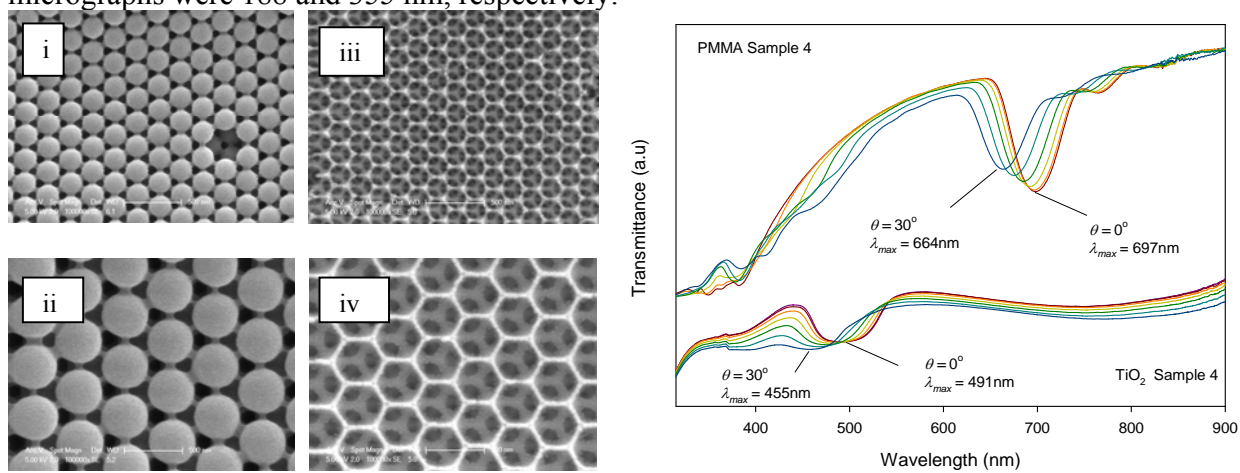

Figure 1(i) and (ii) show SEM micrographs for PMMA colloidal crystals of two different sizes. (iii) and (iv) show the corresponding $TiO_2$ inverse opals fabricated from the PMMA templates. The right panel shows UV-Visible transmission spectra for a PMMA colloidal crystal thin film (Sample 4) and its $TiO_2$ inverse opal replica. UV-Vis data were collected at angles of 0-30° with respect to the surface normal of the *fcc* (111) lattice planes.

Figure 1 (Right) shows UV-Vis transmittance spectra for a PMMA colloidal crystal and its titania inverse opal replica. At normal incidence ($\theta = 0^o$, i.e. along the [111] direction), the PMMA colloidal crystal exhibits a PBG at 697 nm, whereas the titania inverse opal shows a PBG at 491 nm, which reflects the lower solid volume fraction and lower average refractive index of the inverse opal. For both samples, the PBG shifts to shorter wavelengths as the incident angle of light with respect to the *fcc* (111) normal increases, in accordance with the modified Bragg's law given in the introduction. The PMMA colloidal crystals fabricated in this work showed PBGs at 396, 488 604, 697, 818 and 974 nm. The corresponding inverse opals had PBGs at <300, 344, 417, 491, 568 and 692 nm, respectively.

Powder XRD diffraction (Figure 2 Left) revealed that the titania inverse opals were composed of anatase nanoparticles of crystallite size ~11-16 nm. No rutile or brookite diffraction peaks were identified. Line broadening observed for the $TiO_2$ IO's confirmed the presence of nanocrystalline $TiO_2$ particles. Ti L-edge NEXAFS obtained at the Australian synchrotron presented in Figure 2 (Right) show characteristic $L_3$ and $L_2$ edge features which arise from the 2 $p_{3/2} \rightarrow$ 3d and 2 $p_{1/2} \rightarrow$ 3d electronic transitions of the octahedrally coordinated $Ti^{4+}$ ion in anatase [9]. It is worth noting that the various $TiO_2$ polymorphs can be readily distinguished by NEXAFS but not by XPS.

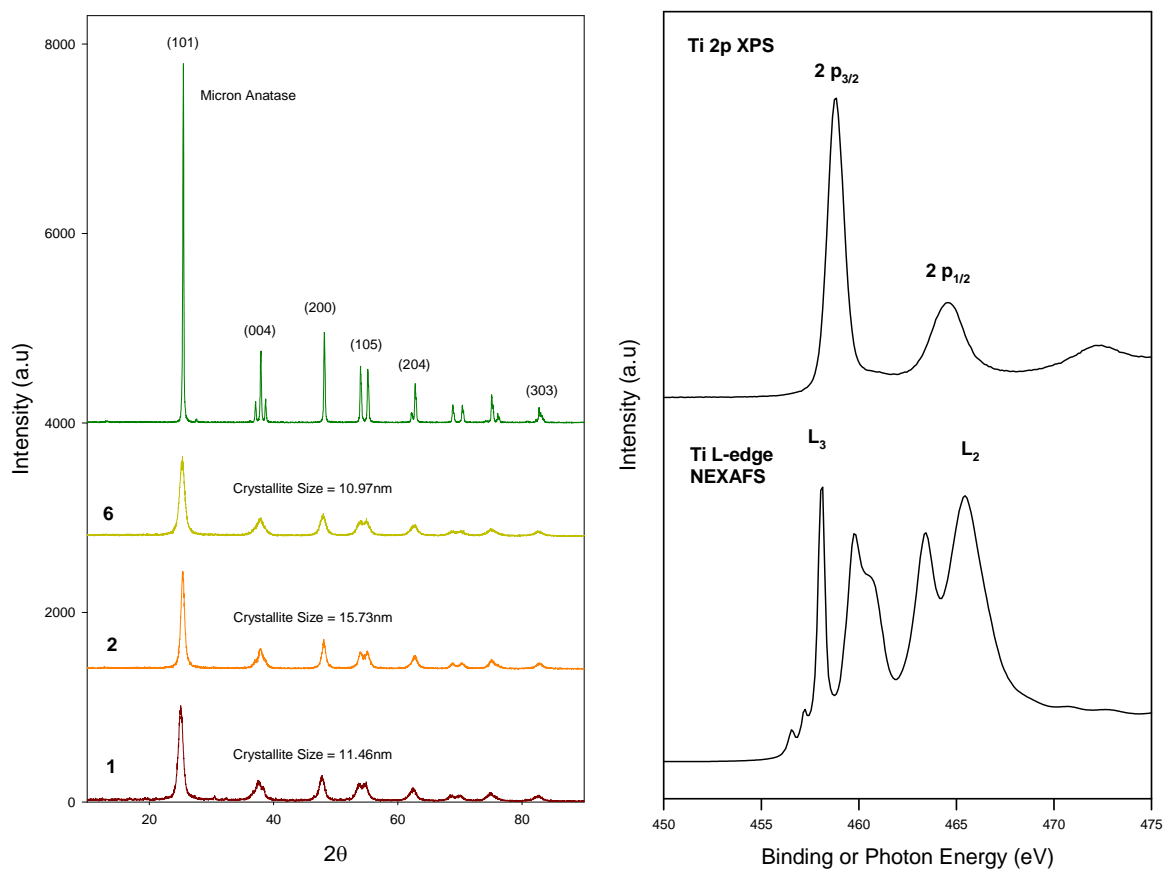

Figure 2 Left: Powder X-ray diffraction patterns for several $TiO_2$ inverse opal powders and a micron sized anatase $TiO_2$ powder. Right: Ti 2p XPS and Ti L-edge NEXAFS for a $TiO_2$ inverse opal.

The gas phase photo-oxidation of ethanol over the titania samples obeyed first order kinetics. The photocatalytic data is summarised in Table 1 below. The results show that Sample 2, which had a PBG at 344 nm, showed the highest rate constant ($1.48 \times 10^{-3}$ s$^{-1}$) amongst the photocatalysts tested. In this sample, the red edge of the PBG overlaps with the electronic absorption edge of TiO$_2$ (388 nm, ~3.2 eV). The low surface area anatase reference powder had the lowest rate constant ($4.49 \times 10^{-4}$ s$^{-1}$). The high surface area sample had a rate constant similar to Sample 1, a titania inverse opal operating outside the 'slow photon' regime. When the rate constants were normalized with respect to surface area, Sample 2 maintained the highest rate constant over the inverse opal samples ($1.44 \times 10^{-3}$ s$^{-1}$m$^{-2}$), confirming that slow photon photocatalytic enhancement, caused by overlap of the electronic absorption edge and red edge of the PBG, enhanced the rate constant.

Table 1. Summarized photocatalytic activity data for TiO$_2$ inverse opal powders and several anatase reference powders. The photocatalytic activities of the samples were evaluated under UV irradiation against the photodecomposition of gaseous ethanol.

| Sample | PBG position (nm) | Area (m$^2$g$^{-1}$) | Rate constant, k'(s$^{-1}$) | Raw Enhancement factor | Normalized rate constant k'$_N$ (s$^{-1}$m$^{-2}$) | Enhancement factor |
|---|---|---|---|---|---|---|
| TiO$_2$ IO #1 | <300.00 | 74.1 | 1.22×10$^{-3}$ | 1.51 | 1.09×10$^{-3}$ | 1.07 |
| TiO$_2$ IO #2 | 344.95 | 68.9 | 1.48×10$^{-3}$ | 1.83 | 1.44×10$^{-3}$ | 1.41 |
| TiO$_2$ IO #4 | 491.40 | 61.9 | 1.03×10$^{-3}$ | 1.28 | 1.11×10$^{-3}$ | 1.09 |
| TiO$_2$ IO #6 | 568.20 | 52.7 | 8.07×10$^{-4}$ | 1 | 1.02×10$^{-3}$ | 1 |
| HSA anatase | - | 108.3 | 1.12×10$^{-3}$ | 1.52 | 7.57×10$^{-4}$ | - |
| LSA anatase | - | 10.4 | 4.49×10$^{-4}$ | 0.56 | 2.89×10$^{-3}$ | - |

Normalized rate constant k'$_N$(s$^{-1}$m$^{-2}$) with respect to area. HSA; high surface area. LSA; low surface area.

**Acknowledgements**
The authors would like to thank Bruce Cowie for his assistance at the SXR beamline of the Australian Synchrotron and the New Zealand Synchrotron Ltd. for financial assistance.

# Crystallographic Orientation of a Palladium-defect Pair in Germanium from Perturbed Angular Correlation Spectroscopy

Adurafimihan A. Abiona[a,*], William J. Kemp[a], Aidan P. Byrne[b],
Mark C. Ridgway[c] and Heiko Timmers[a]

[a] *School of Physical, Environmental and Mathematical Sciences, The University of New South Wales, Canberra Campus, ACT 2602, Canberra, Australia.*
[b] *Department of Nuclear Physics, Research School of Physics and Engineering, Australian National University, Canberra, ACT 200, Australia.*
[c] *Department of Electronic Materials Engineering, Research School of Physics and Engineering, Australian National University, Canberra, ACT 200, Australia.*

[*] *On leave from Centre for Energy Research and Development, Obafemi Awolowo University, Ile-Ife, Nigeria*

Time differential perturbed angular correlation spectroscopy was performed for different orientations of crystalline germanium following diluted doping of the material with the hyperfine interactions probe $^{100}$Pd/$^{100}$Rh. Results are interpreted in order to identify the crystallographic orientation of a palladium-defect pairing that is most pronounced following annealing at 500 °C. Orientation in the <100> direction can be excluded. The data are consistent with a substitutional palladium atom pairing with a vacancy in a nearest-neighbours configuration and the pair being oriented in the <111> direction. This is similar to the palladium-vacancy pairing known to occur in n-type silicon.

## 1. Introduction

Palladium (Pd) has been shown to induce Metal Induced Crystallisation (MIC) of germanium at a lower temperature than many other metals, such as, for example, Cu, Ni, Co or Pt [1-3]. Following MIC, Pd atoms may be left behind in crystalline Ge. Consequently the performance of devices based on this material may be affected. With time differential perturbed angular correlation (TDPAC) spectroscopy using the probe $^{100}$Pd/$^{100}$Rh, the local lattice environment of palladium point defects can be studied.

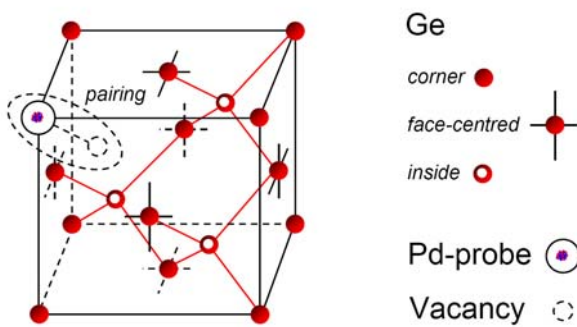

Fig. 1. Illustration of a Pd-vacancy pair in the diamond lattice of germanium oriented in the <111> direction.

Previous TDPAC measurements with the $^{100}$Pd/$^{100}$Rh probe on intrinsic germanium identified a non-zero electric field gradient (EFG) at the probe location [4, 5]. The EFG may be caused by the pairing of a Pd-atom with a neighbouring defect, most likely a vacancy. This is illustrated in Fig. 1 for a <111> orientation.

It has been shown [5] that the defect pairing is most pronounced after annealing the germanium at 500 °C. Further annealing at 700 °C dissociates the pair. The measured quadrupole coupling constant $\nu_Q$ associated with the observed EFG is close to that which has been observed for $^{100}$Pd/$^{100}$Rh in highly doped n-type silicon by Dogra *et al.*

[6-10]. This may suggest that both effects are of similar origin. In silicon the effect has been attributed to the formation of Pd-vacancy pairs, with the Pd being substitutional and the vacancy located in the <111> crystallographic direction. This directionality has been inferred from TDPAC orientation measurements [9, 10]. Equivalent orientation measurements for $^{100}$Pd/$^{100}$Rh in intrinsic germanium were outstanding and have now been performed. Initial results are presented and discussed in this paper.

## 2.    Experimental Details

The $^{100}$Pd/$^{100}$Rh probe nuclei were synthesized with the 14 UD Pelletron accelerator at the Australian National University in Canberra via the fusion evaporation reaction $^{92}$Zr($^{12}$C, 4n)$^{100}$Pd [11]. The $^{12}$C beam energy was 70 MeV. Synthesized probe nuclei were recoil-implanted with energies of several MeV into a germanium specimen cleaved from a (100) germanium wafer. Cleaving occurred on (110) planes.

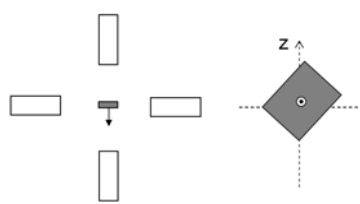

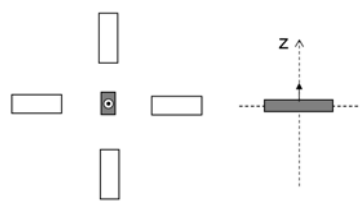

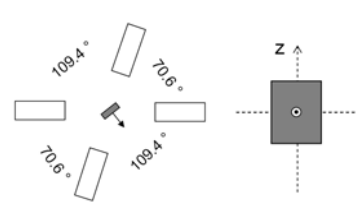

Fig. 2. Respective orientations of the detector array and the germanium specimen (grey). The arrow points in the <100> direction.

The as-implanted germanium specimen was annealed for 20 min at 500 °C in argon flow inside a tube furnace. The TDPAC spectroscopy [12, 13] was performed using a conventional set-up of four cylindrical NaI scintillation detectors forming a planar array. Detectors provided both start and stop signals to a time-to-analogue converter (TAC). The coincidence time distributions of the two γ-rays in the 84 keV - 74 keV γ – γ cascade in $^{100}$Rh were recorded for all detector combinations using NIM-standard electronics [4]. These two γ-rays populate and depopulate the intermediate $2^+$ state in $^{100}$Rh.

Three measurements were performed with detectors aligned with the <100>, the <110> and the <111> crystallographic directions, respectively. This is illustrated in Figure 2. All time distributions obtained were corrected for statistical background events. The distributions for the 90° (70.6°) detector combinations and those for the 180° (109.4°) detector combinations were averaged, thus compensating for differences in detection solid angle and detector efficiency. Finally, using the conventional prescription, the ratio function $R(t)$ was extracted from the data [12].

A modulation of $R(t)$ may be associated with a precession of the γ - γ anisotropy about the direction of the EFG [12,13]. Varying the respective orientation of detector axes and EFG direction modifies the modulation of the ratio function. The comparison of ratio functions $R(t)$ measured for different specimen orientations therefore permits the determination of the crystallographic direction of the electric field gradient.

## 3. Results and Discussion

Figure 3 shows the ratio functions $R(t)$ measured for the three specimen orientations. The data consistently show at $t = 110$ ns the first minimum of the modulation pattern observed previously [4, 5] with an associated quadrupole coupling constant of $\nu_Q = 10.6$ (2) MHz. However, in contrast to the previous measurements, the modulation is less pronounced. The fraction of $^{100}$Pd probe atoms pairing with the defect is thus smaller. The local maximum expected near $t = 220$ ns is only barely visible. This is due to damping of the ratio function caused by non-unique probe environments. It is likely that the annealing procedure applied in the present work either insufficiently cured implantation-induced lattice modifications or, alternatively, has led to the dissociation of Pd-defect pairs, which was previously observed following annealing at 700 °C.

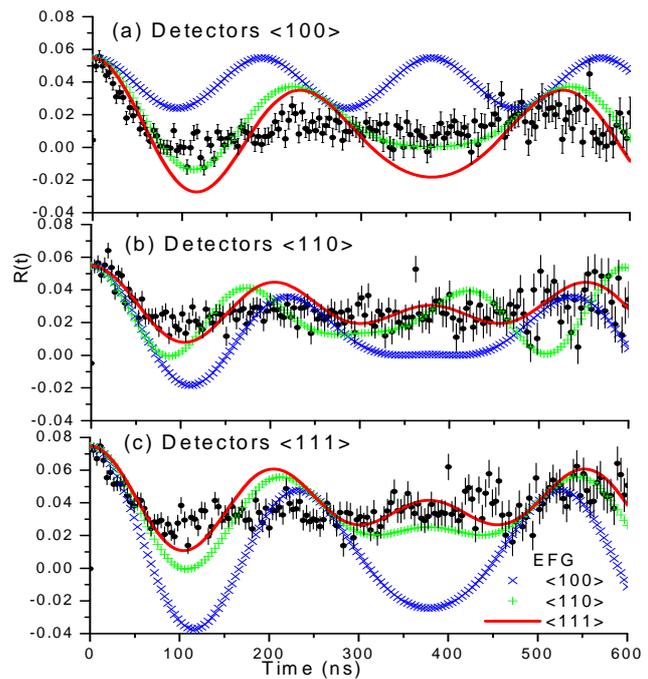

Fig. 3. Measured ratio functions $R(t)$ for $^{100}$Pd/$^{100}$Rh in germanium for the three specimen orientations. Calculations assuming that every palladium probe atom pairs with a vacancy in an otherwise undisturbed lattice are shown for the EFG orientations <100>, <110> and <111>.

It may tentatively be observed that for detector alignment with the <100> direction (Fig. 3a) a modulation exists with two possible maxima near $t = 220$ ns and 500 ns, which are separated by a broad minimum between 300 ns and 400 ns. Figs 3b and c show that detector alignment with the <110> and <111> axes results in a larger mean value of $R(t)$ than apparent in Fig. 3a for alignment with the <100> direction. Also, a local maximum between $t = 300$ ns and 400 ns may exist for both of these two specimen orientations.

Ratio functions can be simulated and fitted using the code *Nightmare* which is based on the routine *NNFit* [14]. Such simulations are also displayed in Figure 3 for all three specimen orientations studied. The calculations assume defect pair alignment, and thus EFG direction, either along a <100> direction, or a <110> direction or a <111> direction, respectively. The simulations further assumed no deviation from axial symmetry ($\eta = 0$), a single 100% probe fraction representing the Pd-vacancy pair and a unique local lattice environment excluding damping. The simulations for EFG direction along a <111> axis correspond to the orientation of the defect pair that is illustrated in Fig. 1.

The calculations show that different respective orientations of detector axes and EFG direction result in different modulation patterns in the corresponding ratio function. In particular, the mean of the ratio function varies. Furthermore, depending on detection geometry, a local extremum near $t = 370$ ns is either pronounced or somewhat attenuated. Careful inspection of the modulation patterns of $R(t)$ suggests that the defect pair is either oriented, as expected, in the crystallographic <111> direction, or possibly in the <110> direction.

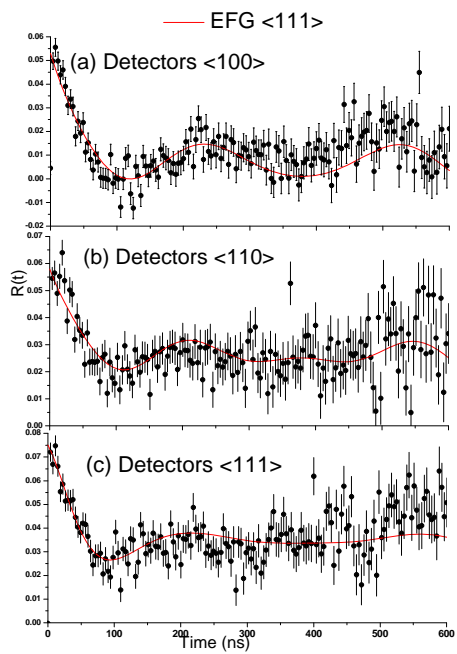

Fig. 4. Measured ratio functions *R(t)* for $^{100}$Pd/$^{100}$Rh in germanium for the three specimen orientations. The data are compared with a calculation that assumes that about 25% of the palladium probe atoms pair with vacancies and that the defect pair and EFG are oriented in the <111> direction.

Figure 4 compares the data with more detailed *Nightmare* calculations that assume defect pair orientation in a crystallographic <111> direction. Agreement is good for an assumed probe fraction of 25% pairing with the defect in an otherwise undisturbed lattice environment and the remainder of probes having diverse non-unique lattice environments.

## 4. Summary and Conclusions

Time differential perturbed angular correlation spectroscopy was performed for different orientations of a germanium specimen that had been recoil-implanted with the hyperfine interactions probe $^{100}$Pd/$^{100}$Rh. The results confirm the quadrupole coupling constant associated with a palladium-vacancy pairing that had been discovered previously. The probe fraction forming this defect-pair in the present specimen was, however, smaller. Tentative interpretation of the data excludes orientation along the <100> direction. Instead the data are consistent with a palladium-defect dumbbell that is oriented in the expected <111> direction. Improved annealing of the sample following probe implantation in future work may strengthen this interpretation.

# Identification of $Fe^{3+}$ Site Coordinations in NAu-2 Nontronite

J.D. Cashion[a], W.P. Gates[b], T.L. Greaves[a1] and O. Dorjkhaidav[a2]

[a] *School of Physics, Monash University, Victoria 3800, Australia.*
[b] *Department of Civil Engineering, Monash University, Victoria 3800, Australia.*

$^{57}$Fe Mössbauer spectra of NAu-2 nontronite show a quadrupole split spectrum with no trace of magnetic ordering down to 5 K. Two quadrupole doublets are due to octahedral *cis* $Fe^{3+}$ and one to tetrahedral $Fe^{3+}$. However, the fourth doublet, with a splitting of 1.28 mm/s is attributed to a new $Fe^{3+}$ species, possibly in a ditrigonal cavity or a polyoxocation species in the interlayer.

1.  **Introduction**

Nontronites are the iron-rich end-members of the smectite group of clay minerals. Structurally, they consist of a sheet of octahedrally coordinated cations ($Fe^{3+}$, $Al^{3+}$, $Mg^{2+}$, $Fe^{2+}$) which is sandwiched between two tetrahedrally coordinated sheets of cations ($Si^{4+}$, $Al^{3+}$, $Fe^{3+}$). The charge imbalance introduced by the lower valency ions in each sheet is compensated by the incorporation of additional cations, usually alkali or alkaline earths, into the interlayer region between two opposing layers.

Iron in the octahedral sheet is coordinated to four oxygen ions and two hydroxyl ions which allows for three different sites, two *cis* and one *trans* with respect to the hydroxyl ions. The smectites are di-octahedral, so that only two of these sites are filled. At the nontronite end, it is accepted that both occupied sites are *cis*, while at the iron-poor montmorillonite end, iron is probably randomly distributed between one *cis* and the *trans* sites. Thus, iron can exist in three different arrangements with respect to its coordinating anions, convoluted with many more arrangements with respect to its nearest and next nearest neighbour cations. In the general case, there is insufficient resolution in $^{57}$Fe Mössbauer spectra to distinguish these possibilities. It was pointed out some while ago that there were cases where the literature assignments were not consistent with the crystallography [1] and that the effect of cation neighbours could be important [2,3]. Recently, positive identifications of Mössbauer parameters with particular iron coordinations have been made [4,5], by selection of clay minerals with particularly simple or related chemistry.

In this paper, we consider the South Australian Uley nontronite, NAu-2, for which the literature structural formula is $M^{+}_{.72}[Si_{7.55}Al_{.16}Fe_{.29}](Al_{.34}Fe_{3.54}Mg_{.05})$ [6]. This is similar to the NAu-1 sample previously analysed [5] whose composition is $M^{+}_{1.05}[Si_{7.00}Al_{1.00}](Al_{0.29}Fe_{3.68}Mg_{0.04})$. The major differences which will affect the Mössbauer spectra are the higher $Si^{4+}$ concentration, the presence of some tetrahedral $Fe^{3+}$ and the resulting, much smaller tetrahedral $Al^{3+}$ concentration in NAu-2. The Mössbauer spectrum of NAu-2 has previously been published [7] and we will return to their assignment of the two principal features of the spectrum to *cis* and *trans* sites in the Discussion.

**2. Experimental**

The sample was carefully purified by saturating initially with $Na^{+}$ and washing by centrifugation to obtain a dispersed fine fraction of < 0.15 μm [6]. Random and orientated powder X-ray diffraction and infrared spectroscopy were used to check the purity. The sample was then reacted and washed three times with 0.1M $CaCl_2$ to produce calcium saturated NAu-2 and then dialyzed to remove excess salts before being finally oven dried at 105 °C.

---

[1] Present address: CSIRO Materials Science and Engineering, Clayton, Vic 3169, Australia.
[2] Present address: Dept. of Safeguards, I.A.E.A., Vienna, Austria

Mössbauer spectra were taken using a $^{57}Co$ in Rh source, mounted on a conventional, symmetrical waveform, constant acceleration drive. All isomer shifts (IS) are quoted relative to α-Fe at room temperature. The spectra were fitted with Lorentzian profiles.

**3. Results**

Spectra were taken at both low and high velocity but only the former will be shown since no evidence was seen for magnetically ordered species. A room temperature spectrum is shown in Fig. 1. It is an unusual spectrum for a clay mineral because of the pronounced outer structure, requiring it to be fitted with four doublets, the parameters for which are in Table 1. The spectra at 78 K and 5 K are almost identical, except for the thermal shift and the parameters for the 5 K spectrum are also given in Table 1. The only changes of note are the increase in the areas of doublets 3 and 4 at low temperatures which should be the more reliable estimates of their proportion. It is usual that the bonding in tetrahedral sites is weaker than that for octahedral sites in the same material. The increase in the outer doublet area also indicates weaker bonding and helps provide evidence for its origin as will be discussed later.

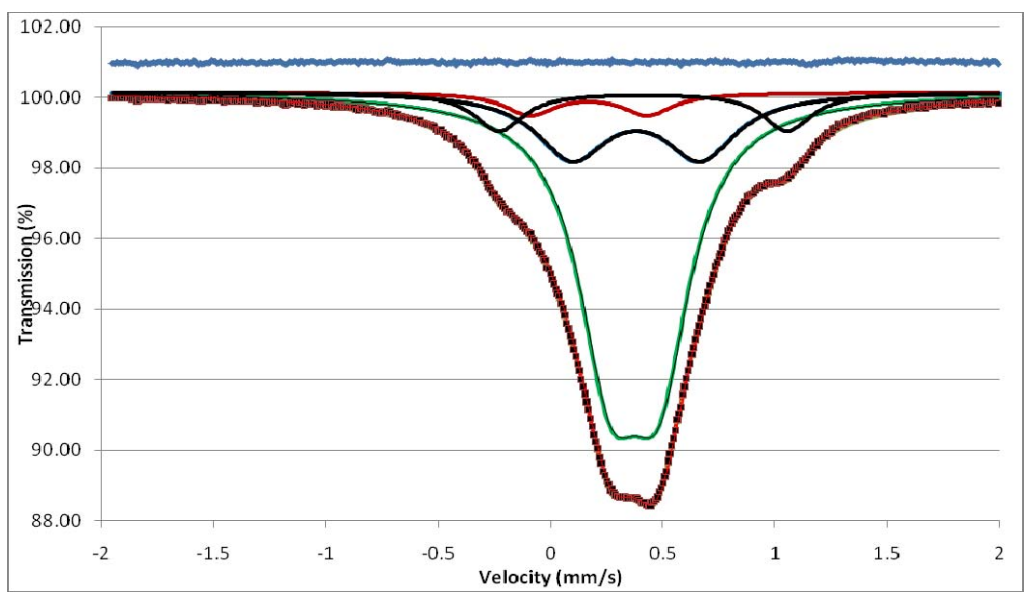

Fig. 1. $^{57}Fe$ spectrum of NAu-2 at room temperature

**4. Discussion**

The dominant doublet has parameters very close to those of the first NAu-1 doublet which was assigned [5] the coordination of three octahedral $Fe^{3+}$ neighbours and then eight tetrahedral $Si^{4+}$ neighbours, or $Fe(300)[Si_8]$ in our terminology, where the number in parentheses refers to the number of octahedral $Fe^{3+}$, $Al^{3+}$ and $Mg^{2+}$ neighbours respectively. This is the most symmetrical coordination and so should produce the smallest quadrupole splitting (QS). It is also the most probable, based on the chemical analysis and random site occupation. The second doublet also has parameters almost the same as those in NAu-1 and is ascribed [5] to the configurations $Fe(300)[Si_7Al]$ and $Fe(300)[Si_7Fe]$, the latter being absent in NAu-1. The higher $Si^{4+}$ concentration in NAu-2 means that substitutions for two $Si^{4+}$ ions have probabilities of <4%. The third doublet is due to tetrahedral $Fe^{3+}$, with an intensity of 9%, consistent with estimates from a variety of other techniques [6].

The first three assignments agree very well with the conclusions of Besson *et al* [8] in their experiments on Garfield nontronite. We note that our values for the relative areas of the two doublets for both NAu-1 and NAu-2 as a function of the $R^{3+}$ tetrahedral substitution fit very well on their Fig. 8.

Table 1. Parameters from least squares fits to the spectra of nontronite NAu-2.

| Temperature | Site | IS (mm/s) | QS (mm/s) | $\Gamma$ (mm/s) | Area (%) |
|---|---|---|---|---|---|
| Room | Oct 1 | 0.37(1) | 0.22(1) | 0.35(1) | 68(4) |
|  | Oct 2 | 0.37(1) | 0.58(1) | 0.37(1) | 20(5) |
|  | Tet | 0.16(1) | 0.53(1) | 0.27(1) | 5(1) |
|  | Doublet 4 | 0.41(1) | 1.28(1) | 0.26(1) | 8(1) |
| 5 K | Oct 1 | 0.49(1) | 0.22(1) | 0.31(1) | 57(2) |
|  | Oct 2 | 0.51(1) | 0.55(1) | 0.30(1) | 21(2) |
|  | Tet | 0.29(1) | 0.47(1) | 0.26(1) | 9(1) |
|  | Doublet 4 | 0.52(1) | 1.25(1) | 0.25(1) | 13(1) |

Assigning the outer doublet is not so straightforward. Jaisi *et al* [7] tentatively ascribed it to *trans*-$Fe^{3+}$, even though it is recognised that nontronites are normally *trans* vacant [9]. It is noteworthy that, along with the tetrahedral $Fe^{3+}$, it was the most easily reduced of the $Fe^{3+}$ sites in their experiments and there seems little reason for the *trans* site to be so much more susceptible to reduction than the *cis* sites. The QS of 1.28 mm/s is very large for $Fe^{3+}$ in clay minerals. Examination of the spectra of related clay minerals such as montmorillonites, which are known to have *trans* occupancy and a large $Mg^{2+}$ concentration, which widens the splitting [4], shows that the QS of the outer doublets can reach 1.2 mm/s (e.g. [10]) but are usually broadened and featureless, as was Garfield nontronite [8]. Consequently, we would argue that this feature is not due to a *trans* configuration but a definite and new configuration.

We have noted two experiments which produced similar parameters as a well-defined doublet. Bakas *et al* [11] observed a doublet in successive oxidation-reduction experiments on a montmorillonite with parameters IS = 0.34(1) mm/s, QS = 1.27(5) mm/s which became magnetically ordered at low temperatures and which they attributed to superparamagnetic maghemite. We are confident, with the benefit of 17 years of hindsight, that their species was not superparamagnetic maghemite, but its true origin is unknown. The concentration of our species is much smaller and this could account for the lack of magnetic ordering. Lego *et al* [12] observed similar parameters in three Czech bentonites, with the Hroznětín outer doublet being well defined and ascribed generically as being octahedrally coordinated.

We can define the following requirements to help in identification of the configuration: (1) the species is not ion-exchangeable [6]; (2) it must be in a very distorted site and have relatively weak bonding; (3) there is no low temperature splitting of the spectrum, so it is not due to any of the usual ferric (oxyhydr)oxides; (4) it must be able to contribute to the structural collapse on reduction, during which it will suffer significant dissolution.

One possibility is that the ions responsible for this signature are in the interlayer. They cannot be isolated $Fe^{3+}$ ions because the resulting interlayer charge would be much too large [6]. There are various structures which can be in the interlayer, of which the best known is pillaring through the incorporation of the $Al^{3+}$ or $Fe^{3+}$ based Keggin molecule, $[FeO_4 Fe_{12}(OH)_{24}(OH_2)_{12}]^{7+}$. This only has 7+ rather than the 39+ of the metal ions which brings the effective charge contribution of the $Fe^{3+}$ to an acceptable level. There is little Mössbauer data on pillared clays, but Gangas *et al* [13] showed that pillaring in the SWa-1 nontronite and an unnamed montmorillonite can produce magnetic ordering by providing magnetic bridges to overcome the magnetic frustration of the honeycomb octahedral sheet. However, only low temperature spectra were shown, without fits, so it is not possible to discern the QS of the unordered component. Whether magnetic ordering occurs will depend on the concentration of pillars. Aouad et al [14] observed pillaring in a saponite and a montmorillonite which they attributed to akaganéite and whose parameters do not match ours. It has been shown ([15] and refs. within) that $Fe^{3+}$ in Fe and Fe/Al pillared montmorillonite can be reduced by various liquid and vapour treatments, sometimes resulting in the collapse of the interlayer structure. Of course, it is not essential that the Keggin molecule is the cause, any polyoxocation incorporating several $Fe^{3+}$ ions, with some anions to reduce the total charge, and satisfying the previous criteria, would also be acceptable.

Another possibility is that the $Fe^{3+}$ ions are in the ditrigonal cavity. There have been many observations of $Li^+$ and $Ca^{2+}$ there, but the only 3d elements observed have been $Cu^{2+}$ [16] and $Cr^{3+}$ [17]. However, there seems no reason why $Fe^{3+}$ could not also be accommodated. Thus, with the lack of prior observations, there are no known Mössbauer parameters for this site, but it would be very distorted, the ions would be isolated and, importantly, it does seem to satisfy all the requirements listed above. A final possibility of $Fe^{3+}$ being on clay particle edges seems unlikely as one would expect observations of such a site in many other clay minerals.

**5. Conclusions**

We have shown that three of the components of the Mössbauer spectrum of NAu-2 are the same as those identified in other nontronites. However, the final component is due to a new species which we assign to probably $Fe^{3+}$ in a ditrigonal cavity or possibly a naturally occurring polyoxocation pillaring molecule. We note that, although we disagree with Jaisi et al [7] over the origin of the outer doublet, this does not affect their overall results, nor the interpretations of the results of their bioreduction experiments.

**Acknowledgements**

We are grateful for the support of the Australian Research Council, Monash University and SmecTech Research Consulting and to T W Turney for helpful discussions.

# Continuous-Wave Terahertz Spectroscopy as a Non-Contact Non-Destructive Method for Characterising Semiconductors

E. Constable and R. A. Lewis

*Institute for Superconducting and Electronic Materials, University of Wollongong, Wollongong NSW 2522, Australia.*

Using the technique of terahertz photomixing, a continuous-wave terahertz source is used to characterise various semiconductors in the frequency range from 0.06 to 1.0 THz. By directly analysing the interference pattern of the transmission through semiconductor wafers using Fabry-Pérot theory, information regarding the carrier concentration, sample thickness and refractive index is obtained without physically altering the sample. The continuous-wave technique enables measurements to be made at much lower frequencies than achievable with traditional pulsed-wave terahertz techniques.

## 1. Introduction

Semiconductors are often characterised by their resistivity at room temperature [1] however standard dc measurement techniques require the attachment of electrodes to the sample. Optical characterisation of semiconductors offers a solution to this problem. Terahertz (THz) frequency radiation features absorptions associated with excitations of phonons and free carriers in semiconductors [2]. Early measurements in the THz frequency regime relied on specialised measurements using synchrotron radiation [2]. The advent of THz time domain spectroscopy (TDS) allowed for a cheap and effective way of characterising semiconductors using THz radiation. The results shown in this paper demonstrate the ability of an alternative THz source known as a two-colour THz source to characterise semiconductor wafers. The two-colour system is generally much cheaper and simpler to operate than a TDS system and also has the benefit of higher resolution and sensitivity at lower frequencies.

## 2. Methodology

2.1. Two-colour continuous-wave THz system

The transmission spectra of the semiconductor samples are taken using a two-colour continuous-wave THz spectrometer. Two tuneable near infrared laser diodes (lasers 1 & 2 of Fig. 1) operating at wavelengths around 853 nm are used to pump a low-temperature-grown gallium arsenide (LTG GaAs) photomixer. The laser diodes are tuned to have a frequency difference within 1 THz. The mixed signal incident on the photomixer has a beat frequency equal to this frequency difference. The desirable properties of LTG GaAs allows the beat frequency to oscillate charge carriers within the chip [3]. A log-periodic antenna built into the chip couples this oscillation to free space producing the continuous-wave THz radiation.

A spectrum is taken with the sample positioned at the THz beam focus shown in Figure 1. The transmitted signal is refocused onto a Schottky diode detector. A reference spectrum is then taken with no sample in the beam path. A simple ratio of the two spectra gives the transmittance, *T*.

2.2. Theory

Applying the Drude model, the relative permittivity of the sample is given as $\varepsilon = \varepsilon_1 - i\varepsilon_2$, where:

$$\varepsilon_1 = \varepsilon_\infty - \frac{ne^2\tau^2}{m_e\varepsilon_0(1+\omega^2\tau^2)}, \qquad \varepsilon_2 = \frac{ne^2\tau}{m_e\varepsilon_0\omega(1+\omega^2\tau^2)},$$

$n$ is carrier concentration, $m_e$ is the effective electron mass and $\tau$ is the collision time. Using Fabry-Pérot theory the transmittance of the sample for *s*-polarisation can be expressed as:

$$T = \left|\frac{(1+\Gamma)(1-\Gamma)}{e^{i\theta} - \Gamma^2 e^{-i\theta}}\right|^2,$$

where:

$$\theta = \left(\frac{2\pi d}{\lambda_0}\right)(\varepsilon - \sin^2\theta_i)^{1/2}, \quad \Gamma = \frac{(Z-1)}{(Z+1)} \quad \text{and} \quad Z = \frac{\cos^2\theta_i}{(\varepsilon - \sin^2\theta_i)^{1/2}}$$

Here $\theta_i$ is the angle of incidence, $d$ the thickness of the sample, and $\lambda_0$ is the free-space wavelength of the incident radiation [2]. Often $\varepsilon^{1/2}$ is referred to as the complex refractive index, where $\varepsilon^{1/2} = \eta + i\kappa$ [4]. Here $\eta$ is the real refractive index and $\kappa$ is the extinction coefficient. If absorption effects are ignored the complex and real refractive indices are equivalent.

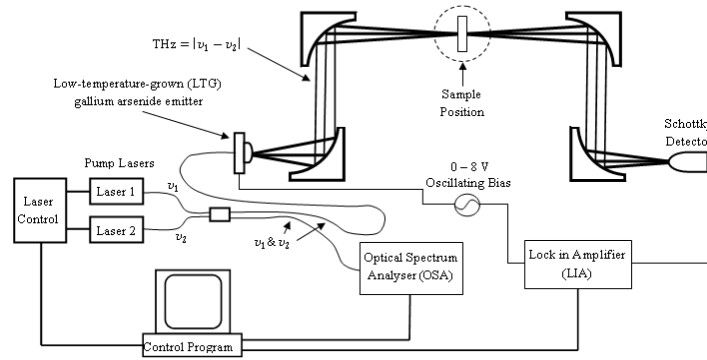

Fig. 1. Diagram of two-colour THz spectrometer setup for measuring semiconductor transmittance.

### 3. Results and discussion
3.1. Silicon wafer carrier concentrations

Figure 2 shows the spectrum of the transmitted radiation through a silicon sample and the reference spectrum with no sample in the beam path. The Fabry-Pérot oscillations in the silicon spectrum are clearly evident. The sharp absorption dips seen at 0.55 THz and 0.78 THz are part of the well-known H2O molecule rotational series and occur as a result of water vapour in the laboratory air.

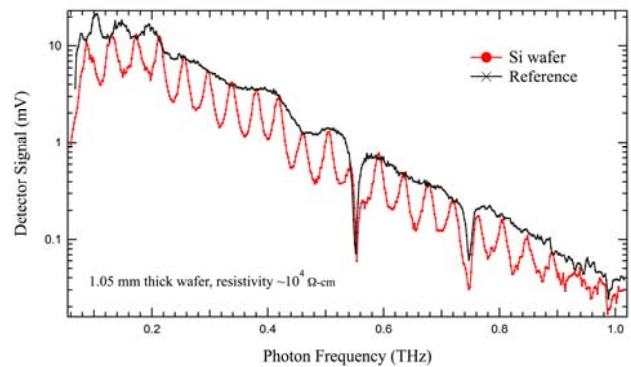

Fig 2. Spectrum of silicon wafer and background reference from 0.06 to 1.0 THz.

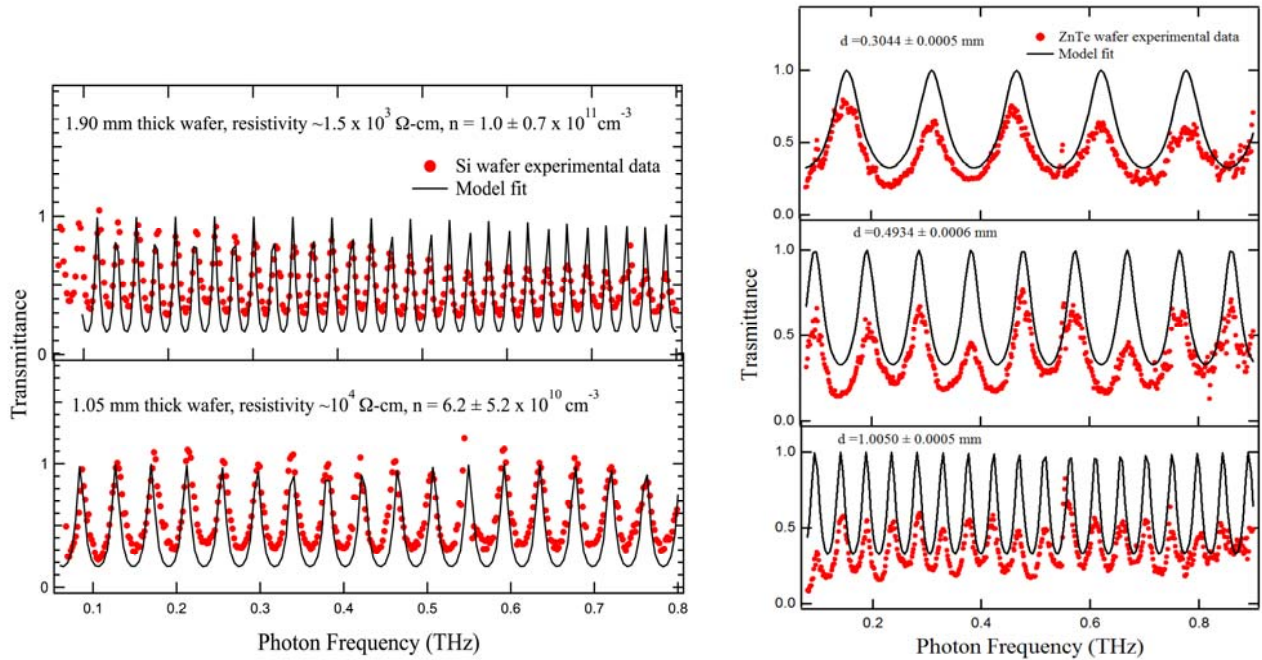

Fig 3. Experimental transmittance fitted using the Drude model for two silicon samples of varying resistivity (left) and a series of ZnTe samples of varying thickness (right).

The transmittance of two high-resistivity silicon wafers is shown in Figure 3. The experimental transmittance is fitted from 0.06 to 0.8 THz using a least squares method with the carrier density as the fitting parameter. The electron mobility can be considered, to a good approximation, independent of the carrier density. We then take a constant value of $\tau = 0.2$ ps calculated from the known electron mobility of 1350 cm$^2$/Vs, a permittivity of $\varepsilon_\infty = 11.7$ and an effective mass of $m_e = 0.26 m_0$ in the fit [1]. As can be seen in Figure 3, the model shows good agreement with the experimental data indicating carrier concentrations of $n = (6.2 \pm 5.2) \times 10^{10}$ cm$^{-3}$ and $n = (1.0 \pm 0.7) \times 10^{11}$ cm$^{-3}$ for the $1.0 \times 10^4$ Ω-cm and $1.5 \times 10^3$ Ω-cm samples respectively. These values agree with expected concentrations for samples of high resistivity [1, 2, 5]. The high level of uncertainty in the results can be explained by the lack of sensitivity the transmittance has to changes in the carrier concentration at the low concentration levels found in the high resistivity samples used. For a given thickness a wide range of concentrations could easily fit the data. It is expected that for samples of lower resistivity the carrier concentration will play a more significant role in the absorption and therefore would be able to be determined more accurately from a model fit of the transmittance [5]. It should be noted that the period of the oscillations in the transmittance is dependent to first order on the sample thickness. Therefore, the accuracy of this method is strictly limited by the accuracy to which the sample thickness is known.

3.2. Zinc telluride wafer thicknesses

By ignoring absorption effects the fitting function is simplified by fixing $\varepsilon^{1/2}$ equal to the refractive index. Then, for a known refractive index, the characterisation method may be used for high accuracy measurement of a sample's thickness by fitting for $d$. Figure 3 shows the experimental and fitted transmittance for a series of zinc telluride (ZnTe) samples of varying thickness. A constant refractive index of 3.188 is used [6]. The poor fit for the amplitudes of the oscillations is a result of neglecting the absorption effects. As $d$ is only dependent on the period of oscillation this method is acceptable when determining $d$. The thickness of each ZnTe sample is determined to within 95% confidence and agrees closely with the nominal thicknesses given by the manufacturer.

3.3. Zinc selenide refractive index

Alternatively, if the sample thickness is known to a high accuracy the refractive index can be determined directly from the transmittance by fixing $d$ and fitting for $\varepsilon^{1/2}$. Figure 4 demonstrates this with the refractive index of a 0.76 mm thick zinc selenide (ZnSe) wafer determined to be $3.008 \pm 0.002$.

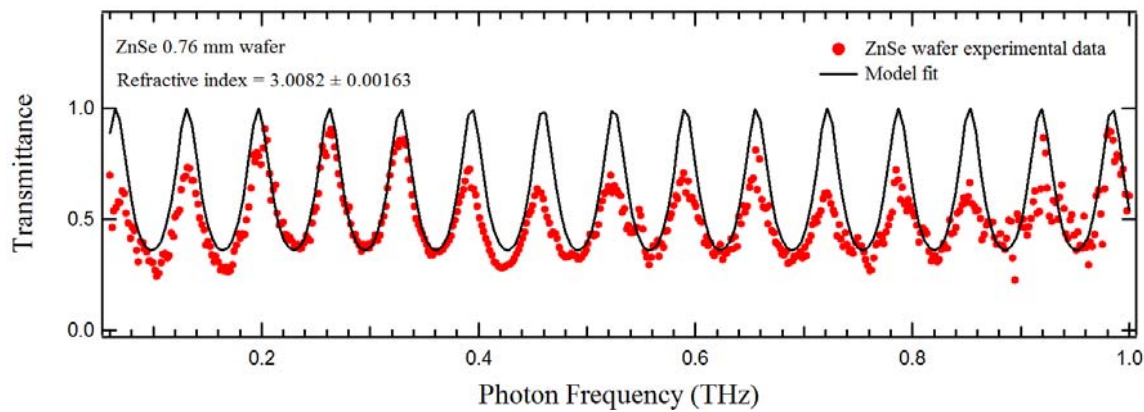

Fig. 4. Experimental transmittance of ZnSe sample fitted with Drude model for determining refractive index.

## 4. Conclusion

In summary we have shown that a two-colour THz spectrometer may be used to measure the transmittance of high-resistivity semiconductors allowing a non-contact and non-destructive method of determining their carrier concentrations, thicknesses and refractive indices. However determining accurate values of carrier concentrations using this method is limited to samples of a lower resistivity with known sample thicknesses. Therefore, this method is likely to be most useful in industry quality control where less accurate measurements are sufficient for the nominal parameters given with a product. The two-colour THz spectrometer is a top candidate for this type of application due to its low start-up and running costs when compared to other THz spectrometer setups, such as time domain spectroscopy.

**Acknowledgments**

This work is supported by the Australian Research Council and by the University of Wollongong.

# Shape versus Size Relationship of Prosthetic Ultra-high Molecular Weight Polyethylene Wear Particles

Laura G. Gladkis [1,2], Jennifer M. Scarvell [2], Paul N. Smith [2] and Heiko Timmers [1]

[1] *School of Physical, Environmental and Mathematical Sciences, University of New South Wales at ADFA, Canberra, ACT 2600*
[2] *Trauma and Orthopaedic Research Unit, The Canberra Hospital, PO BOX 11, Woden, ACT 2606*

Atomic force microscopy (AFM) is used to characterize in detail ultra-high molecular weight polyethylene (UHMWPE) wear debris from a low contact stress (LCS) knee prosthesis, actuated with a state of the art knee simulator. The size and shape of debris particles is quantified in all three spatial dimensions. The three-dimensional AFM information indicates that for the prosthesis and the conditions studied here debris particles are mostly oblate, with the smallest particles having the greatest degree of deformation.

## 1. Introduction

Ultra high molecular weight polyethylene (UHMWPE) is the selected material for tibial inserts in total knee arthroplasty due to its low wear rate, excellent abrasion resistance and high mechanical stability [1, 2]. In spite of the adequate material properties, wear is observed because of the highly dynamic stresses experienced in a knee joint, therefore affecting the clinical performance of the joint. Particle debris created from the tibial insert interacts with many kinds of white blood cells and immune cells, starting a cascade of events that ends with prosthesis loosening and therefore revision surgery [3, 4].

The abundance and morphology of UHMWPE debris particles influences their bio-response. Irregularly-shaped particles have been shown to have an increased biological reactivity [5, 6, 7] compared with regular-shape debris. Our previous study [8] has demonstrated that for the prosthesis studied here but using a constant load actuator, micro-scale particles tend to be more spherical, whereas nanoscale particles have a tendency to be deformed with oblate, flake-like shapes.

Fractionation of different debris particle sizes has often been achieved by filtration, and images of the debris on the filter are typically obtained by scanning electron microscopy (SEM) [5, 6, 9]. With this technique, the two dimensional projection of a particle is determined, and some qualitative contrast information about the particle height can be inferred. With atomic force microscopy (AFM), in contrast, the size and geometry of polyethylene wear particles can be measured in all three spatial dimensions with a precision on the nanometre scale. AFM imaging has rarely been applied to characterize UHMWPE debris [10, 11, 12]. However, Scott *et al.* [11] have demonstrated consistency with scanning electron microscopy (SEM).

In order to expand our previous study [8], debris particles were created under realistic conditions using a PROSIM knee simulator. The size and shape of particles has been measured directly in three spatial dimensions with AFM.

## 2. Materials and methods
2.1. PROSIM knee simulator

A pristine knee prosthesis was actuated with a PROSIM single-station knee simulator (Simulation Solutions Ltd) in order to obtain UHMWPE wear debris particles. The prosthesis was a standard size, cobalt-chrome on UHMWPE, low contact stress (LCS) mobile bearing knee system (DePuy/Johnson & Johnson); see Fig.1 (a). A walking regime was simulated by following ISO standards (see Fig. 1 (b) and (c) for typical flexion-extension and axial load parameters). The simulation was performed for 3.7 days, which roughly equates to 6 months of walking. Since the focus of this work is on the spectrum of debris particles and their characterization, deionised water (Millipore Milli-Q) was used rather than bovine serum. Moreover, with water as lubricant the extraction is simplified.

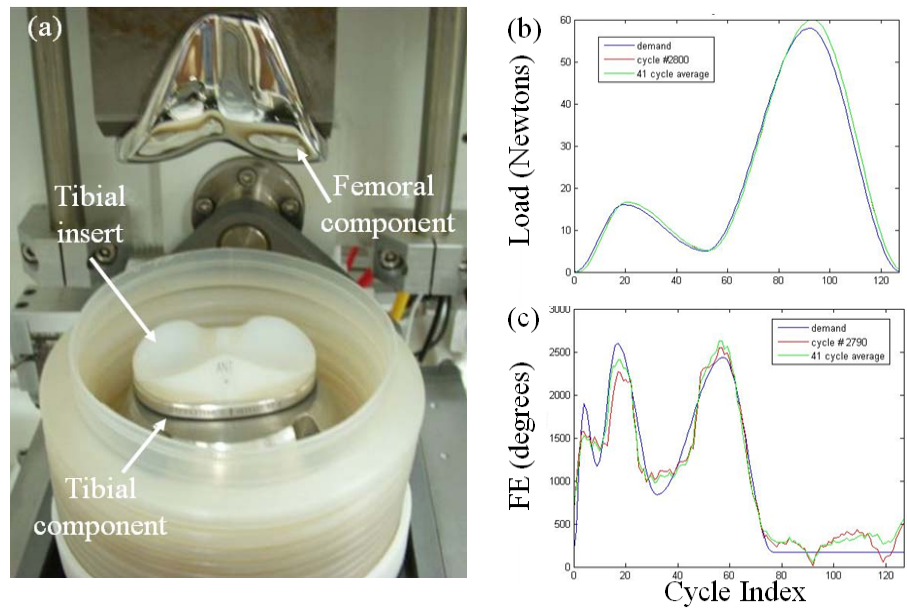

Fig. 1. (a) Photo of the LCS Knee prosthesis used in this experiment. (b) Flexion-Extension and (c) Axial load corresponding to a walking simulation (ISO standards).

2.2. Sample preparation

The water lubricant containing UHMWPE debris particles was collected at the end of the PROSIM simulation. Twelve particle size fractions were obtained following a novel filtration protocol [8]. The nominal filter pore diameters were 10, 5, 3, 1, 0.8, 0.6, 0.4, 0.3, 0.2, 0.1, 0.05 and 0.025 μm. Following filtration, each filter with debris particles was left to dry over 24 hours, stored on a covered Petri dish. In preparation for AFM characterization, a $10 \times 10$ mm$^2$ piece was cut out of the centre of the filter and attached with double-sided adhesive tape onto a mirror-finished silicon substrate. SEM images of wear debris particles on the 20-10 μm and 3-5 μm fraction filter medium are shown in Fig. 2 (a), (b). It is apparent that the particles tend to rest on their largest surface.

2.3. Atomic force microscopy

For each fraction, 5-10 images of debris particles on filter were taken with an AFM, NT-NTEGRA PRIMA (manufactured by NT-MDT) using semi-contact mode. The AFM tip scans a small area on the flat substrate and directly probes the height variations due to debris particles resting on the substrate. Height measurements with this technique have a precision in the nm-range. The lateral $x$- and $y$-resolution of the instrument is better than 20 nm, and the height resolution ($z$-coordinate) is better than 5 nm. In addition to images of filters with debris particles, several images of pristine filters were recorded for comparison. The instrument software package allows for automated particle size measurements. However, it was found that an artificial digitization of the data can occur. Therefore, the length, width and height of each particle in an image have been measured individually by using the length and the cross-sectioning tools of the software.

## 3. Results

In this work the length, width and height of a particle have been defined as previously published [8]. These three parameters were measured for all particles on the AFM images. Most of the particles tended to be non-spherical, with most of them having three significantly different measurements for length, width and height.

Figure 2 (c), (d) shows two typical images for the 0.6-0.8 μm and 0.2-0.3 μm fraction respectively. It is apparent that some of the particles are greatly elongated and fibril-like. Particles less elongated (almost spherical) are also present.

Our previous work [8] established that the sphericity $S$ of a particle [13, 14] is an appropriate quantity to describe their shape. $S$ is a measure of how spherical an object is, and it is defined as the ratio of the surface area of a sphere (with the same volume as the given object) to the surface area of the object. Values of $S$ close to unity correspond to spherical shapes, whereas small values represent deformed particles. The volume of the particle was calculated assuming their shape can be approximated by an ellipsoid, the formula used was:

$$V = (4\pi/3) \times [(L/2) \times (W/2) \times (H/2)]$$

where $L, W, H$ refer to the length, width and height respectively. The sphericity $S$ for UHMWPE debris particles are displayed in Fig. 3 as a function of the ellipsoidal volume. Figure 3 (a) corresponds constant load actuator results [8], (b) to PROSIM knee simulator data.

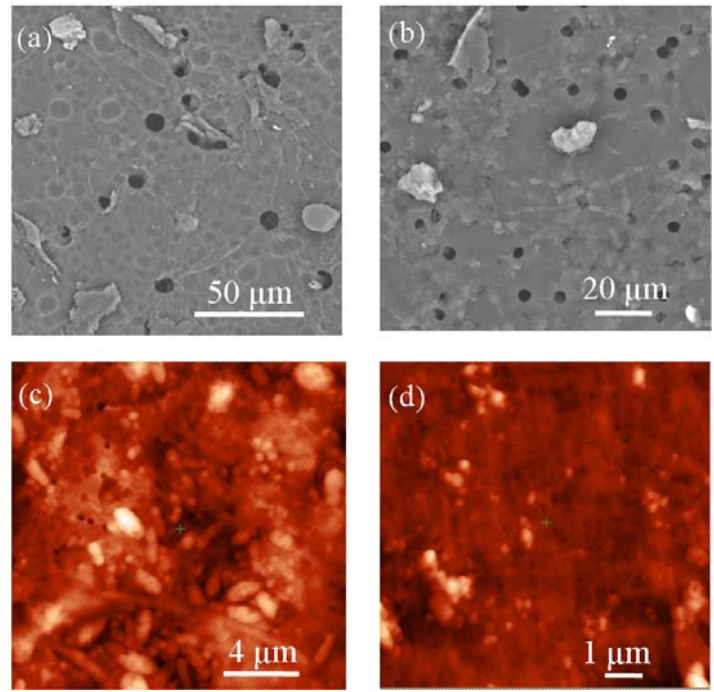

Fig. 2. SEM (top) and AFM (bottom) images of UHMWPE debris on filter paper. The filtered particle fractions shown here are: (a) 10-20 μm, (b) 3-5μm, (c) 0.6-0.8 μm and (d) 0.2-0.3 μm.

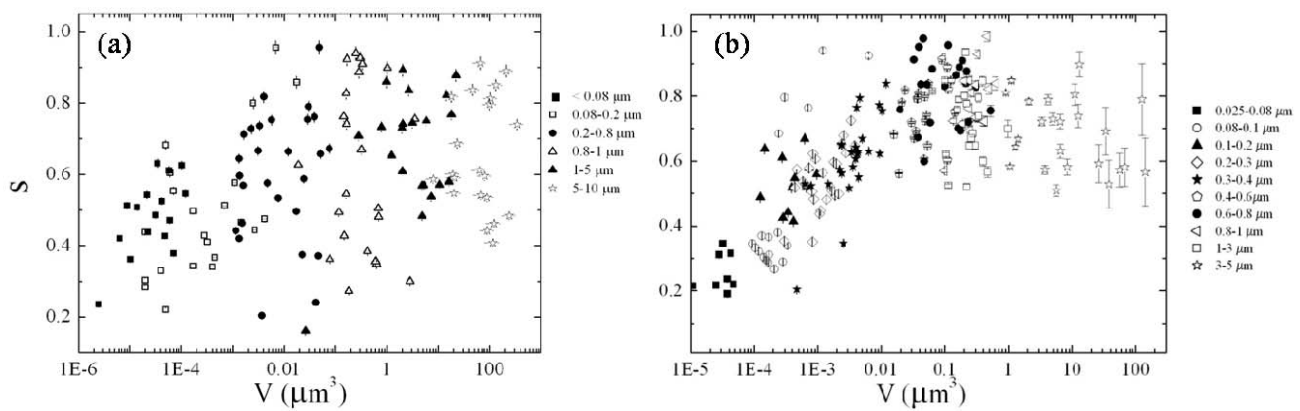

Fig. 3. Scatter plot of sphericity $S$ as a function of particle volume for a walking simulation with a (a) constant load actuator [8] and (b) PROSIM knee simulator. It is apparent that particle sphericity increases with particle size.

In this representation, which takes into account measurements of all three spatial dimensions, an interesting relationship between shape and size is observed. It can be seen for both situations that small particles are strongly deformed, whereas larger particles tend to be less deformed with values of sphericity close to 0.7. On the filter paper, particles tend to precipitate by resting on a large area of their surface. Since the height tends to be significantly smaller than both length and width, which tend to be similar, most particles have oblate shapes. It was observed that the oblate nature of the particles persist for all filtration fractions.

The possibility that small debris particles may be deformed, as they are in this wear study, may have a bearing on their bioactivity.

## 4.  Conclusions

UHMWPE wear debris was produced by actuating a knee prosthesis with a state of the art knee simulator. Atomic force microscopy has been found to be well suited for measurements of all three spatial dimensions of wear particles. For the knee prosthesis and the wear conditions studied here debris particles are mostly oblate, flake-like particles, with the smallest particles having the greatest degree of deformation. Accurate shape information may contribute to a better understanding of the wear mechanisms in prostheses and the bioactivity of UHMWPE particles.

**Acknowledgments**

The authors are grateful to Ian Peterson for giving them access to the NTEGRA Atomic Force Microscope in the School of Engineering and Information Technology at UNSW at ADFA. Thanks to Jake Warner for providing with the knee prosthesis image, DePuy/Johnson & Johnson are acknowledged for the knee prosthesis provided.

# A Frustrated Three-Dimensional Antiferromagnet: Stacked $J_1 - J_2$ Layers

C.J. Hamer[a], Onofre Rojas[a,b] and J. Oitmaa[a]

[a] *School of Physics, The University of New South Wales, Sydney NSW 2052, Australia.*
[b] *Universidade Federal de Lavras, CP3037, 37200-000, Lavras MG, Brazil.*

We study a frustrated three-dimensional antiferromagnet of stacked $J_1 - J_2$ layers. The intermediate 'quantum spin liquid' phase, present in the two-dimensional case, narrows with increasing interlayer coupling and vanishes at a triple point. Beyond this there is a direct first-order transition from Néel to columnar order. Possible applications to real materials are discussed.

1. **Introduction**

The study of frustrated quantum antiferromagnets remains an active field, characterized by a strong interplay between theory and experiment. An archetypal model, which has been extensively studied, is the '$J_1 - J_2$ model', where $S = ½$ spins are located on the sites of a square lattice, with 1st and 2nd neighbour antiferromagnetic interactions $J_1$, $J_2$ [1]. This model shows two kinds of magnetic order at zero temperature, conventional Néel order for small $J_2$ and 'columnar' order for large $J_2$. In the intermediate region $0.4 < J_2/J_1 < 0.6$, strong frustration destroys long-range magnetic order and leads to a disordered 'quantum spin liquid' phase.

It has been argued [2] that the layered materials $Li_2VOSiO_4$ and $Li_2VOGeO_4$ are well represented by this model with $J_2 \gg J_1$, i.e. in the columnar phase. However electronic structure (LDA + U) calculations suggest that the coupling between layers, $J_3$, is by no means negligible. This means that $J_3$ ought to be included in any fitting to experimental data. It also raises the question of the nature of the overall phase diagram of a model of coupled $J_1 - J_2$ layers. The present work is motivated by this question. We employ the well-established technique of linked-cluster series expansions at temperature $T = 0$ [3].

In Figure 1 we show the structure of the model, and a schematic phase diagram. Our calculations confirm the form of this phase diagram.

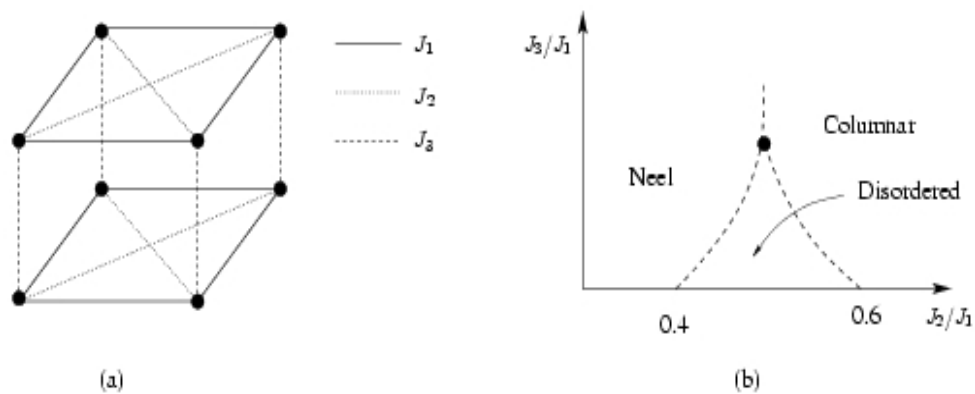

Fig. 1. (a) Tetragonal unit cell of the model; (b) Conjectured phase diagram at $T = 0$. The transition lines are (probably) first-order and the solid circle is thus a triple point.

## 2. Method and Results

2.1 Ground State Properties

Expansions about Néel and columnar ordered Ising ground states yield estimates of the ground state energy and magnetization for any choice of the parameters $J_1$, $J_2$, $J_3$. Figures 2 and 3 show these quantities versus $J_2/J_1$ for two values of the interplane coupling $J_3$.

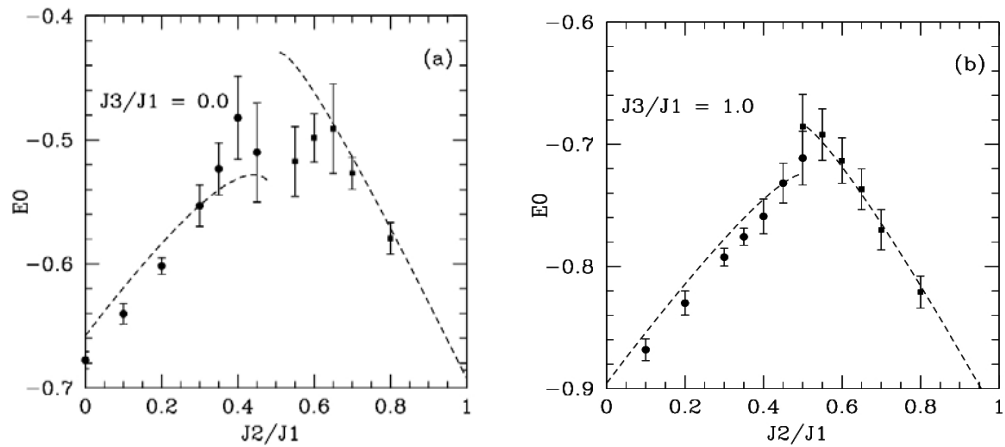

Fig. 2. Series estimates of the ground state energy per site for (a) $J_3/J_1 = 0.0$, and (b) $J_3/J_1 = 1.0$. Filled and open circles are from Néel and columnar phase expansions, respectively. The dashed lines are linear spin-wave theory predictions.

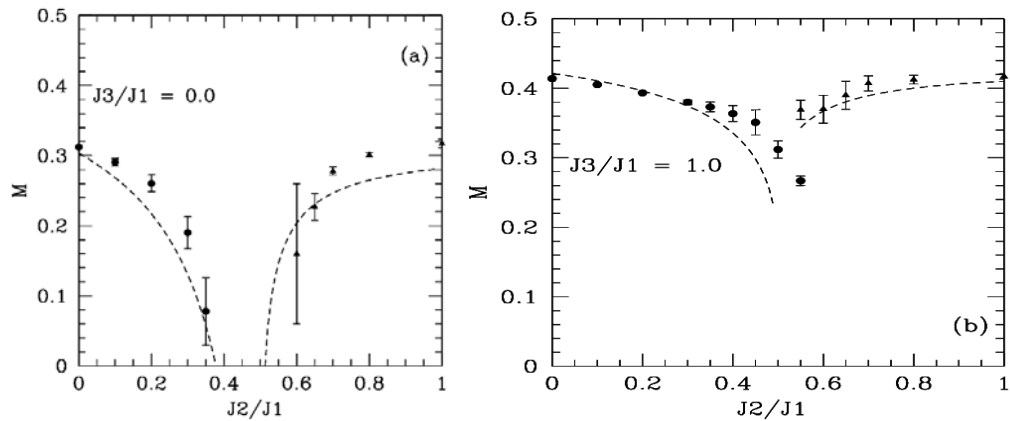

Fig. 3. Series estimates of the magnetization for (a) $J_3/J_1 = 0.0$, and (b) $J_3/J_1 = 1.0$. Filled and open circles are from Néel and columnar phase expansions, respectively. The dashed lines are linear spin-wave theory predictions.

We note a striking difference between the two cases. For $J_3/J_1 = 0$ the existence of an intermediate phase is clearly seen, particularly from the magnetization curves. On the other hand, for $J_3/J_1 = 1.0$, the energy branches clearly meet, albeit with different slopes, indicative of a direct first-order transition. The magnetization curves (Fig. 3b) show that the magnetization in each phase remains finite at the transition.

Such calculations, for a range of $J_3$ values, leads us to conclude that the intermediate phase shrinks to zero at a point $J_2/J_1 = 0.54 \pm 0.03$, $J_3/J_1 = 0.16 \pm 0.03$, consistent with the schematic phase diagram in Fig. 1.

2.2   Magnon Excitations

Our series technique is also able to determine the energies of magnon excitations of the model. Fig. 4 shows a set of typical results, for the Néel phase, for $J_3 = 0.5$, and two values of $J_2$.

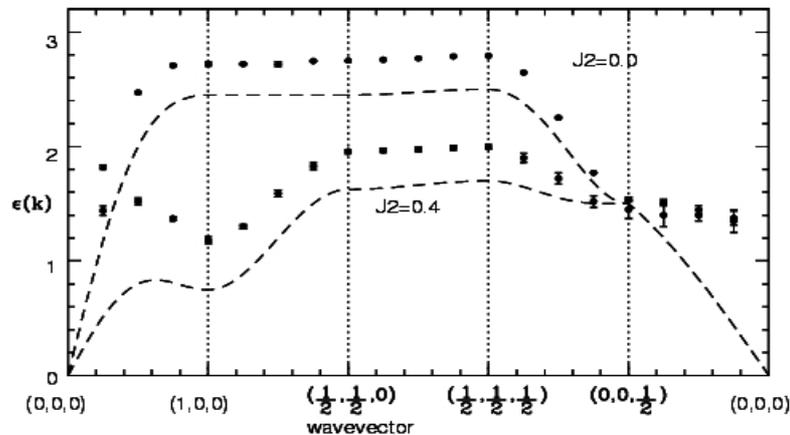

Fig. 4. Magnon dispersion curves along symmetry directions in the Brillouin zone, for $J_3 = 0.5$ and two values of $J_2$. The dashed lines are linear spin-wave predictions.

As can be seen in Fig. 4, we are able to identify fine details of the dispersion curves. In particular, we see a pronounced dip forming at wavevector $(\pi,0,0)$ as the phase boundary is approached, and a pronounced shoulder near $(0,0,\pi/2)$. Linear spin-wave theory is in qualitative agreement with the presumably more accurate series results, but generally underestimates the excitation energies.

We have also calculated dispersion curves in the columnar ordered phase. Details and examples will be given in a future paper [5].

**3.   Conclusions**

A frustrated spin ½ model antiferromagnet, consisting of frustrated $J_1 - J_2$ antiferromagnetic layers coupled by an antiferromagnetic $J_3$, which is believed to provide a good description of the layered materials $Li_2VOSiO_4$ and $Li_2VOGeO_4$ has been studied by series expansion methods.

We find that the magnetically disordered spin-liquid phase, which has been identified as a prominent feature in the single layer $J_1 - J_2$ model, becomes narrower with increasing interlayer coupling $J_3$, and vanishes at a triple point, beyond which there is a direct first-order transition between Néel and columnar magnetic order. The location of the triple point is estimated as $J_2/J_1 = 0.54 \pm 0.03$, $J_3/J_1 = 0.16 \pm 0.03$. This is a little lower than found in an earlier calculation [4]. We also compute dispersion curves for magnon excitations, which will provide a stronger test of the applicability of the model, when and if inelastic neutron scattering data become available for these materials.

Details of these calculations will be given in a forthcoming paper [5]. In the three-dimensional model the magnetically ordered phases will persist to finite temperatures, up to some critical surface $T_c(J)$. This will be the subject of future work.

The model studied here is also of possible relevance for understanding the magnetic properties of the recently discovered iron pnictide superconductors, but this remains controversial [6].

**Acknowledgments**

We are grateful for the computing resources provided by the Australian Partnership for Advanced Computing (APAC) National Facility.

# Structural Properties of Compounds in the $M$PS$_{3-x}$Se$_x$ Family

B. Hillman[a], L. Norén[a] and D.J. Goossens[ab]

[a] *Research School of Chemistry, Australian National University, Canberra, 0200, Australia..*
[b] *Research School of Physics and Engineering, Australian National University, Canberra, 0200, Australia.*

The families of layered materials $M$PS$_3$ and $M$PSe$_3$ where $M$ = Mn, Fe, Ni, Zn etc shows a wide range of fascinating behaviour, magnetic and structural. The structure of the $M$PS$_3$ compounds is monoclinic and the in-plane coordination number is 3. $M$PSe$_3$ compounds, on the other hand, are rhombohedral. We explore the effect of replacing sulphur with selenium, $M$PS$_{3-x}$Se$_x$, and the structural properties of some compounds in this family.

## 1.   Introduction

In this study we investigate both $M$PS$_3$ and $M$PSe$_3$ which are layered quasi two-dimensional materials which have been studied extensively for their interesting structural and magnetic properties [1, 2, 3]. Little if any research has been done on the intermediate compounds of composition $M$PS$_{3-x}$Se$_x$ (where $M$ = Mn, Fe and Ni). Because Se is larger than S, if the samples for which $0 < x < 3$ form an isostructural series, the possibility of a systematic study of the behaviour of the two-dimensional magnetism found in these materials arises, particularly as a function of interplanar spacing. This study investigates the structural effect of the substitution of selenium for sulphur ($x$ = 0, 1, 1.5 & 2) and determines the point of transition from the monoclinic sulfide structure (*C* 2/*m*) to the rhombohedral structure of the selenide (*R*-3).

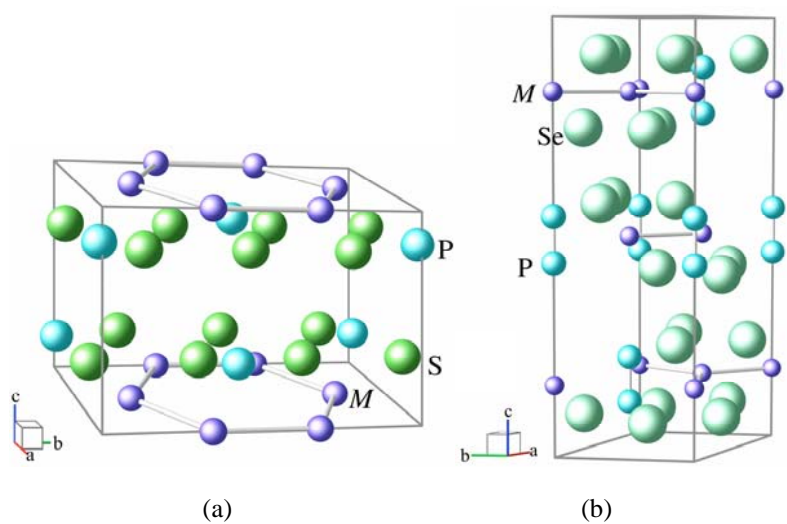

Fig. 1. a) Unit cell of $M$PS3. b) Unit cell of $M$PSe$_3$. The metal honeycomb lattice is indicated on the sulfide structure. $M$ = metal ion, P = phosphorus, S = sulphur. The structures shown in a) and b) were constructed using data from [4] and [5], respectively, and produced using the *Balls & Sticks* software package [6].

As can be seen in Fig. 1, the unit cell of both $M$PS$_3$ and $M$PSe$_3$ contains a layer of metal ions arranged in a honeycomb lattice (indicated by a hexagon on Fig. 1a). The metal layer lies between layers of S or Se in which the atoms are positioned in a distorted octahedral arrangement around the metal ions [4,5]. Adjacent S, or Se, layers are separated by a Van der Waals gap. Each metal ion hexagon has a phosphorus dimer at the centre.

## 2. Sample Preparation

The compounds were synthesised from powders of metal sulfide ground with phosphorus, sulphur and selenium powder in stoichiometric ratios and pressed into 6 mm pellets. These were then placed into quartz tubes which were evacuated to $10^{-3}$ Torr and sealed. The samples were heated to 400ºC and held there for a week, then ramped to 700ºC and held for a further seven days. This was done to prevent explosions caused by overpressure from the sulphur. Once the compounds were sintered X-ray powder diffraction was used for structural characterisation performed with a Siemens D5000 X-ray diffractometer.

Analysis of the composition and homogeneity of the samples was undertaken using a Hitachi S-4300 scanning electron microscope. The compounds were found to be homogeneous and single phase, with compositional deviations within a few percent. In some cases, most notably the iron compounds, more Se than expected was present, suggesting that S had been lost while heating. The $FePS_{1.5}Se_{1.5}$ sample was the most extreme case, with an observed composition of $FePS_{1.4}Se_{1.6}$. The $NiPS_2Se$ and $FePS_3$ sample were within 5% of the desired composition and all other samples were within 2%.

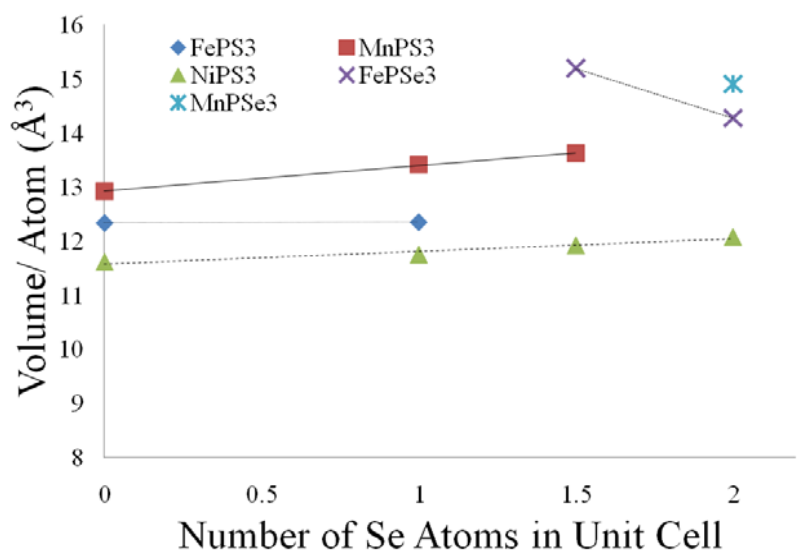

Fig. 2. The change in volume as the number of Se atoms in the unit cell increases for each structure. The error bars are smaller than the data points.

## 3. X-Ray Diffraction

The diffraction peaks shift towards a smaller 2θ as the concentration of selenium increases and the lattice parameters expand. The change in structure is evident when comparisons are made between diffraction patterns. The number of peaks and relative positions indicate which structure the intermediate compounds form.

The structural characterisation was carried out with reference to the known crystal structures of $M$PS$_3$ and $M$PSe$_3$ [2,4,5]. Due to massive preferred orientation of the plate-like crystallites, Rietveld refinement of the X-ray data was not possible. The lattice parameters for the samples were determined using Le Bail fitting [7]. As the samples are highly layered the plate-like crystals preferentially align such that the (00*l*) planes are parallel to the sample surface, which enhances the (00*l*) reflections and diminishes the (*hk*0) reflections. This allows reliable determination of lattice parameters, but not of atomic coordinates and displacement parameters.

As expected, on replacing S with Se the lattice parameters expand in order to facilitate the greater size of the Se atom. This can be seen in Fig. 2 and Table 1 for the monoclinic sulfide structure, particularly with reference to $NiPS_{3-x}Se_x$, which is monoclinic across the full composition range. There is insufficient data for the selenide-structure samples to indicate a trend.

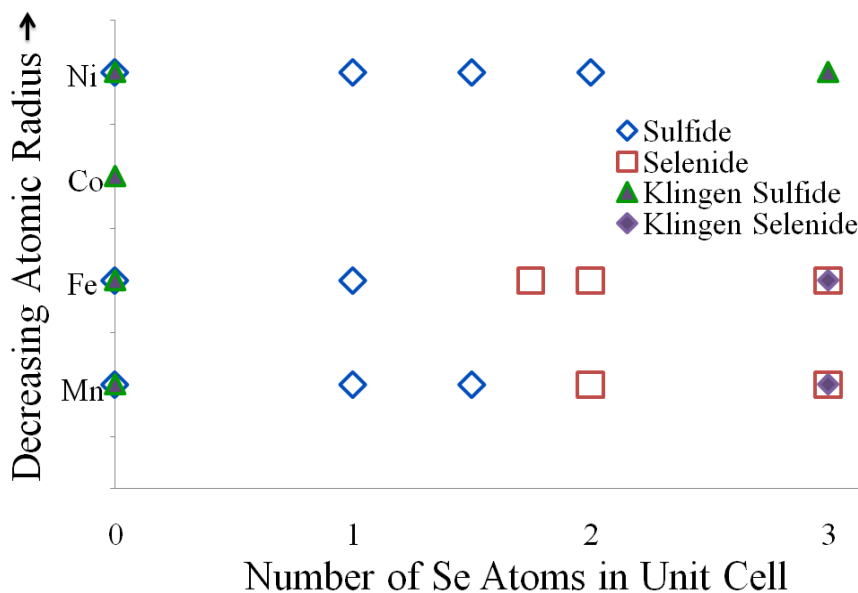

Fig. 3. The change in structure as the number of selenium atoms in the unit cell increases for each set of compounds. Results from this experiment and literature results from Klingen [1] are shown.

## 4. Discussion

As may be seen in Fig. 3, the nickel compounds do not undergo any structural change as Se level increases, unlike the Mn and Fe samples. The point of transition requires further investigation to determine precisely. Figure 3 shows a phase diagram for this system, and shows that both $FePS_{3-x}Se_x$ and $MnPS_{3-x}Se_x$ show a transition from the sulphide to selenide structure as a function of composition, with no evidence of a mixed-phase region.

Table 1. Structural parameters for the $MPS_{3-x}Se_x$ materials as a function of S:Se ratio. Cell volumes greater than 450 Å$^3$ indicate the selenide structure has been formed.

| Sample | $a$ (Å) | $b$ (Å) | $c$ (Å) | $\beta$ (°) | Volume (Å$^3$) |
|---|---|---|---|---|---|
| $FePS_3$ | 5.958(4) | 10.31(1) | 6.730(4) | 107.25(7) | 394.93 |
| $FePS_2Se$ | 5.864(3) | 10.335(6) | 6.837(1) | 107.37(5) | 395.45 |
| $FePS_{1.5}Se_{1.5}$ | 6.420(5) | 6.420(7) | 19.593(7) | | 699.34 |
| $FePSSe_2$ | 6.214(7) | 6.214(2) | 19.642(8) | | 656.90 |
| $MnPS_3$ | 6.009(3) | 10.59(1) | 6.805(3) | 107.26(6) | 413.66 |
| $MnPS_2Se$ | 6.086(5) | 10.704(8) | 6.933(2) | 108.04(6) | 429.22 |
| $MnPS_{1.5}Se_{1.5}$ | 6.1457(3) | 10.699(8) | 6.988(3) | 108.30(8) | 436.22 |
| $MnPSSe_2$ | 6.3027(2) | 6.3027(3) | 19.944(7) | | 686.11 |
| $NiPS_3$ | 5.808(4) | 10.095(5) | 6.630(2) | 106.94(7) | 371.85 |
| $NiPS_2Se$ | 5.788(3) | 10.108(9) | 6.739(1) | 107.40(8) | 376.27 |
| $NiPS_{1.5}Se_{1.5}$ | 5.788(4) | 10.198(9) | 6.770(3) | 107.15(6) | 381.80 |
| $NiPSSe_2$ | 5.787(4) | 10.24(1) | 6.841(2) | 107.30(7) | 386.77 |

## 5. Conclusions

The work outlines the first study of the dependence of the structure of $MPS_{3-x}Se_x$ materials on the S:Se ratio (Table 1). The point of transition from the monoclinic sulphide structure to the rhombohedral selenide structure lies between $FePSSe_2$ and $FePS_{1.4}Se_{1.6}$ for iron-based materials and MnPS$_{1.5}$Se$_{1.5}$ and MnPSSe$_2$ for Mn-based. There was no evidence of layering of the S and Se, which is to say that, from these experiments, they appear to substitute equally into all sites. From these limited data the transition from monoclinic to rhombohedral structure appears to be direct, rather than via a mixed phase region. Investigation of compositions closer to the critical is expected to prove of interest. The composition at the transition could be determined more precisely by more closely investigating the region of the phase diagram where the transition occurs. Samples of other metals could be investigated to further complete the structure phase diagram. For example, cobalt is of particular interest as it is a member of the first row transition metal series and lies between Ni and Fe in size.

**Acknowledgements**
We thank Dr Hua Chen of the CAM at ANU for assistance with electron microscopy, Dr M. Sterns of RSC at ANU for help with XRD. We acknowledge the financial support of the ARC and AINSE.

# Quadrupole Coupling Constants for $^{100}$Rh in Transition Metals from Perturbed Angular Correlation Spectroscopy

William J. Kemp[a], Adurafimihan A. Abiona[a,*], Patrick Kessler[b],
Reiner Vianden[b] and Heiko Timmers[a,c]

[a] *School of Physical, Environmental and Mathematical Sciences, The University of New South Wales, Canberra Campus, ACT 2602, Canberra, Australia*
[b] *Helmholtz-Institut für Strahlen- und Kernphysik, Nußallee 14-16, 53115 Bonn, Germany*
[c] *Department of Nuclear Physics, Research School of Physics and Engineering, Australian National University, Canberra, ACT 200, Australia*

[*] *On leave from Centre for Energy Research and Development, Obafemi Awolowo University, Ile-Ife, Nigeria*

Time differential perturbed angular correlation spectroscopy of $^{100}$Rh in zinc, rhodium, antimony, hafnium and rhenium has confirmed expectations for zinc and rhodium and provided the first measurements of quadrupole coupling constants for the three other transition metals with preliminary values of $v_Q$ = 4.3 MHz, 5.7 MHz, and 3.2 MHz, respectively. The three results appear to be consistent with published results for zirconium and ruthenium.

## 1. Introduction

The nuclear quadrupole moment $Q$ has fundamental importance in nuclear structure physics and it is a crucial parameter for hyperfine interaction studies of condensed matter. Nuclear quadrupole moments may be calculated from measurements of the quadrupole coupling constant $v_Q$ in solids according to:

$$Q = \frac{h}{e}\frac{v_Q}{V_{zz}} \qquad (1)$$

where $h$ and $e$ are Planck's constant and the positive elementary charge, respectively. This requires, however, that the principal component $V_{zz}$ of the Electric Field Gradient (EFG) at the probe site is well known. Theoretical calculations of the EFG for solids have traditionally been difficult. Recent theoretical advances, such as the *WIEN2k* code [1], can now provide accurate predictions of electric field gradients in solids [2, 3]. In order to exploit such progress, a more complete compilation of quadrupole coupling constants than presented in Ref. [4] and improved experimental accuracy would be advantageous. The list of measurements of quadrupole coupling constants with time differential perturbed angular correlation spectroscopy (TDPAC) using the probe $^{100}$Pd/$^{100}$Rh is particularly sparse. The current accepted value of the nuclear quadrupole moment of the relevant 74 keV $2^+$ state in $^{100}$Rh ($t_{1/2}$ = 215 ns) is $Q$ = 0.076(20) barn [5]. It may be noted that the relative uncertainty of this value is more than 25%. This project aims to provide several new TDPAC measurements of $\nu_Q$ for the $^{100}$Pd/$^{100}$Rh probe while verifying existing results. Here preliminary results are presented for the transition metals zinc, rhodium, antimony, hafnium and rhenium.

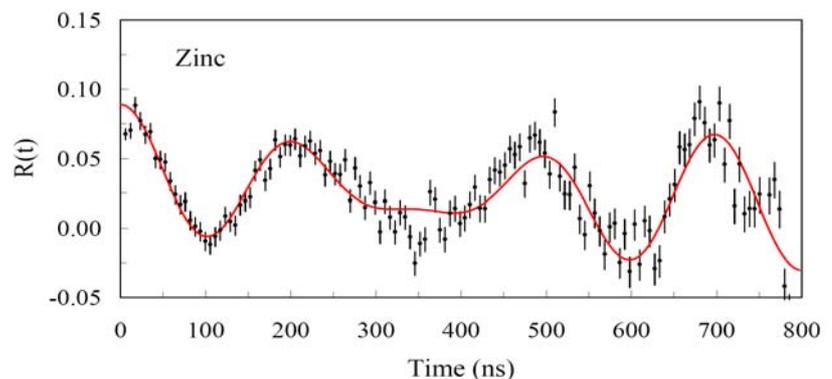

Fig. 1. Measured ratio function $R(t)$ for $^{100}$Rh in zinc. The fitted curve is shown in red.

## 2. Experimental Details

The 14 UD Pelletron accelerator at the Australian National University in Canberra was used to synthesize the $^{100}$Pd/$^{100}$Rh probe nuclei via the fusion evaporation reaction $^{92}$Zr($^{12}$C, 4n)$^{100}$Pd. The 70 MeV $^{12}$C beam was incident on the zirconium production target over 20 h at a beam current of 1 µA. Synthesized probe nuclei recoil-implanted at forward angles into polycrystalline samples of zinc, rhodium, antimony, hafnium and rhenium with recoil energies of several MeV [6].

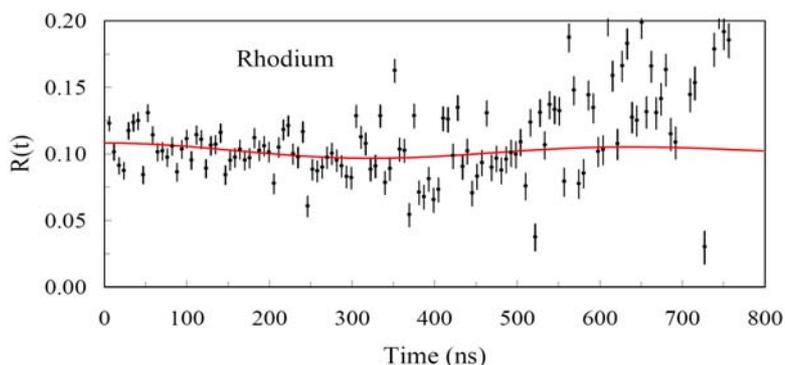

Fig. 2. Measured ratio function $R(t)$ and fit for $^{100}$Rh in rhodium.

As-implanted samples were studied with TDPAC spectroscopy [7] using a conventional set-up of four conical BaF$_2$ scintillation detectors forming a planar array with start/stop detector combinations at angles of 90° and 180°. For each detector combination the coincidence time distribution of the two gamma-rays in the 84 keV – 74 keV γ – γ cascade, populating and depopulating the intermediate $2^+$ excited state in $^{100}$Rh, were recorded using NIM-standard electronics. All time distributions obtained were corrected for statistical background events. The distributions for the 90° detector combinations and those for the 180° detector combinations were averaged, thus compensating for slight differences in detection solid angle and detector efficiency. Finally, using the conventional prescription, the ratio function $R(t) = A_{22}G(t)$ was extracted from the data [8]. The ratio function reflects the gamma-gamma anisotropy $A_{22}$ of the probe and its time-dependent modulation $G(t)$ due to hyperfine interactions. The ratio functions were fitted using the code *Nightmare* which is based on the routine *NNFit* [9]. All fits assumed EFGs with no deviation from axial symmetry employing an axial asymmetry parameter $\eta$ = 0. The non-physical rise of some measured ratio functions for large times is probably due to a saturation effect in the time-to-analogue converter (TAC). It is most pronounced for rhenium and antimony and is being investigated further.

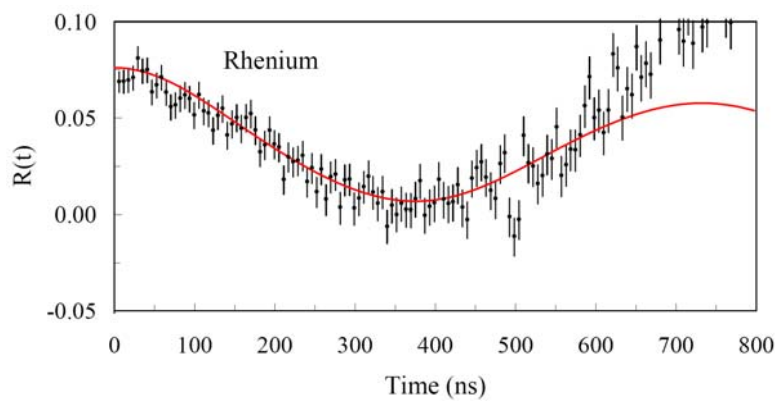

Fig. 3. Measured ratio function $R(t)$ and fit for $^{100}$Rh in rhenium.

## 3. Results

Figure 1 shows the measured ratio function $R(t)$ for $^{100}$Rh in zinc. Zinc has a hexagonal lattice. It is apparent that $R(t)$ is modulated as would be expected for a unique EFG with negligible damping. The fit reproduces the observed modulation pattern well with a quadrupole coupling constant of $v_Q$ = 11.4 MHz. This result agrees with the accepted value [4].

Rhodium has a cubic lattice. As there is no EFG for substitutional probe integration in cubic lattices, the ratio function should show no modulation, but the constant value of the experimental anisotropy. Figure 2 confirms that this is the case.

Rhenium has a hexagonal lattice. Ignoring the spurious effect beyond 600 ns, the data can be well fitted. The fit shown in Figure 3 assumes a quadrupole coupling constant of $v_Q$ = 3.2 MHz. The rhenium sample was also measured following annealing in argon in a tube furnace at 800°C. After annealing the measured coupling constant was $v_Q$ = 3.7 MHz.

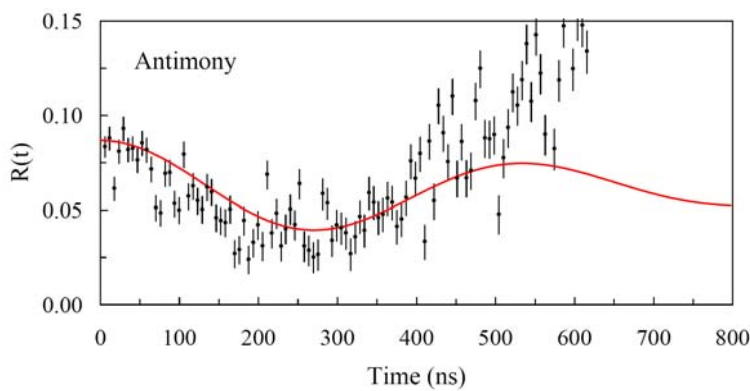

Fig. 4. The measured ratio function $R(t)$ and a fit in red for $^{100}$Rh in antimony.

Antimony has a rhombohedral lattice. Figure 4 shows that the data may be fitted below 400 ns, beyond which the non-physical rise of the data may be attributed to an artifice of the spectrometer. The fit assumes a quadrupole coupling constant of $v_Q$ = 4.3 MHz.

Hafnium has a hexagonal lattice. The measured ratio function in Figure 5 shows a slight modulation of the anisotropy. Some damping is apparent which is likely due to implantation induced lattice damage resulting in a non-unique EFG. The fit suggests $v_Q$ = 5.7 MHz.

**4.　Summary and Conclusions**

The measured quadrupole coupling constants for the $2^+$ state in $^{100}$Rh have been compiled in Table 1 together with accepted values for zinc, zirconium and ruthenium. The measured value for zinc agrees with expectation. The new measurements for antimony, hafnium and rhenium are similar to accepted values for other transition metals such as zirconium and ruthenium. For example, the measured quadrupole coupling constant of $v_Q$ = 5.7 MHz is close to the value known for the isoelectronic metal zirconium of $v_Q$ = 8.1(4) MHz. Moreover, the elements ruthenium and rhenium which are in neighbouring groups in the periodic table both have relatively small quadrupole coupling constants.

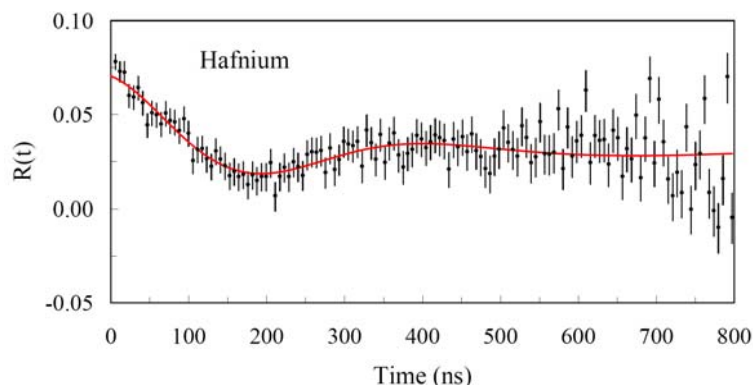

Fig. 5. The measured ratio function $R(t)$ for $^{100}$Rh in hafnium. The fit with the code *Nightmare* is shown in red.

Table 1. The quadrupole coupling constants measured in this work in comparison with literature values.

| transition metal | atomic number Z | axial ratio $c/a$ | measured $\nu_Q$ | accepted $\nu_Q$ (from Ref. [4]) |
|---|---|---|---|---|
| Zinc | 30 | 1.856 | 11.4 | *11.4 (2)* |
| *zirconium* | *40* | *1.593* | - | *8.1 (4)* |
| *ruthenium* | *44* | *1.584* | - | *1.0 (3)* |
| antimony | 51 | (rhombohedral) | 4.3 | - |
| hafnium | 72 | 1.582 | 5.7 | - |
| rhenium | 75 | 1.615 | 3.2 | - |

Besides confirming the new measurements, without the spurious effect in the ratio function for large times, future work, including measurements following annealing, will aim at establishing uncertainties for the measured quadrupole coupling constants. Measurements of $\nu_Q$ for additional transition metals may allow trends to be observed, which would test theory.

# Proximity Effect of High Energy Ion Tracks in Amorphous SiO$_2$

J.W. Leslie, B. Hillman, and P.Kluth

*Department of Electronic Materials Engineering, Research School of Physics and Engineering, Australian National University, Canberra, 0200, Australia.*

We have studied the effect of proximity of ion tracks in amorphous SiO$_2$ on their chemical etching behaviour. The ion tracks were generated in 2 micrometer thick SiO$_2$ by irradiation with 185 MeV Au ions and subsequent etching using hydrofluoric acid. Scanning electron microscopy reveals a decrease in the etched track radius below a distance of adjacent tracks of approximately 0.2 microns. Long range strain generated during track formation is a possible explanation for this behaviour.

## 1. Introduction

As highly energetic heavy ions pass through a material, they can disturb the structure in their paths, leaving behind cylindrical damage zones a few nanometres in radius and microns in length, called 'ion tracks'. The formation of ion tracks is governed by the interaction of the ion with the target electrons [1,2]. The swift heavy ions excite large numbers of electrons, resulting in rapid localised heating facilitated by electron-phonon coupling. The temperature can far exceed the melting temperature of the material, which can lead to local modifications in a narrow region of a few nanometres around the ion trajectory. The subsequent rapid cooling can quench in the modified material forming an ion track.

SiO$_2$ has been chosen because it is both technologically relevant and well documented. Additionally, fission tracks in quartz and silicate minerals are often used in geochronology and for dating of archaeological objects. In amorphous SiO$_2$ the density difference between the track and the unaffected material is of the order of 3%. This small variation, combined with the amorphous nature of both the track and background material renders direct observation of the tracks difficult [3].

Often the material in the ion track shows enhanced chemical etching when compared to undamaged material. This can lead to conical etch pits in the micrometer size range that are easily imaged using scanning electron microscopy (SEM). In this paper, we have used SEM to study the effect of proximity on the size of etched ion tracks in amorphous SiO$_2$. The technique developed in this paper provides another means for the continuing research to develop a more detailed understanding of ion track properties.

While the work itself is of fundamental nature, it has implications for areas such as nanofabrication, nuclear physics, geochronology, archaeology, and interplanetary science. In particular the materials system under investigation (SiO$_2$/Si) is of high relevance for the semiconductor industry.

## 2. Experiment
### 2.1 Sample preparation

The ion tracks were produced in commercially available 2 µm thick layers of a-SiO$_2$, thermally grown on Si(100) substrates. The irradiation was performed at the ANU Heavy Ion Accelerator Facility by irradiation with Au ions at 185 MeV, with a fluence of $1 \times 10^9$ ions/cm$^2$. The irradiation was performed at room temperature with the incident ion at normal to the sample surface. Subsequently the samples were etched in 5% HF for 5 minutes at room temperature.

The etching is accelerated in the ion track region, which leads to the formation of conical holes centred around the tracks. The unetched tracks have a diameter of the order of ten nanometres [3] and the etched diameter is typically hundreds of nanometres. These etched holes can be directly observed by using SEM.

2.2 SEM analysis

Example SEM images of the etched ion tracks are shown in Figure 1. The holes created by etching the tracks appear as the dark regions. Figure 1a shows a top view image, such as those used in the analysis. Figure 1b shows a side view, displaying the conical nature of the holes. It is readily apparent that isolated tracks leave approximately equal sized holes, while adjacent tracks in close proximity leave smaller holes.

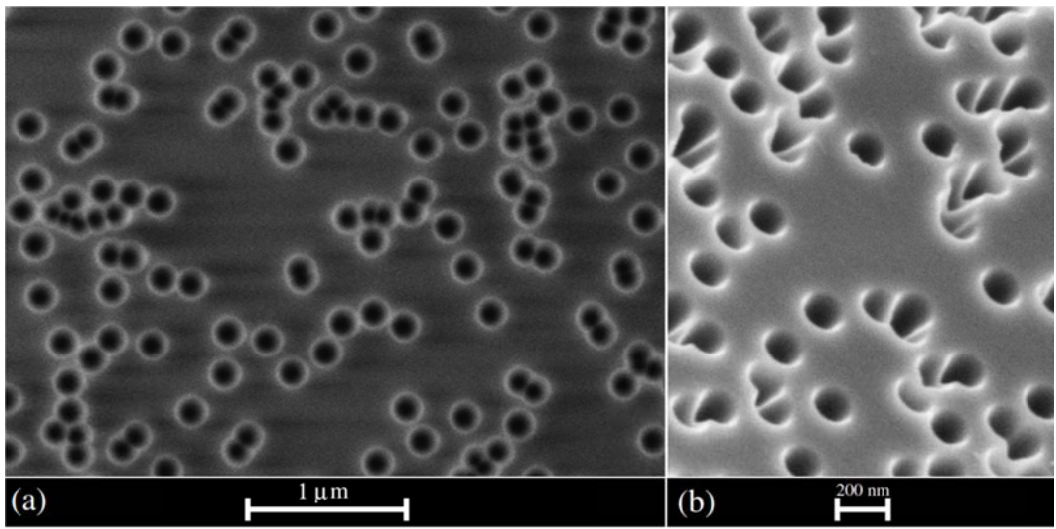

Fig. 1. (a) SEM top view image of etched ion tracks in a-SiO$_2$, (b) SEM side view image

Analysis of the images was performed using built in functionality of the open source software package *ImageJ* [4]. First the image was converted to a binary form based on brightness thresholds, which are set to select the dark central spots of the etched tracks. The bright rings around these holes enable reproducible and accurate thresholding. Images in binary form lend themselves well to automated particle analysis and particle counting.

The specific measured sizes of the individual etched tracks are somewhat dependent on the thresholding conditions, however, provided they are consistent, the analysis technique provides a means to compare the track sizes across many images. Additionally the etching conditions, while uniform across the sample, will also arbitrarily affect the size of the etched tracks.

The automatic particle analysis returns the x-y coordinates of the centre of each track, as well as its area. The holes were assumed to be approximately circular, allowing the radius to be approximated from the area. For each hole, the Pythagorean distance was found to the nearest neighbour.

3. **Results and discussion**

Figure 2 shows a plot of the nearest neighbour distance as a function of the etched track radius. Two approximately linear regions are apparent from Figure 2. In the first region, above a nearest neighbour distance of 0.2 µm, the average etched track radius is constant at about 0.08 µm. At nearest neighbour track distances below 0.2 µm, a linear decrease of the etched track radius is apparent, suggesting that adjacent tracks separated in this distance range influence their respective etching behaviour. Importantly, this effect occurs before the etched tracks overlap. A possible explanation for this behaviour can be modification of the etching rate by strain induced from tracks in the vicinity. Ion tracks in amorphous $SiO_2$ have been described as frozen in pressure waves which produce local strain fields in the material [5].

For etched track radii below approximately 0.06 µm there appears to be another region of greater slope in Figure 2. This is likely governed by tracks influenced by multiple close neighbours, which is not taken into account in the single nearest neighbour graph. At the given low irradiation fluence, however, the number of tracks influenced by multiple neighbours is small and thus not significant in the presented statistics.

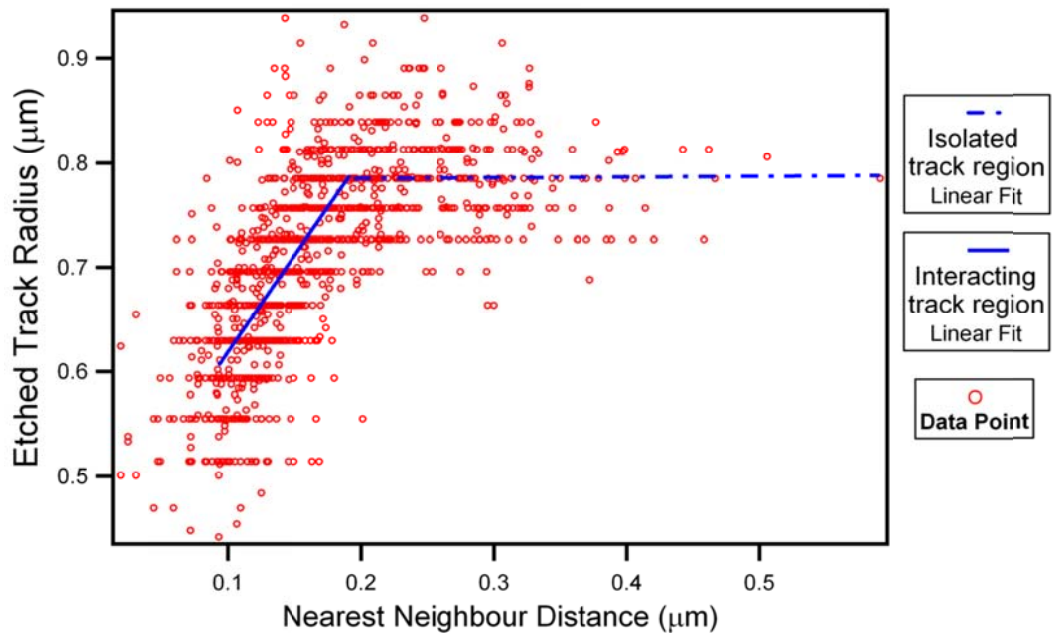

Fig. 2. Etched track radius plotted as a function of the nearest neighbour distance, and linear least squares fits of the two major regions.

The analysis was performed on images at various magnifications. At higher magnifications the size of the etched tracks can be measured more accurately, however fewer tracks can be measured in each image. At lower magnifications we could measure far more tracks simultaneously, however the pixel size constrains how accurately the track size can be measured. These pixel size limitations are responsible for as the horizontal lines in the data points in Figure 2. The data points from both the high and low magnifications agree, and allow a compromise between the number of measurements, and the accuracy of these measurements.

From Figure 2 we are able to observe the effects of proximity of neighbouring ion tracks at distances far greater than the actual unetched track diameter of approximately 10 nm [3]. The specific etching conditions chosen are somewhat arbitrary, and more experiments with varying etching times and irradiation fluences are currently in progress. Such experiments will be able to yield more detailed information about the strain generated during high-energy ion irradiation, and thus a better understanding of the ion track generation process.

## 4. Conclusions

We have analysed SEM images of etched ion tracks in amorphous $SiO_2$ to study the effect of their proximity on the etch radius. The tracks were found to influence each other at distances below 0.2 µm. This is beyond both the actual track diameter as well as the etched track diameter, which are approximately 10 nm and 1.8 µm, respectively. This indicates that this influence is unlikely to be a result of the particular etching conditions used, and is likely the result of strain in the material created by the formation of the track. Extending the investigations of this project to different implantation and etching conditions, as well as different materials, will yield more detailed information about the strain generated during track formation, and ultimately yield a better understanding of the ion track properties.

**Acknowledgments**

The authors acknowledge the Australian Research Council for financial support and thank the staff at the ANU Heavy Ion facility for their technical assistance. We also would like to thank Dr Darren Goossens for his encouragement and for suggesting the Wagga 2011 conference.

# Possible Lubrication and Temperature Effects in the Micro-scratching of Polyethylene Terephthalate

Yanyan Liu and Heiko Timmers

*School of Physical, Environmental and Mathematical Sciences,
University of New South Wales at ADFA, Canberra, ACT 2600, Australia*

Possible effects of water lubrication and environmental temperature on micro-scratching of PET have been studied. Micro-scratching was conducted by a purpose-built micro-scratcher with silicon cubic corner tips. Two conditions (dry scratching and scratching in water) have been investigated. Micro-scratching in water was performed at three different water temperatures (9°C, 20°C, 73°C). Zum Gahr's formalism was used to quantify the micro-scratches. All scratches show evidence of ploughing by the tip. Micro-scratches obtained in water at 20°C tend to have a negative Zum Gahr parameter. This may be due to the polymer being more elastic under these conditions causing a significant upward relaxation of the groove walls after the tip action.

1. **Introduction**

Micro-scratching is a powerful materials characterisation technique which provides information that aids the understanding of wear properties of materials such as polymers. Previous work on single asperity of polymers includes that of Ducret *et al.* [1] and Sinha *et al.* [2]. Wear mechanisms at the micro- and nano-scale are expected to be considerably different from those at the macro-scale [3]. This is due to the increased role of van der Waals forces that control the formation and rupture of the junction between counterface and polymer. Some studies show that the tribological behaviour of polymers exhibits a strong dependence on the imposed friction conditions [4]. Friction-induced heat results in a temperature increase in the surface layer during sliding wear [5]. More specifically Briscoe *et al.* [6] have defined the surface layer of a polymer being worn in two zones: (i) the interface zone and (ii) the cohesive zone. This distinction depends on temperature. Temperature increases in the surface layer may influence the wear mode and the deformation of the polymer during sliding wear [4, 7]. Such temperature-dependent effects suggest that the environmental temperature may also affect the tribological behaviour of polymer surfaces [8].

Conceptually the effects of micro-scratching can be discussed using a parameterisation developed by Zum Gahr. Stround and Wilman [9] found out that only a proportion of the volume of a polymer scratch groove is due to the removal of wear debris. Predominantly the groove volume is due to plastically displaced material which forms pile-up ridges alongside the groove. Based on the formalism first described for ductile metals, a parameter $f_{ab}$ may be defined that quantifies the degree of ploughing observed for a micro-scratch. The parameter relates the groove dimensions to the dimensions of the pile-up ridges with ideal micro-ploughing corresponding to $f_{ab} = 0$ and ideal micro-cutting corresponding to $f_{ab} = 1$ [10]. In a novel approach we have chosen silicon cubic corner tips to model single asperities in the micro-scratching of polymers. Such tips are readily available and may be used with well-defined geometries in either flat-on or edge-on mode. Polyethylene terephthalate (PET) provides an interesting test-case of lubrication and environmental temperature effects in micro-scratching.

## 2. Experimental details

2.1. Sample preparation

PET (Goodfellow Cambridge Limited) was used without any surface modification due to its low pristine surface roughness measured with AFM as 10 ± 4 nm. The amorphous 1 mm thick PET sheet was cut into 10 mm × 10 mm pieces. Samples were cleaned ultrasonically in an alcohol bath and dried in flowing nitrogen. Before and after micro-scratching samples were characterised using AFM and, following coating with a thin silver film to suppress charging, with SEM.

2.2. Scratching experiments

The micro-scratcher is illustrated in Fig. 1. The micro-scratcher consists of three parts: (i) a head with scratching tip, (ii) a micrometer screw, and (iii) a polymer sample holder. The normal load applied was 150 µN. The scratch length varied between 150–300 µm. The sliding velocity was 300 µm/s. The tip attacked in flat-on mode. Different environmental conditions were used. Firstly, a series of dry micro-scratches were performed. Secondly, three series of micro-scratches were performed with water lubrication at water temperatures of 9°C, 20°C and 73°C, respectively. In each series about 10 scratches were conducted.

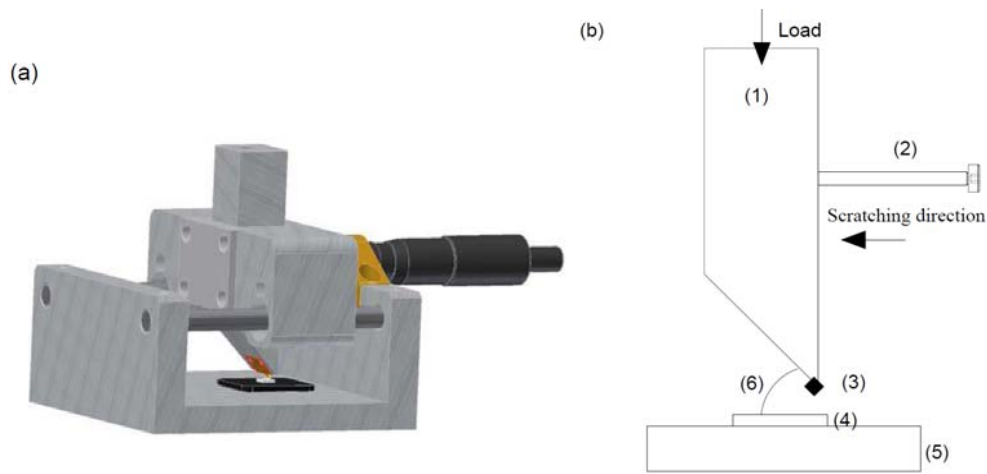

Fig. 1. (a) Illustration of the micro-scratcher; (b) Schematic of the micro-scratching beam: (1) load beam, (2) micro-meter screw, (3) silicon cubic corner tip, (4) PET sample, (5) sample holder, (6) 45° attack angle.

The scratching tips with 90° cubic corner geometry were cleaved from a commercial (100) silicon wafer along (110) crystal planes. The tip, after being cleaned ultrasonically, was imaged by SEM to verify correct geometry within the requirements of the experiment, see Fig. 2(a). Following micro-scratching the tip was imaged again to confirm that it was not modified by its action on the polymer. No polymer debris was observed on the tip.

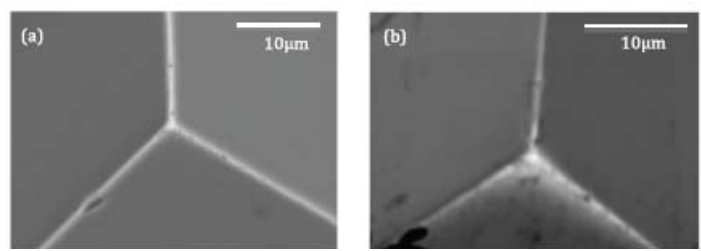

Fig. 2. (a) silicon cubic corner tip before scratch; (b) silicon cubic corner tip after scratch.

## 3. Results

Figure 3 displays SEM and AFM images of dry scratches. It is apparent that the micro-scratching gives reproducible results. All scratches show the characteristic symmetric ridges alongside the scratch groove, which are the result of ploughing action by the tip. The length of the scratch groove is 250 µm. No debris particles are evident in or near the groove.

Figure 4 shows SEM images of micro-scratches obtained under the four environmental conditions studied. The commencement of the scratching action results in slightly different morphological features. The following uniform motion of the tip, however, results in very similar grooves. Within the reproducibility of the scratching no significant differences of the scratching grooves are apparent.

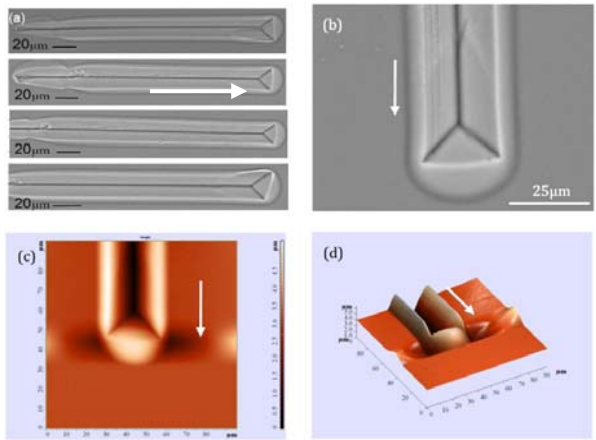

Fig. 3. (a) Reproducibility of microscratching of PET, (b,c,d) the end of a scratch groove, (b) SEM, (c) two-dimensional AFM image, (d) three-dimensional AFM image. The arrows indicate the scratching direction.

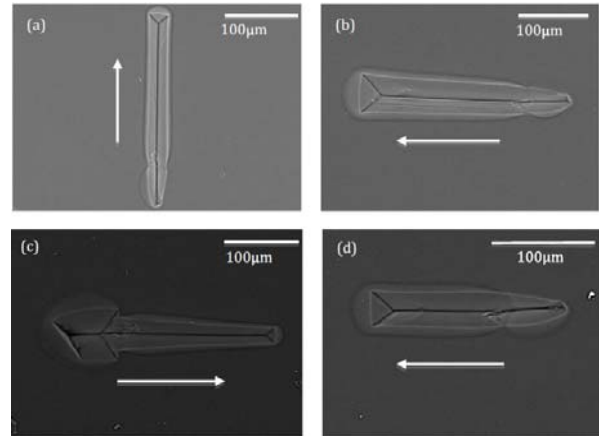

Fig. 4. (a) dry scratching, (b) scratching in 9°C water, (c) scratching in 20°C water, (d) scratching in 73°C water.

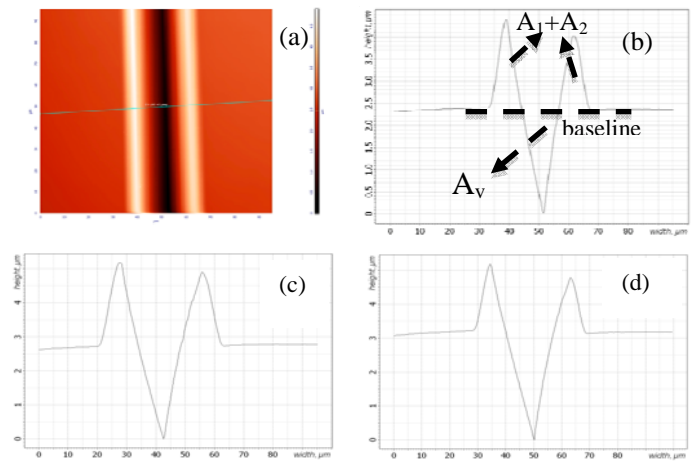

Fig. 5. Examples of measured Zum Gahr cross-sections. The groove after dry scratching in (c) is considerably wider than that after scratching in water at 20°C in (d).

Zum Gahr's cross-sectional profile analysis was used to quantify the scratch grooves. In this analysis the Zum Gahr parameter is defined as follows:

$$f_{ab} = \frac{A_v - (A_1 + A_2)}{A_v} \qquad (1)$$

Here $A_v$ is the cross-sectional area of the scratch groove and $(A_1+A_2)$ are the combined cross-sectional areas of the pile-up ridges. The results of the Zum Gahr analysis are shown in Figure.6. Dry scratching and scratching in water at 9°C and at 73°C give similar values of $f_{ab}$. Mean values are (0.21 ± 0.03), (0.16 ± 0.04), and (0.21 ± 0.04), respectively, and agree within uncertainties. In contrast the mean value of $f_{ab}$ for microscratching in water at 20°C is negative with $f_{ab} = -0.25 \pm 0.05$.

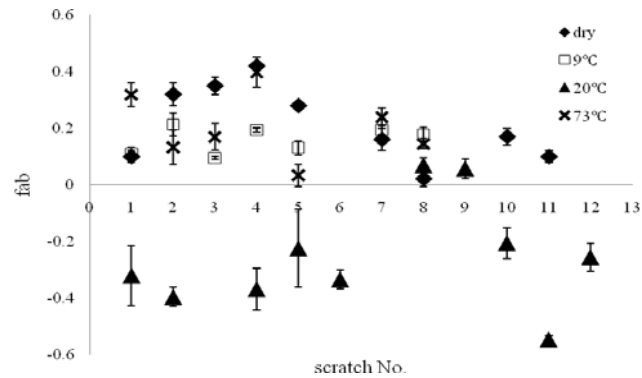

Fig. 6. Zum Gahr parameter $f_{ab}$ values obtained for the different micro-scratching conditions.

## 4.　Conclusions

The observation of ridges alongside the scratch groove indicates that micro-scratching of PET with cubic corner tips ploughs the material rather than cutting it. Scratching in water at 9°C and at 73°C does not significantly alter the properties of the micro-scratch when compared with dry scratching. All three conditions give positive values of $f_{ab}$ which agree within uncertainties.

Interestingly, for a water temperature of 20°C the Zum Gahr parameter is negative. The negative value of the parameter may be explained as an upward relaxation of the compression experienced by the groove walls once the ploughing tip has passed. This reduces the cross-sectional area of the groove, adds material to the ridges and thus results in a negative Zum Gahr parameter. The observation suggests that the material is particularly elastic under this environmental condition and it warrants further investigation.

# Studies of Magnetic Nanoparticles Formed in SiO$_2$ by Ion Implantation

A.E. Malik[a,b], W.D. Hutchison[a], K. Nishmura[c] and R.G. Elliman[b]

[a] *School of Physical, Environmental and Mathematical Sciences, The University of New South Wales at ADFA, Canberra ACT 2600.*
[b] *Electronic Materials Engineering Department, Research School of Physics and Engineering, Australian National University, Canberra, ACT 0200, Australia.*
[c] *Graduate School of Science and Engineering, University of Toyama, Toyama, Japan*

Nanoparticles of elemental cobalt and CoPt alloys have been synthesised, via ion implantation and a range of thermal annealing conditions, within 100 nm silica thin films thermally grown on silicon substrates. The size and spatial distributions of the nanoparticles are determined from transmission electron microscopy of sample cross-sections and the crystal structure from glancing angle X-ray diffraction analysis. These results are correlated with magnetisation measurements on the same range of samples.

1.  **Introduction**

Ion implantation is a versatile technique for studying the synthesis and properties of nanosized metal particles in different host materials [1]. In particular, it is a non-equilibrium process that can be used to introduce metallic species into a host material at concentrations well above their equilibrium solubility limit. It also provides the flexibility to choose particular metallic species or combinations of species for investigation. Such nanoparticles have many potential applications in areas as diverse as biotechnology, magnetic fluids, catalysis, magnetic resonance imaging and data storage. The latter application typically requires a high density of small magnetic nanoparticles located in the near-surface region of a protective matrix or thin-film. The ability to pattern arrays is also an advantage offered by the implantation technique.

Cobalt, a well known magnetic material for data storage applications, has been paid special attention for the past few years, particularly with regard to the evolution of magnetic properties with the cluster sizes, and super-paramagnetic behaviour [2]. Co-Pt alloy films with a face-centred tetragonal structure have long been considered the bench mark for ultrahigh density magnetic recording media [3,4]. In this paper we expand on work presented at Wagga 2010 [5] by examining the formation of nanoparticles of Co and nominally 50% Co 50% Pt alloy, in silica via implantation, for a range of subsequent thermal treatments. Magnetisation measurements and analysis of the structure and particle size distributions of these samples were undertaken and comparisons made between the two different kinds of nanoparticles and four different anneal conditions.

2.  **Sample preparation**

Starting substrates consisted of a 100 nm SiO$_2$ layer thermally grown on (100) oriented Si wafers. Separate regions of the silica layer were implanted with 50 keV Co ions with fluences of $6\times10^{16}$ ions/cm$^2$ or, for the Co-Pt alloy, sequentially implanted with 50 keV Co and then 100 keV Pt ions each with a fluence of $3\times10^{16}$ ions/cm$^2$. According to the Monte-Carlo ion-range simulation code *SRIM 2007* [6], the average range of both Co and Pt ions in SiO$_2$ is ~43 nm at these respective energies. After implantation the samples were variously annealed in a nitrogen atmosphere at either 500°C, 700°C or 900°C for one hour, or 900°C for two hours. Implantation fluences and depths were confirmed with Rutherford backscattering spectrometry (RBS) using 2 MeV He$^+$ ions. The size and spatial distributions of the nanoparticles were measured by transmission electron microscopy (TEM) of sample cross-sections, and the crystal structure of the nanoparticles was determined from glancing angle x-ray diffraction measurements. Magnetisation was measured at room temperature using a SQUID magnetometer. For the latter measurements, the substrates were thinned to ~ 100 μm to reduce the diamagnetic contribution from silicon.

### 3.　Results and Discussion

A systematic structural and compositional investigation was carried out on the samples using TEM. Figure 1(a) shows bright-field, cross-sectional images for samples implanted with Co, while the estimated Co particle size distributions are in Fig. 2. Clearly the sizes of the nanoparticles increases with increasing anneal temperature and time. The average size increasing from ~2 nm for the 500°C anneal to ~19 nm at 900°C (2 hours). The diffusion of Co with the increasing temperature also leads to formation of some features at the Si substrate interface and the surface. The TEM images of Fig. 1(b) are for the Co-Pt samples with the corresponding 4 anneal conditions. Size distributions for these samples are in Fig 3, where again we see larger particles following the higher temperature anneals. The addition of

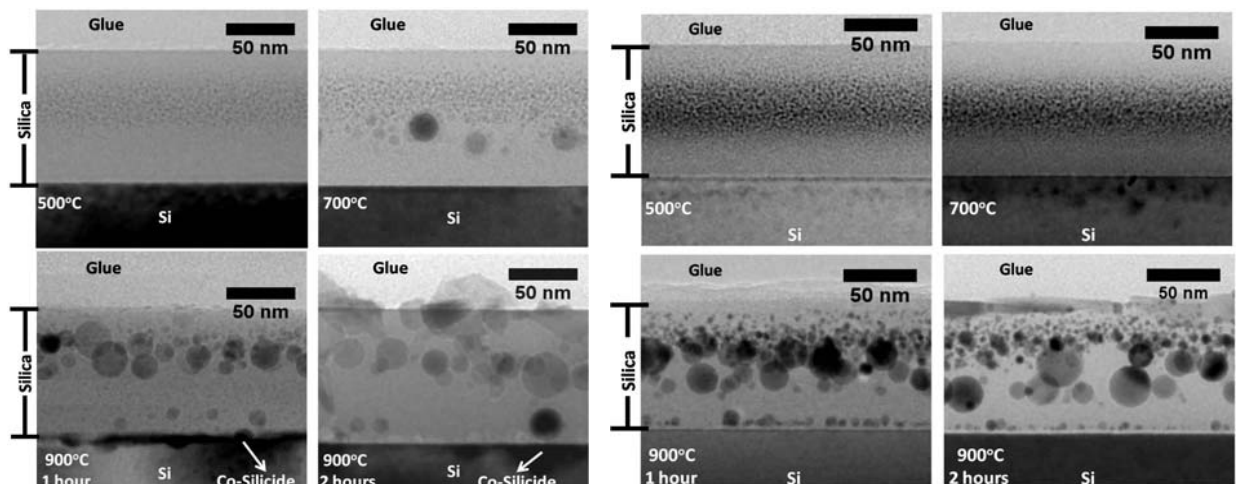

Fig. 1. TEM images of Co (left) and Co-Pt (right) implanted silica samples. Samples were annealed at 500°C, 700°C, 900°C (1hour) and 900°C (2 hours) respectively.

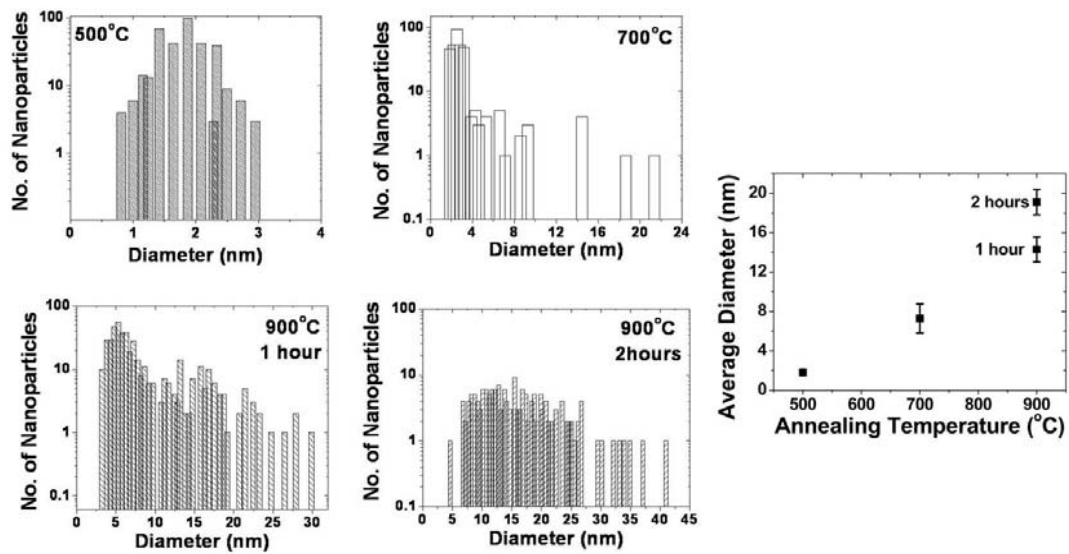

Fig. 2. Size distributions of the Co implanted silica samples annealed at 500°C, 700°C, 900°C (1hour) and 900°C (2 hours) respectively.

Pt has reduced the amount of Co diffusion, although some diffusion is still evident with Co oxide formation at the surface and $CoPt_3$ formation elsewhere indicating differential movement of Co and Pt.

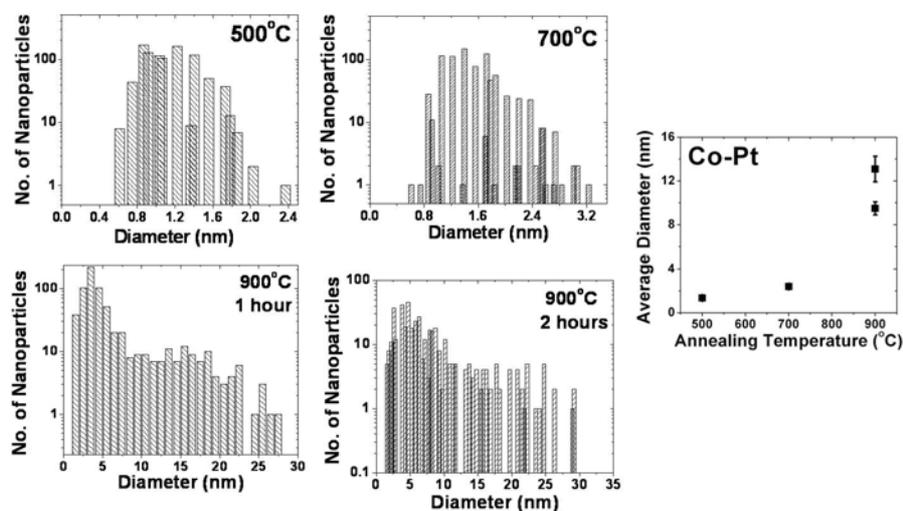

Fig. 3. Size distributions of the CoPt implanted silica samples annealed at 500°C, 700°C, 900°C (1hour) and 900°C (2 hours) respectively.

All the samples were also examined with grazing incident x-ray diffraction. The XRD on Co nanoparticles annealed at 500°C revealed a single hcp phase (the small nanoparticles give broad 101 and 200 peaks). At 700°C these peaks are narrower with the addition of a more intense peak at 44.2° corresponding to the fcc phase. As the annealing temperature increases, Co nanoparticles grow larger in size and the dominant phase changes from hcp to fcc. The 900°C sample shows only fcc (not hcp) peaks. Also there are indications of Co-silicide ($CoSi_2$) and Co oxide formation at 900°C. XRD for the alloy shows the same trend of narrower peaks for higher anneal temps corresponding to increased particle sizes. The main phase present is fct Co-Pt with a hint of $CoPt_3$ at 900°C and 2 hr. Formation of some Co oxide at the surface (seen via the TEM) explains the depletion of Co required for the latter.

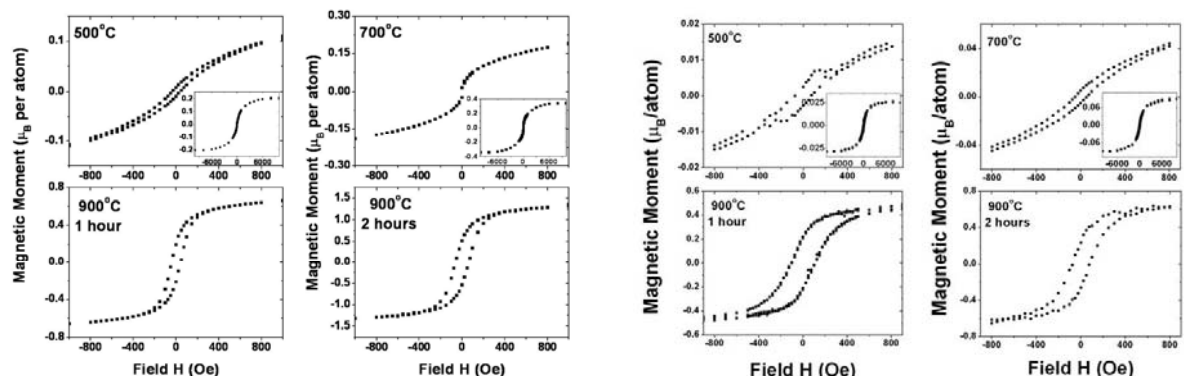

Fig. 4. Magnetisation measurements for Co (left) and Co-Pt (right) implanted silica samples.

SQUID magnetometry was used to show and characterise the magnetic behaviour of the nanoparticles. Hysteresis loops were collected at room temperature and up to ±10000 Oe. The resulting data for all the Co and CoPt cases are shown in Fig. 4, wherein the magnetic moment per atom was calculated based on the nominal implant fluence and the main curves are truncated at ±900 Oe (Cases where saturation is not achieved at this field have an inset). In general, the lower temperature anneals result in largely paramagnetic/ super paramagnetic responses to applied field while the 900°C anneals are largely ferromagnetic. Saturation and remanent magnetisation ($M_s$ and $M_R$) increase with anneal conditions (particle size) for both types of nanoparticles and values are plotted in Figs. 5 and 6 for Co and CoPt, respectively. The trend with coercivity ($H_c$) is also generally an increase with hotter anneal. Although there is a minimum at 700°C for the Co case due to initial formation of fcc Co which is magnetically softer than the hcp phase it replaces, and for Co-Pt, $H_c$ reduces for the longer anneal at 900°C which links to increased formation of nonmagnetic Co oxides and $CoPt_3$.

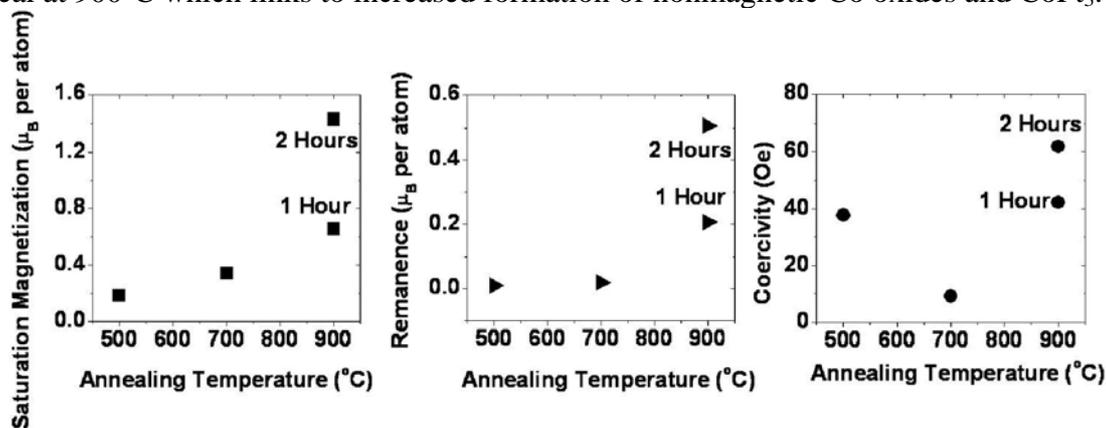

Fig. 5. Summary of $M_s$, $M_R$ and $H_c$ for the Co implanted silica samples with different anneal conditions.

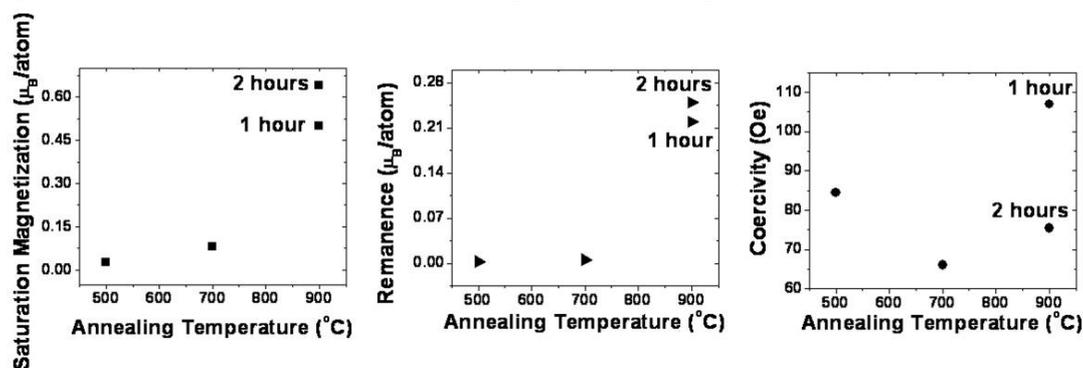

Fig. 6. Summary of $M_s$, $M_R$ and $H_c$ for the CoPt implanted silica samples with different anneal conditions.

### 4.  Conclusions:

Our results provide further evidence that ion implantation is a viable approach for the formation of patterned arrays of magnetic nanoparticles in a silica matrix as may be useful for magnetic storage and other applications. For the examples, Co and Co-Pt, explored here it is clear that the size and magnetic nature of the resulting particles depend critically on anneal conditions. The elemental Co nanoparticles produced the superior saturation magnetisation when annealed at 900°C for 2 hours. However, the Co-Pt alloy nanoparticles produce a higher coercivity, when annealed for 1 hour at 900°C.

# High Temperature Thermodynamics of the Multiferroic $Ni_3V_2O_8$

J. Oitmaa[a] and A. Brooks Harris[b]

[a] *School of Physics, The University of New South Wales, Sydney NSW 2052, Australia.*
[b] *Dept of Physics and Astronomy, University of Pennsylvania, Philadelphia PA 19104, U.S.A.*

We study a Heisenberg model proposed to describe the magnetic properties of the multiferroic material $Ni_3V_2O_8$. Using high-temperature expansions we compute the specific heat and zero-field susceptibility as functions of temperature. This will allow a comparison with experimental measurements.

1. **Introduction**

The nickel vanadate $Ni_3V_2O_8$ is a much studied material [1], with transitions at temperatures of 9.1K, 6.3K, 3.9K between various magnetically ordered phases, including a 'multiferroic' phase which shows simultaneous ferroelectric and magnetic order. The material itself is rather complex, with 6 magnetic $Ni^{++}$ $S=1$ ions per orthorhombic unit cell, forming a structure of coupled 'Kagome staircase' planes, shown in Fig. 1.

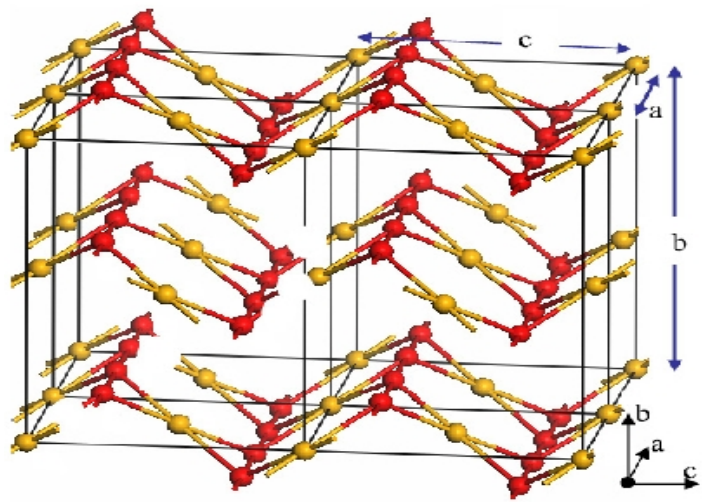

Fig. 1. Magnetic sites in $Ni_3V_2O_8$ showing the two types of Ni sites: 'spine' sites (dark circles) and 'cross-tie' sites (light circles). Figure reproduced from [1] with permission from IOP Publishing Copyright 2008.

An important feature of the magnetic interactions in this material is frustration, which is responsible for the complex sequence of ordered phases. The highest temperature ordered phase is incommensurate with the lattice, with moments predominantly on the spine sites. The ordering wavevector is largely determined by competition between nearest and next-nearest exchange interactions along the spines ($J_1$ and $J_2$ in Fig.2). The multiferroic phase sets in at a lower temperature 6.3K.

A first principles electronic structure calculation (LDA+U) [1] has identified as many as 12 different exchange constants, of which 5 appear to be dominant. These are shown as $J_1$, $J_2$, … $J_5$ in Figure 2. A set of proposed values (in meV), given in [1], is also shown. Of course, there is considerable uncertainty in these values.

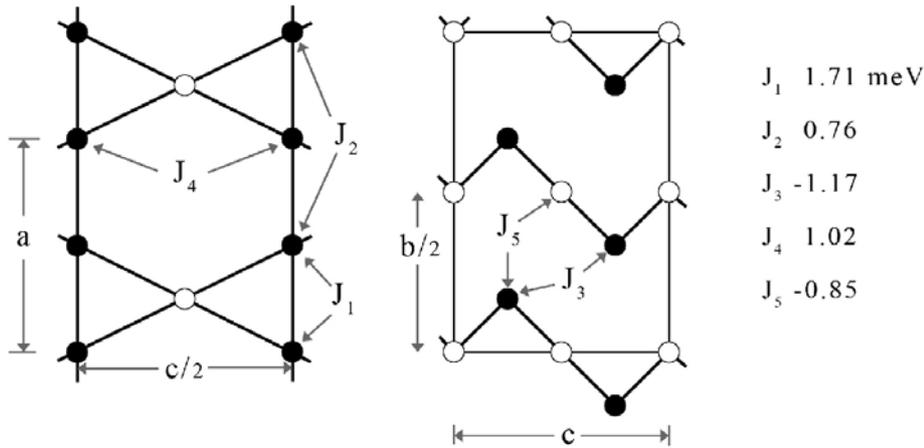

Fig.2. Dominant exchange interactions in $Ni_3V_2O_8$, and estimated values

The aim of the present work is to use a 5-parameter Heisenberg model and to calculate the specific heat and magnetic susceptibility in the paramagnetic phase, using high-temperature series expansions, and to compare with experimental data [2].

## 2. The calculation

We assume an isotropic Heisenberg model

$$H = \sum_{<ij>} J_{ij}\, \mathbf{S}_i \cdot \mathbf{S}_j - g\mu_B B \sum_i S_i^z$$

where the $S_i$ are spin-1 operators, the first summation is over all exchange coupled pairs, with $J_{ij}$ one of the five possible exchange parameters, and the final term is the interaction with an external magnetic field $B$. We derive high temperature expansions for the model by a standard linked-cluster method [3]. This approach has a long history, going back to the 1950's. However, we know of no previous such work with so many different exchange constants.

The expansions for the logarithm of the partition function $Z$ per site, and for the zero-field susceptibility, take the form

$$\ln Z/N = \ln 3 + \sum_r a_r \beta^r$$

$$k_B T \chi / (g\mu_B)^2 = 2/3 + \sum_r c_r \beta^r$$

where $\beta = 1/k_B T$ is the expansion variable and the $a_r$, $c_r$ are numerical coefficients whose values depend on the $J$'s. The first few coefficients are, explicitly

$$a_2 = 4\,(J_1^2 + J_2^2 + J_3^2 + J_4^2 + 2J_5^2) / 9$$

$$a_3 = 2\,(J_1^3 + J_2^3 + J_3^3 + J_4^3 + 2J_5^3 - 8J_1^2 J_2 - 8J_1 J_5^2) / 27$$

$$c_1 = -16\,(J_1 + J_2 + J_3 + J_4 + 2J_5) / 27$$

$$c_2 = 20\,(J_1^2 + J_2^2 + J_3^2 + J_4^2) / 81 + 104 J_5^2 / 81 + 128\,(J_1 J_2 + J_1 J_3 + \ldots + J_4 J_5) / 81$$

We have carried out the calculation through 10th order for the zero-field free energy and through 7th order for the susceptibility. This involves 72577 and 364916 distinct clusters with five possible bond types, which are embeddable on the $Ni_3V_2O_8$ lattice. The resulting series can then be analysed by standard Padé approximant methods.

## 3. Results

In Table 1 we show the coefficients of the specific heat and susceptibility series for the particular $J$ values given in Fig. 2.

Table 1. Series Coefficients for Specific Heat and Susceptibility

| $N$ | $C/k_B$ | $\chi$ |
|---|---|---|
| 0 | $0.653866666667 \times 10^1$ | $0.666666666667 \times 10^0$ |
| 1 | $-0.106630202444 \times 10^2$ | $-0.367407400740 \times 10^0$ |
| 2 | $0.938305051214 \times 10^0$ | $-0.114674567901 \times 10^1$ |
| 3 | $-0.407665087950 \times 10^2$ | $0.503738067490 \times 10^1$ |
| 4 | $0.480120021243 \times 10^3$ | $-0.143967012941 \times 10^2$ |
| 5 | $-0.228030693775 \times 10^4$ | $0.375923499903 \times 10^2$ |
| 6 | $0.861441411583 \times 10^4$ | $-0.845223517854 \times 10^2$ |
| 7 | $-0.342195552001 \times 10^5$ | $0.173902466284 \times 10^3$ |
| 8 | $0.147719535268 \times 10^6$ | |

As is apparent, the series terms alternate in sign. This indicates a dominant singularity on the negative $\beta$ axis, and makes the analysis somewhat problematic. Indeed, it is not possible to determine the position of the critical point with any accuracy. However at higher temperatures (smaller $\beta$) both quantities can be evaluated from Padé approximants, with a fair degree of precision. The results are shown in Fig. 3.

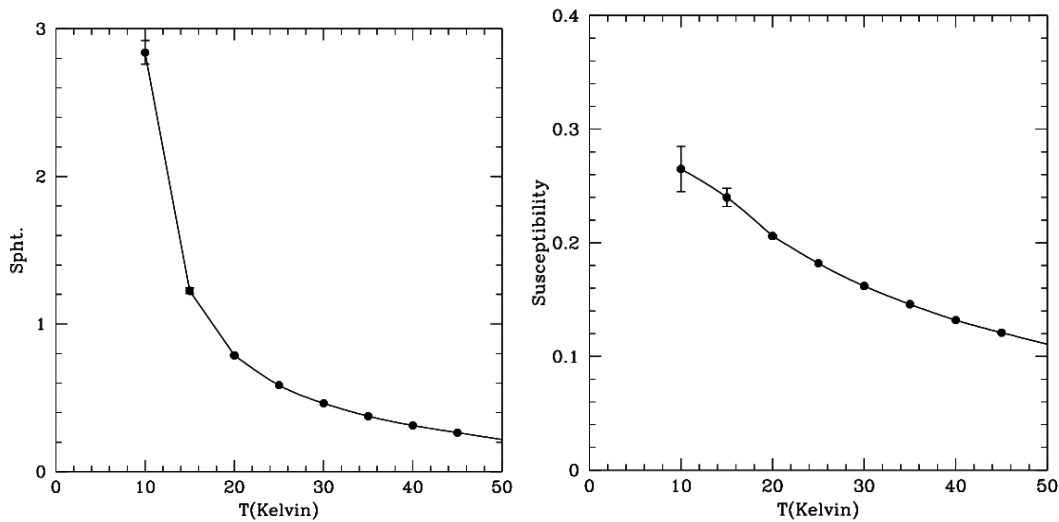

Fig.3. Specific heat and susceptibility for $Ni_3V_2O_8$ versus temperature, obtained from the 5-parameter Heisenberg model. The data points are averages of high-order Padé approximants, and the error bars indicate confidence limits, based on the degree of consistency between different approximants.

**4.  Discussion**

We have computed the high-temperature specific heat and magnetic susceptibility of a 5-parameter Heisenberg model, believed to apply to the multiferroic material $Ni_3V_2O_8$. The specific heat shows evidence of a divergence around 10K, in good agreement with the actual transition temperature of 9.1K in $Ni_3V_2O_8$. We are not able to estimate the transition temperature of the model directly, as the specific heat series is not well suited to this. To make such an estimate would require a calculation of the magnetic structure factor. Our goal remains to compare our results in detail with the experimental data, and to attempt to constrain the values of the exchange parameters. This remains work in progress.

Our approach is applicable to the high temperature paramagnetic phase, and is able to probe the highest temperature transition, but is, of course, unable to probe the lower temperature transitions or ordered phases. This is, theoretically, a very difficult problem. An isostructural material $Co_3V_2O_8$ also shows interesting magnetic properties, and a similar study could be undertaken for that case.

**Acknowledgments**

We are grateful for the computing resources provided by the Australian Partnership for Advanced Computing (APAC) National Facility.

# Does the Quantum Compass Model in Three Dimensions have a Phase Transition?

J. Oitmaa and C.J. Hamer

*School of Physics, The University of New South Wales, Sydney, NSW 2052, Australia.*

We investigate the existence of a phase transition to a low-temperature orientational or nematic phase in the Quantum Compass model on the simple cubic lattice. We conjecture, on the basis of our results, that there is no such finite temperature transition. This is in contrast to the square lattice, where such a transition has been found.

## 1. Introduction

'Quantum Compass' models are spin models in which the nearest-neighbour exchange coupling has the form $J^\alpha S_i^\alpha S_j^\alpha$ where $\alpha = (x,y,z)$ depends on the direction of the particular link or bond, as shown in Fig. 1. This then implies a coupling between the spin space and the physical space of the lattice. Such models were first introduced and used to describe orbital ordering in transition metal compounds [1]. They are also of interest in quantum information theory, as models of $p + ip$ superconducting arrays [2].

In the present work we consider the isotropic (all $J$'s equal) spin-½ model on the square and simple cubic lattices, and examine the existence of a phase transition between the high-temperature disordered phase and a low-temperature phase with orientational or 'nematic' order. Such a transition has been found for the square lattice, using Quantum Monte Carlo methods [3] but there has been no previous study of the three-dimensional (3D) case.

We employ a standard method of high-temperature series expansions [4], where long series are computed in powers of $x = J/k_B T$ for thermodynamic quantities, and analysed via Padé approximant and other methods to identify a singular point $x_c$ corresponding to the expected phase transition. Special features of this model allow rather long series to be obtained (up to $x^{20}$ for the 3D case). Details of the calculation are given in a forthcoming paper [5].

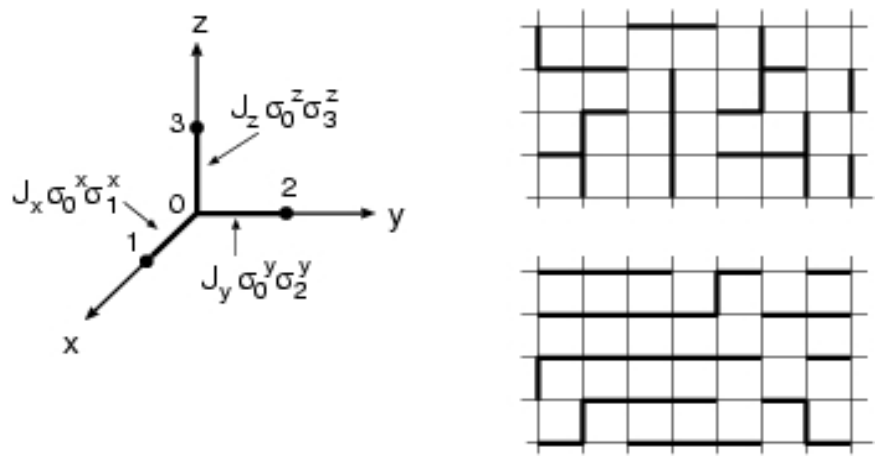

Fig. 1. (Left) The exchange couplings of the model; (Right) A schematic of the orientational or 'nematic' phase. The solid bonds have lower than average energy.

## 2. Method and Results

The Hamiltonian of the model, in standard notation, is

$$H = J_x \sum_{<ij>}^{(x)} \sigma_i^x \sigma_j^x + J_y \sum_{<ij>}^{(y)} \sigma_i^y \sigma_j^y + J_z \sum_{<ij>}^{(z)} \sigma_i^z \sigma_j^z$$

where the $\sigma_i^\alpha$ are Pauli spin operators ($\alpha = x,y,z$ ; $\sigma_i^\alpha = 2S_i^\alpha$ ). By expanding the Boltzmann factor $e^{-\beta H}$ in powers of $\beta = 1/k_B T$, and evaluating the resulting traces of spin operators, we obtain a conventional high-temperature expansion for the free energy

$$-\beta f = \ln 2 + \sum_r a_r (J_x, J_y, J_z) \beta^r$$

From the free energy expansion we can compute the specific heat. However, this is usually not the best quantity for identifying a phase transition, as it has only a weak singularity. For magnetic phase transitions the existence and/or location of a critical point is usually determined from the susceptibility, which has a much stronger singularity. In the present model, including a field term in the Hamiltonian $H = H_0 - hD$ where $D$ is the nematic order parameter:

$$D = 2J_z \sum_{<ij>}^{(z)} \sigma_i^z \sigma_j^z - J_x \sum_{<ij>}^{(x)} \sigma_i^x \sigma_j^x - J_y \sum_{<ij>}^{(y)} \sigma_i^y \sigma_j^y$$

We obtain a similar high-temperature expansion for the generalized susceptibility

$$\beta\chi = \lim_{h \to 0} \partial^2/\partial h^2 (\ln Z/N) = \sum_{r=2}^{\infty} c_r (J_x, J_y, J_z) \beta^r$$

We have derived the series to order $\beta^{16}$ for the isotropic model (all $J$'s equal). The series for the susceptibility, which contains only even powers of $\beta$, is:

$\beta\chi$ = 6 − $6\beta^2$ + $20\beta^4$ − 81.0476190476... $\beta^6$ + 367.149206349... $\beta^8$ − 1787.51576719... $\beta^{10}$ + 9155.75874989... $\beta^{12}$ − 48688.4786086... $\beta^{14}$ + 266451.000791... $\beta^{16}$ + ...

As is apparent, the terms are of alternating sign, indicating that the dominant singularity lies on the negative $\beta^2$ i.e., the imaginary $\beta$ axis. Thus we need to use Padé approximants or similar methods to analyse the series beyond its radius of convergence. The series have proved difficult to analyse, because of the complex singularity structure. However, for the two-dimensional (2D) case we do find a critical point, consistent with the Monte Carlo result [3]. On the other hand, we find no evidence for a critical point in the 3D case.

The difference is seen more strikingly in Fig. 2, which shows estimates of the inverse susceptibility, obtained from Padé approximants, versus temperature $T$.

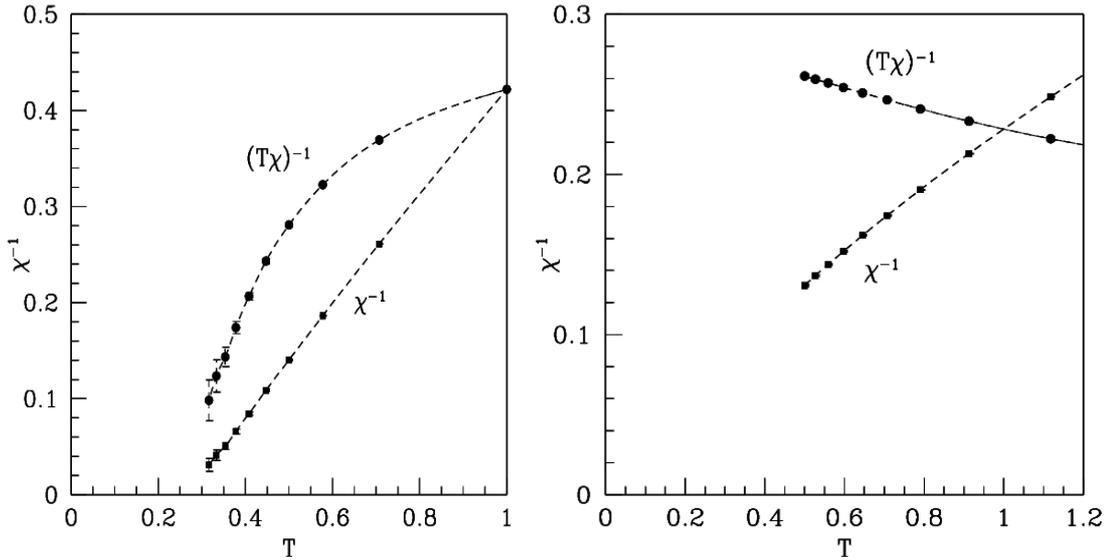

Fig.2. Inverse susceptibility versus temperature for the (a) 2D and (b) 3D models. The lines are guides for the eye.

In the 2D case the inverse susceptibility clearly approaches zero, heralding a phase transition, at a temperature $k_B T/J \sim 0.25$. This is consistent with, albeit less precise than the Quantum Monte Carlo result [3]. This clearly demonstrates that the method works. On the other hand, for the 3D case, $\chi^{-1}$ appears to approach zero, if at all, very close to or at $T = 0$. The $(\beta\chi)^{-1}$ points are, in fact, monotonically increasing, indicating that $\chi$ is increasing less rapidly than $1/T$ with decreasing temperature.

## 3.  Discussion

The question of the existence of a thermodynamic phase transition in the Quantum Compass model in 3D models remains an important and open question. The present work is, to our knowledge, the first attempt to address this problem using the technique of high-temperature series expansions, a well established method in other contexts.

We have found a striking qualitative difference between the previously studied 2D Quantum Compass model, which has been shown to have a phase transition to a low-temperature phase with orientational or nematic order, and the 3D model. Our results lead us to conjecture that there is no finite $T$ transition in the 3D model. At first sight this seems surprising, since for normal magnetic phase transitions an increase in spatial dimensionality reduces the effect of thermal fluctuations and thereby leads to an increase in critical temperature. In the present model the bond interactions along different spatial directions compete with each other in 'pulling' the spins in different directions, and the additional direction in going from two- to three dimensions leads to an increase in the tendency to disorder.

The series have proved difficult to analyse, as they are dominated by singularities on the imaginary temperature axis. This is, perhaps, a reflection of the peculiar 'one-dimensional' nature of the couplings in the model. Further insight into this complex singularity structure may suggest other, more effective, ways of analysing the series.

**Acknowledgments**
We are grateful for the computing resources provided by the Australian Partnership for Advanced Computing (APAC) National Facility.

# Scaling of Critical Temperature and Ground State Magnetization near a Quantum Phase Transition

J. Oitmaa and O.P. Sushkov

*School of Physics, The University of New South Wales, Sydney NSW 2052, Australia.*

We investigate a simple model antiferromagnet which shows a Quantum Phase Transition between a conventional Néel antiferromagnetic phase and a dimerized 'Valence Bond Solid' phase. Both the critical temperature and the ground state magnetization in the Néel phase approach zero at the critical point, and are found to scale according to power laws, with exponents 0.5, as expected from general considerations.

1.  **Introduction**

    The subject of Quantum Phase Transitions (QPT's) in condensed matter systems remains a frontier area of research [1]. These are transitions, at temperature $T = 0$, in the nature of the ground state of a strongly correlated quantum system. At QPT's large quantum fluctuations play the same role as do large thermal fluctuations at normal phase transitions, and analogous universal scaling laws are to be expected. In real materials QPT's can be induced by pressure, by applied magnetic fields, or by disorder.

    Theoretical studies of QPT's are largely based on various simplified models, particularly low dimensional antiferromagnets, where the system can be tuned through a QPT by varying a particular parameter g in the Hamiltonian. Most of the models hereto studied have been two-dimensional. Examples include antiferromagnets with strong and weak bonds, with or without frustration, and bilayer systems, where the QPT separates a conventional Néel antiferromagnetic phase from a dimerized phase with only short-range correlations and no magnetic order. In such systems the magnetic order present in the ground state does not extend to finite temperatures.

    In the present work we study a three-dimensional model where magnetic order persists to some critical temperature $T_c(g)$. We expect $T_c(g)$ to vanish as $g \rightarrow g_c$, and we are interested in comparing the vanishing of $T_c$ with the vanishing of the ground state magnetization $M_0$. The magnetization is expected to vanish as $(g_c - g)^\beta$, where the critical exponent $\beta$ is expected to have the value 0.5, corresponding to a classical thermally driven transition in four spatial dimensions, which is mean-field like (with possible logarithmic corrections). Does $T_c(g)$ scale with $(g_c - g)$ in the same way?

2.  **The model and results**

    Our model, shown in Fig. 1(a), is a spin-½ tetragonal antiferromagnet with bonds of strength $J$ and $gJ$. For $g = 1$ we have an isotropic cubic antiferromagnet, which will have reduced Néel order in the ground state ($M_0 = 0.42$) and a critical temperature $k_B T_c/J = 1.89$. On the other hand, for $g \gg 1$, the strong bonds will form spin-singlet dimers, leading to a phase with short-range correlations and no magnetic order. A QPT will separate these two phases, as shown schematically in Fig. 1(b).

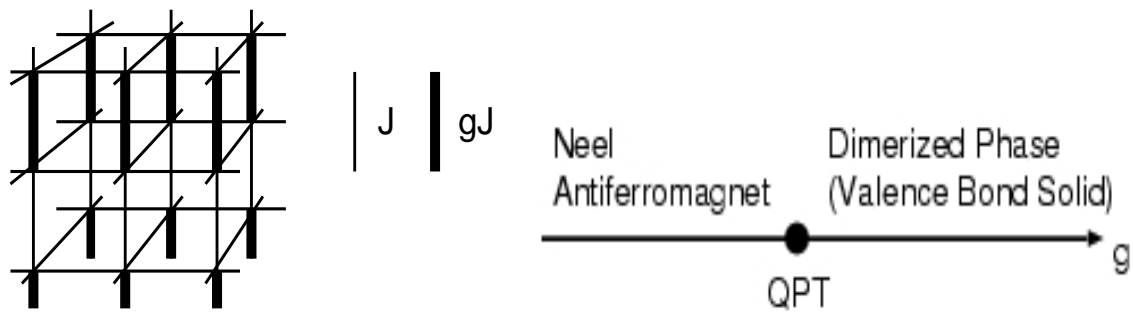

Fig. 1. (a) The model, with thin lines denoting *J* bonds and thick lines denoting *gJ* bonds;
(b) Schematic phase diagram for the model at temperature $T = 0$.

This model has been studied previously [2], in connection with magnetic-field induced QPT's, and the critical point located at $g_c = 4.013 \pm 0.003$ (in our units), using Quantum Monte Carlo (QMC) methods. However, the critical temperature in the Néel phase has not, to our knowledge, been previously studied.

Our calculations are based on series expansion methods [3], and involve a number of separate parts:

- In the ground state we compute the energy, starting from both Néel and dimer limits. The crossing point will determine the position of the QPT. The energy curves are shown in Fig. 2. As is apparent, the two curves meet smoothly, as expected for a second-order QPT, or Quantum Critical Point (QCP). It is not possible to locate $g_c$ accurately from this crossing.
- We compute the magnetization at $T = 0$, in the Néel phase. This is also shown in Fig. 2. As is again apparent, the magnetization decreases rather sharply to zero near $g \sim 4.0$. Although the error bars become rather large near the QCP, from a Padé approximant analysis we estimate the critical point at $g_c = 4.05 \pm 0.05$, consistent with, but less precise than the QMC result. The merging of the energy curves is consistent with this value.
- Finally we compute high-temperature expansions for the Néel susceptibility. This is not the physical susceptibility, but the response to a 'staggered' field. This susceptibility is expected to have a strong divergence at the critical point, and can be used to estimate the critical temperature $T_c(g)$. This curve is also shown in Fig. 2. It is clear that the critical temperature also drops sharply to zero at the QCP. However, these series are rather short and, consequently, the error bars near the QCP large.

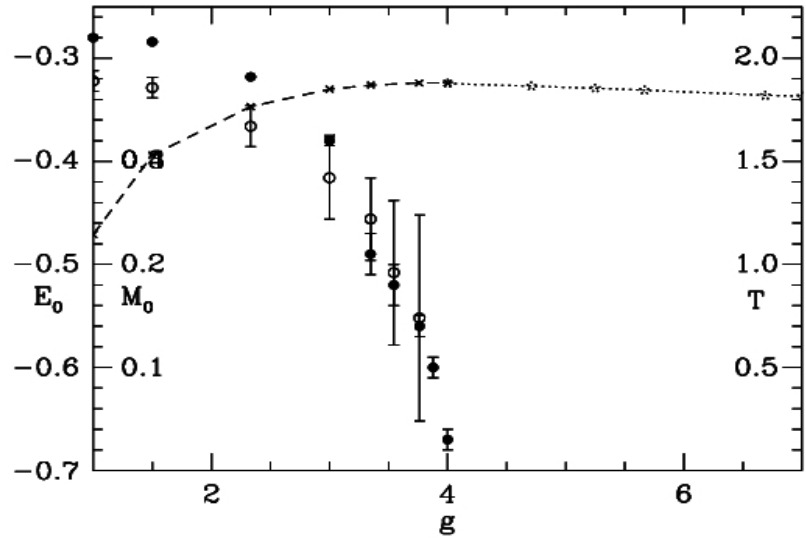

Fig. 2. Variation of the ground state energy $E_0$ in the Néel and dimer phases (dashed and dotted lines – left scale), the ground state magnetization $M_0$ in the Néel phase (solid circles – inner left scale), and the critical temperature $T_c$ (open circles – inner right scale) with the tuning parameter $g$. The 'temperature' $T$ plotted is the dimensionless quantity $k_B T_c / J_{av}$.

To investigate the scaling behaviour of $M_0$ and $T_c$ in more detail, we first plot $M_0^2$ versus $g$. This is shown in Fig. 3 (left). Within the error limits the points lie on a straight line, confirming the scaling law $M_0 \sim (g_c - g)^{0.5}$, and yielding the more precise estimate $g_c = 4.02 \pm 0.02$. In Fig. 3 (right) we show log-log plots of both $M_0$ and $T_c$ versus $(g_c - g)$. From the figure we see that the $M_0$ data fall well on a straight line with slope 0.5, consistent with the plot in Fig. 3 (left). It is not possible to estimate $T_c$ as close to the QCP, and consequently there are fewer points and larger error bars. However, the points are also consistent with a power law, with the same exponent 0.5.

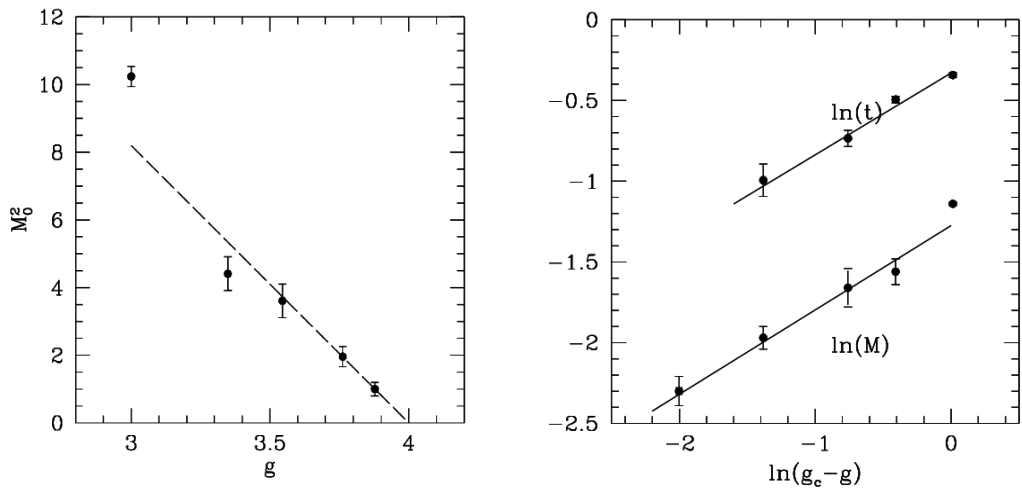

Fig. 3. (Left) Plot of $M_0^2$ versus $g$; (Right) Log – log plots of magnetization $M_0$ and critical temperature $T_c$ versus $(g_c - g)$. The solid lines have slope 0.5.

### 3. Discussion

We have studied a three-dimensional spin model which has a Quantum Phase Transition between a conventional Néel ordered antiferromagnetic phase and a dimerized 'valence bond solid' phase which has no long-range magnetic order. The quantum critical point is located at $g_c = 4.01 \pm 0.01$. The ground state magnetization in the Néel phase is found to vanish as $g \to g_c$ with a power law with exponent ½, as expected from general considerations. The critical temperature also vanishes as $g \to g_c$, and, within the limited precision of our results, appears to follow a power law with the same exponent ½.

Our model is not directly applicable to any real physical system. However, we believe that the results are universal. Recent work [4] on the material TlCuCl$_3$, where the QPT is induced by pressure, shows precisely the same kind of behaviour.

**Acknowledgments**

We are grateful for the computer resources provided by the Australian Partnership for Advanced Computing (APAC) National Facility.

# Ultra-high Molecular Weight Polyethylene Prosthetic Wear Debris Transport in a Wheel-on-Plate Model System

J.A. Warner[a], L.G. Gladkis[a], A. E. Kiss[a], J. Young[b], P. N. Smith[c], J. Scarvell[c], C. C. O'Brien[d] and H. Timmers[a]

[a] *School of Physical, Environmental and Mathematical Sciences, The University of New South Wales, Canberra Campus, ACT 2600, Australia.*
[b] *School of Engineering and Information Technology, The University of New South Wales, Canberra Campus, ACT 2600, Australia.*
[c] *Trauma and Orthopaedic Research Unit, The Canberra Hospital, PO BOX 11, Woden, ACT 2606, Australia.*
[d] *Graduate School of Biomedical Engineering, University of New South Wales, Sydney NSW 2052, Australia.*

The transport pathways of ultra-high molecular weight polyethylene wear debris particles in orthopedic prostheses are of interest due to common prosthesis failure by osteolysis. Such debris transport has been studied in a wheel-on-plate model system which isolates the sliding wear mode. For the model system radioisotope tracing has demonstrated the existence of two-way debris transfer between the wheel rim and the polyethylene plate with only little debris particles being dispersed in the water lubricant. It is likely that this dispersion commences where the wheel rim exits the wear region. Dye tracing has elucidated the existence of a lubricant vortex at this point and also where the wheel rim enters the lubricant. These vortices can effectively distribute the particles. Furthermore debris pathways alongside and underneath the wheel are suggested from which polyethylene debris may be re-introduced to wear processes as third body particles. Dye tracing results have been confirmed by computational fluid dynamics simulations.

1. ## Introduction

Microscopic wear debris particles generated from Ultra-High Molecular Weight Poly-Ethylene (UHMWPE) tibial inserts in knee or hip prostheses have been shown to induce harmful biological responses ultimately leading to osteolysis [1]. The pathways of such polyethylene debris, connecting its tribological origin at the wear interfaces (Fig. 1a) with typical sites of osteolysis, such as the contact surfaces of prosthesis and bone, are not known. In addition some debris pathways may result in the reintegration of debris in the articulating wear surfaces acting there as third body particles. This may accelerate wear [2]. On the other hand reintegration of debris particles into the prosthesis removes them from pathways leading to sites of osteolysis.

The identification and quantification of debris particle transport and its pathways in prostheses may aid the understanding of the origins of osteolysis in joint replacement patients. It also has the potential to inform advanced prosthesis designs. Considering the complexity of the tribology and the fluid dynamics of the lubricant in actual prostheses, this work has focused on a model system as a first approach to the problem. A wheel-on-plate model system which combines a rotating steel wheel with a UHMWPE plate has been devised to isolate the sliding wear mode of prostheses. The model system has been studied experimentally with radioisotope tracing and dye tracing, as well as simulated theoretically with computational fluid dynamics (CFD). The radioisotope tracing results have previously been published [3,4]. Here we report on the dye tracing experiments and the fluid dynamics simulations.

It may be noted that while the model system has some similarities with actual prostheses, such as the metal-polyethylene interface and the dominance of sliding wear, pronounced differences also exist. Prostheses feature multi-directional articulation, whereas in the model system the articulation is one-directional. In the system chosen the rim surface area of the steel wheel was an order of magnitude larger than the contact area on the UHMWPE plate. In actual prostheses the surface areas of the articulating parts are in contrast very similar. Moreover instead of synovial fluid or bovine serum, water was used as lubricant in the model system. Any extrapolation of the results presented here to the situation in actual prostheses has to be therefore made with care. However, the model system has potential to gradually be refined to provide a better approximation and give more accurate insight.

2. **Experimental and Simulation Details**

The wheel-on-plate model system studied [3] (see Fig.s 1b, 2a and 2b) consists of a stainless steel wheel which rotates under load with its rim being in articulation with a flat UHMWPE plate. The plate dimensions are 10 mm ×10 mm × 3 mm. The wheel radius is 25 mm, the rim width is 10 mm and the rotational frequency chosen in these experiments was 0.125 Hz. The static vertical loads were 1 kg, 3 kg and 5 kg. De-ionized water lubrication was used, however, other lubricants, such as bovine serum would be possible in future work.

In the radioisotope tracing experiments several polymer plates were ion-implanted with the $^{111}$In/$^{111}$Cd radioisotope tracer at a depth of 200 nm [3,4]. Two characteristic gamma-ray lines of this tracer were subsequently used to follow and quantify UHMWPE wear debris in the model system as the sliding wear progressed.

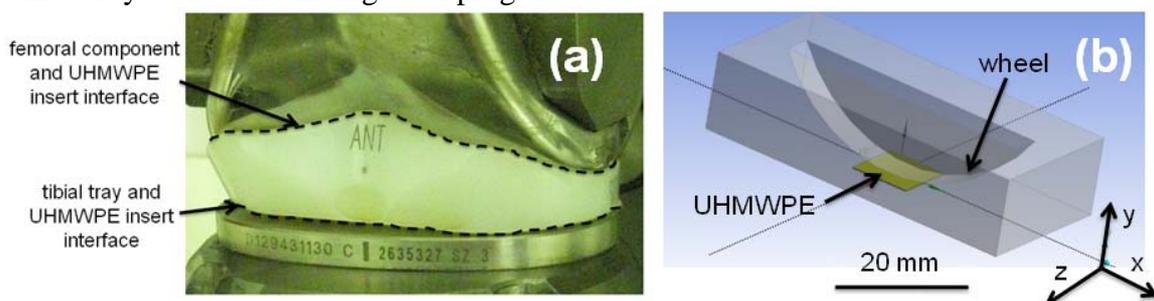

Fig. 1. (a) Close-up of the femoral component of a knee prosthesis sitting on top of the UHMWPE insert, itself supported by the tibial tray underneath. The dashed lines indicate the two separate wear interfaces. (b) Cross-sectional drawing of the geometry of the wheel-on-plate model system as applied in the CFD simulations. The wheel (dark grey) rotates clockwise in the water lubricant (light grey).

In the dye tracing experiments a drop of ink was introduced at a certain location in the water lubricant. Immediately afterwards the rotational motion of the wheel was commenced and the establishment of flow patterns was observed. The visualization of the lubricant flow via the dye was video-recorded. The Computational Fluid Dynamics (CFD) simulations of the model system were performed with the commercial package *ANSYS Fluent 12.1*. [5]. The parameterisation used is shown in Fig. 1b. The sides and the bottom of the lubricant enclosure were modeled as stationary walls, the wheel as a moving wall, and the free surface of the lubricant was not directly simulated but rather modeled as a shear-free wall. A symmetry plane was employed (the *x-y* plane in Fig. 1b) to reduce the size of the computational domain. The mesh consisted of 1.1 million cells, with prism boundary layer cells extending from the wheel and walls, and unstructured polyhedral cells in the remainder of the domain. Using the *ANSYS Fluent* software the unsteady incompressible Navier-Stokes equations were solved, with second order upwind spatial discretization, second-order implicit time stepping, and the SIMPLE algorithm for pressure-velocity coupling. The flow was assumed to be fully laminar. The tangential wheel velocity was 19.6 mm/s consistent with the rotational velocity employed in the experiments.

## 3.    Results

Under load, the sliding wear of the rotating steel wheel of the model system gradually works a cavity into the UHMWPE plate which closely reflects the geometry of the wheel. Following experimentation polyethylene debris can be seen on the rim of the steel wheel. Atomic force microscopy has confirmed that the morphology of the rim surface permits such debris to be accommodated. The scanning electron micrograph in Fig. 2 shows that on the polyethylene plate debris piles up where the wheel exits the region of articulation [3].

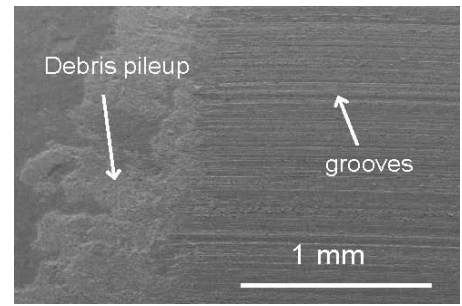

The radioisotope tracing experiments [3,4] showed that UHMWPE wear debris is transferred from the polyethylene plate to the wheel rim and vice versa. Thus a two-way pathway exists which integrates and re-releases polymer debris particles in both wear surfaces. The continued presence of these particles at the wear interface may result in $3^{rd}$ body wear processes. The radioisotope tracing indicated that the two-way transfer pathway dominates debris transport since typically only less than 1% of radioactively labelled debris was dispersed into the water lubricant at the end of the experiment. This dominance is likely due to the fact that in the model system the surface area of the articulating rim is relatively large. However, debris particle dispersion in the lubricant has been observed to continue with time [3]. It may also be expected to play a greater role in actual prostheses, where the polished surface of the metal component has less capacity to accommodate polyethylene debris particles.

Fig. 2. SEM image of the region of the UHMWPE plate where the steel wheel exits the wear cavity recorded after actuation.

The micrograph in Fig. 2 suggests that most of the debris dispersion in the lubricant commences near the exit of the wheel from the wear cavity. Possible pathways of debris particles released to the lubricant at this location have been studied with dye tracing. Figure 3a displays a photograph of the dye tracing experiment 3 secs after the initiation of the wheel motion. It is apparent that a vortex has formed in the lubricant in front of the wheel. Polyethylene wear debris exiting the wear cavity here can become caught in this vortex and thus may be expected to be uniformly distributed in the lubricant.

Figure 3b indicates that lubricant flows backward alongside the wheel with some flow directed towards the point where the rim of the wheel enters the lubricant. Thus a flow mechanism exists which may return dispersed debris particles to the wear process (Figure 3).

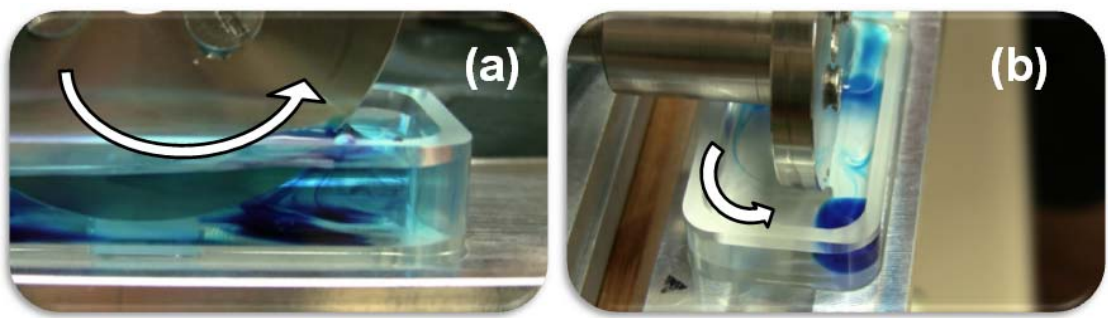

Fig. 3. Photographs of the dye tracing experiments 3 seconds after wheel motion commenced. The arrow indicates the motion of the wheel. (a) Vorticity in the lubricant where the rim of the wheel exits the lubricant. (b) Evidence for flow towards the region where the rim of the wheel enters the lubricant.

Figure 4 shows typical results from the CFD simulations. In Fig. 4a twenty velocity streamlines have been plotted for a time period of 2 seconds. The figure shows a side view with the wheel rotating clockwise. The vortices observed with dye tracing are evident for both the rim exit and the rim entry to the lubricant.

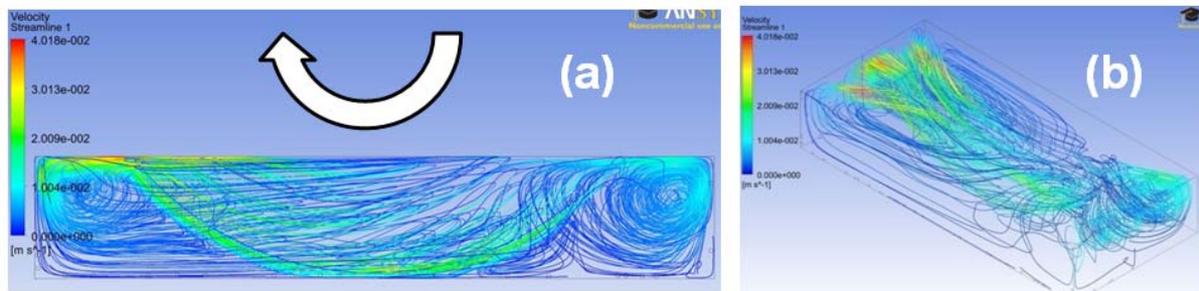

Fig. 4. CFD simulations for the wheel-on-plate model system. (a) Side view of the velocity streamlines, with arrow indicating wheel motion. (b) Isometric view. For both panels, wheel rim exits the lubricant on the left.

Figure 4b shows that the streamlines connect the vortex at the exit of the wheel rim alongside the wheel with the vortex where the rim enters the lubricant. Importantly, this flow continues underneath the wheel passing the wear cavity. Further on at high velocity the flow then reconnects with the vortex at the exit of the rim of the wheel. Thus a continuous flow pattern exists which has the capacity to return dispersed wear particles to the wear interface. Both the dye experiments and the increased flow velocity in the CFD simulations indicate that the vortex near the exit of the wheel rim from the lubricant is more significant. Due to the similar densities of UHMWPE and water (0.93 g/cm$^3$ and 1.00 g/cm$^3$), it is likely that polyethylene debris particles will follow the streamlines.

## 4. Conclusions

The combined information from radioisotope tracing, dye tracing and CFD simulations has identified wear debris transport pathways in a wheel-on-plate model system. The results may inform on similar transport mechanisms in orthopedic prostheses. In the model system polyethylene wear debris particles are transferred between the articulating steel and UHMWPE surfaces. Debris dropping out of this two-way transfer is likely dispersed in the lubricant where the rim of the wheel exits the fluid. Fast flow and a strong vortex at this location may uniformly distribute debris particles in the lubricant. Particles can be transported via a flow alongside the wheel. Importantly a lubricant vortex where the rim of the wheel enters the lubricant may return wear debris to the wear interface as 3$^{rd}$ body particles.

The results suggest that wear particles produced in an orthopedic prosthesis are returned to the wear interface and may undergo a number of wear interactions between the articulating surfaces before eventually they may be dispersed well away from the wear interactions with some of them on pathways towards regions of bioactivity. Further work with this model system may be performed with bovine serum, better representing the actual synovial fluid, and by considering the effects on the flow pattern of time-dependent loading and of reciprocating motion with directional changes.